\documentclass[%
aip,
amsmath,amssymb,
reprint,nofootinbib%
]{revtex4-2}
\usepackage{graphicx,array,booktabs,rotating}
\usepackage{dcolumn}
\usepackage{bm}
\usepackage{mathptmx}
\usepackage{multirow}
\usepackage{afterpage,float,color,xcolor}
\usepackage{hyperref}
\usepackage{etoolbox} 
\usepackage{tabularx,tabulary}
\usepackage[capitalize]{cleveref}
\usepackage{times}
\hypersetup{
	colorlinks,
	linkcolor={blue!100!black},
	citecolor={blue!100!black},
	urlcolor={blue!100!black}
}
\newcommand{\PRLsep}{\noindent\makebox[\linewidth]{\resizebox{0.3333\linewidth}{1pt}{$\bullet$}}\bigskip}
\makeatletter
\newcommand\footnoteref[1]{\protected@xdef\@thefnmark{\ref{#1}}\@footnotemark}
\makeatother

\newcommand{\etal}{\textit{et al.}}
\usepackage[page]{totalcount}
\usepackage{etoolbox,fancyhdr,xcolor}%
\usepackage[displaymath,pagewise]{lineno}

\begin{document}
\preprint{AIP/123-QED}
\title[\textbf{Journal to be decided} (2022) $|$  Manuscript under preparation]{Supersonic flow unsteadiness induced by control surface deflections}

\author{S. K. Karthick}%
\email{skkarthick@ymail.com}
\affiliation{ 
Faculty of Aerospace Engineering, Technion-Israel Institute of Technology, Haifa-3200003, Israel}

\author{D. Bhelave}%
\affiliation{ 
Department of Aerospace Engineering, Indian Institute of Technology, Kanpur-208016, India}
\affiliation{Currently at Bhabha Atomic Research Centre, Trombay, Mumbai-400085, India}

\author{A. De}%
\affiliation{ 
Department of Aerospace Engineering, Indian Institute of Technology, Kanpur-208016, India
}%

\date{\today}

\begin{abstract} 
Control surface deployment in a supersonic flow has many applications, including flow control, mixing, and body-force regulation. The extent of control surface deflections introduces varying flow unsteadiness. The resulting fluid dynamics influence the downstream flow characteristics and fluid-structure interactions severely. In order to understand the gas dynamics, an axisymmetric cylindrical body with a sharp-tip cone at zero angles of attack ($\alpha=0^\circ$) is examined in a free stream Mach number of $M_\infty=2.0$ and Reynolds number of $Re_{D}=2.16 \times 10^6$ ($D=50$ mm). Four static control surface deflection angles ($\theta = \pi/36,\pi/6,\pi/3,\pi/2$, rad) are considered around the base body. The cases are computationally investigated through a commercial flow solver adopting a two-dimensional detached eddy simulation (DES) strategy. Recirculation bubble length, drag coefficient's variation, wall-static pressure statistics, acoustic loading on the model and the surroundings, $x-t$ trajectory and $x-f$ spectral analysis, pressure fluctuation's correlation coefficient on the model, and modal analysis are obtained to understand the flow unsteadiness. At $\theta = [\pi/36]$, the wall-static pressure fluctuations behind the control surface are minimal and periodic, with a mere acoustic load of about 50 dB. At $\theta = [\pi/2]$, a violent periodic fluctuation erupted everywhere around the control surface, leading to a higher acoustic load of about 150 dB (3 times higher than the previous). For $\theta = [\pi/6]$ and $[\pi/3]$, high-frequency fluctuations with small and large-scale structures continuously shed along the reattaching shear layer, thereby causing a broadened spectra in the control surface wake.
\end{abstract}

\maketitle

\section{Introduction}\label{sec:intro}
\noindent
Bodies moving at supersonic and hypersonic speeds experience shock waves and unsteady flows\cite{Deshpande2020,Gaurav2021,Vatansever2021,Tumuklu2019,Hickel2020} around them. Due to the high flow speeds around the vehicle, adverse pressure gradients may be generated, resulting in separated boundary layer flows with shock waves which dictate its motion to a considerable extent\cite{Tumuklu2018,Duck1997,Porter2019,Bourdon1999}. In such a flow regime, the body's structure plays a significant role in governing the flow across it. Even minor structural changes can significantly alter the flow characteristics, the nature of which, if known, can help to obtain the desired flow\cite{Dolling2001,Gaitonde2015,Clemens2014,Settles1979}. Control surfaces such as flaps or aerospikes\cite{Sugarno2022,Narayana2020,Tekure2021} introduce flow separations that could enhance the vehicle performance if the flow dynamics across it is completely understood. Such an understanding has enabled, for instance, the development of re-entry vehicles in the past. Conventionally, all supersonic vehicles had sharp-spiked nose cones to reduce the loss in vehicles' kinetic energy through shock waves and also to reduce sonic booms. However, any re-entry vehicle with such a configuration burned up when it entered the atmosphere at suborbital speeds due to immense vehicle heating, which no known material could withstand. This problem, which was like a barrier in space exploration, was solved by Allen\cite{allenstudy} by using a blunt nose cone instead of a sharp spiked one in the re-entry module, which increased the energy loss to the surrounding air and reduced the heating of vehicle, which was desirable in re-entry vehicles. Control surfaces like spoilers\cite{Stanbrook1957} or flaps\cite{Kattari1955} decelerate or alter the flow characteristics around a missile body. However, they induce a boundary layer separation which may cause drastic variation in surface flow properties, which should be understood to prevent adverse effects on performance\cite{kuehn1959,Wilcox1990,Nilavarasan2022}. 

Supersonic flow over control surfaces placed close to the leading edge produces downstream unsteadiness or wakes. The flow field resembles the case of flows over aerospikes with a strong downstream wake, in addition to the upstream disturbances from flow separation. Amsden and Harlow\cite{Amsden1965} did a numerical analysis of wake generated in flow over a flat projectile at a freestream Mach number of $M_\infty = 1.89$, neglecting real gas and viscous effects, which become more pronounced at higher Mach numbers. The upstream unsteadiness was studied by Hui\cite{Hui1972} using a wedge, mimicking a control surface deflecting with the same wedge angle. The amplitude of flow oscillations was assumed to be small, but it is valid only for small control surface deflection angles. Early studies of the wake behind the control surface were carried out by Kazakov\cite{Kazakov1982}, who applied the method of matched asymptotic expansions to solve Navier strokes equations to investigate unsteady flow over a control surface and reported the change in flow properties in the wake region downstream of the control surface. A similar wake is generated in the case of the spiked bodies. Moreover, downstream shock waves may appear, as reported by Rao\cite{RAO1974}. The study of wakes in aerospikes was further done in the hypersonic flow regime by Sebastian \etal\cite{Sebastian2016}. The researcher varied the ratio of the spike length and the base body ($L/D$) along with the angle of attack ($\alpha$). Moreover, the author reported the drag coefficient variation as a function of both. Further studies by Sahoo \etal\cite{Sahoo2020,Sahoo2021} at $M_\infty = 2$ and zero angle of attack ($\alpha = 0^\circ$) with different $[L/D]$ and $[d/D]$ (spike and base body ratio) ratios reported that the variation of shock-wave unsteadiness was more pronounced with the changes in $[d/D]$ rather than $[L/D]$.

The characteristics of the wake flow and vortex-laden unsteadiness generated due to supersonic flows across spiked bodies and control surfaces depend significantly on the angle of attack-$\alpha$, as reported by David\cite{Degani1992} who studied the vortex characteristics for four different $\alpha$. Intense pressure fluctuations were observed on the ramp (a representation of a control surface) for a freestream flow at $M_\infty = 3$ by Dolling and Murphy\cite{Dolling1983} asides from the wake. They observed the generation of a highly unsteady separation shock wave of high amplitude and low frequency. Additionally, they reported the similarity of the ramp flow behaviour to that observed in three-dimensional blunt fin-induced flows. Similar studies were done for a freestream flow at $M_\infty = 2$ by Park \etal\cite{Park1994} and at $M_\infty = 5$ by Erengil and Dolling\cite{Erengil1991,Erengil1993}, and Gramman and Dolling\cite{GRAMANN1992}, with different geometrical ramp parameters. In all these studies, the ramp was kept stationary relative to the flow.

Park \etal\cite{Park2001} studied the shock-wave characteristics and pressure variations in an oscillating ramp with angle variation of $0-10^\circ$. Similarly, Sanghi \etal\cite{Deshpande2015} studied the behaviour of wall pressure response and boundary layer variations in an oscillating ramp. All these studies helped us understand the flow characteristics over ramps at different speeds. Further, towards using the shock waves and boundary layer separations generated by control surfaces to alter the airflow around the vehicle, Kontis \etal\cite{Saad2012} investigated the use of micro-ramps at hypersonic flow conditions. They reported that the micro ramp immersed in the boundary layer caused an upstream interaction length by delaying the pressure rise, thereby suppressing the Shock/boundary layer interaction (SBLI) effect. Ramps show similar behaviour to control surfaces in the upstream region but lack the downstream wake flow and the supersonic mixing layer. Early attempts were made by Stratford\cite{Stratford1956} to analyse the mixing layer and to calculate the thrust reduction due to the presence of a mixing layer downstream of the control surface.

Some studies focus on the effects of control surface placements to capture the specific flow separations generated. Control surfaces, for instance, have been used in re-entry vehicles to alter the shock-wave structure to reduce localised overheating of vehicles. Raman \etal\cite{Raman2020} reported the creation of streamwise and spanwise vortices while introducing a control surface and demonstrated the formation of a separation zone for higher control surface angles and the associated increment in the blunt cone stability. Another study concerning the control surface oscillations on aircraft wings was done by Kandil and Salman\cite{KANDIL1991}, who captured the induced flow separation using numerical methods. Like oscillating ramps, oscillating control surfaces have also been studied. Coon and Chapmaan\cite{Coon1995} did an experimental study at a freestream flow of $M_\infty = 2.4$ with control surface angle varying in the range of $0-30^\circ$. They measured the pressure oscillations in the turbulent boundary layer and evaluated the oscillation frequencies of the unsteady flow variation.

Most studies above concerning the control surfaces are done with a control surface deflection angle of $30^\circ$ or less. Few studies are done with larger control surface angles, which may occur during off-design operation, accidental deployment during system failures, or while functioning as a supersonic spoiler. Hence, adequate data needs to be collected in those scenarios too. Moreover, the role played by the control surface deployments near the leading edge is scarce as they play a vital role in highly manoeuvrable air-to-air interceptions. In the present study, a numerical investigation on the behaviour of a control surface in the proximity of a sharp leading edge has been done with various control surface deflection angles, which spans as $\theta = [\pi/36,\pi/6,\pi/3,\pi/2]$ (rad). The elaborated objectives of the current study are given below:
\begin{itemize}
    \item To identify the notable changes in the external flow characteristics between the different control surface deflection angles using field statistics of flow variables.
    \item To investigate the variation of the upstream/downstream separation/recirculation bubble length between the cases of different control surface deflections and probe for the existence of scaling laws.
    \item To assess the fluid-structure interaction from the resulting unsteady flow field by monitoring the variations in the drag coefficient for a range of control surface deflection angles.
    \item To contrast the differences in the total pressure loading and fluctuation intensity on the base body and the surroundings due to changes in the control surface deflection angle.
    \item To evaluate the net acoustic load acting on the model and around the control surface by computing the overall sound pressure level for different control surface deflections.
    \item To understand the unsteady dynamics ahead and behind the control surface by performing spectral analysis using $x-t$ and $x-f$ plots construction and verifying the existence of dominant length scales using correlation coefficients.
    \item To prove the presence of dominant spatiotemporal structures using the modal analysis tools like Proper Orthogonal Decomposition (POD) and Dynamic Mode Decomposition (DMD) and justify the observations from the previous analysis.
\end{itemize}

The outcome of the present study provides an estimation of the expected unsteady loads during the deployment of control surfaces. Such a result directly affects the design and development of rockets and missiles. Moreover, identifying the underlying dominant spatiotemporal modes helps arrive at suitable control devices to overcome the adversaries of control surface-induced unsteadiness.

The rest of the manuscript is organised in the following manner. In Sec. \ref{sec:prob_stat}, a brief description of the problem statement is given. In Sec, \ref{sec:num}, details of the adopted numerical methodology is elaborated. In Sec. \ref{sec:res_disc}, vital simulation results and the corresponding discussions from different analyses are explained. In Sec. \ref{sec:conclusions}, some of the important conclusions from the present study are enumerated.

\begin{figure*}
	\includegraphics[width=0.8\textwidth]{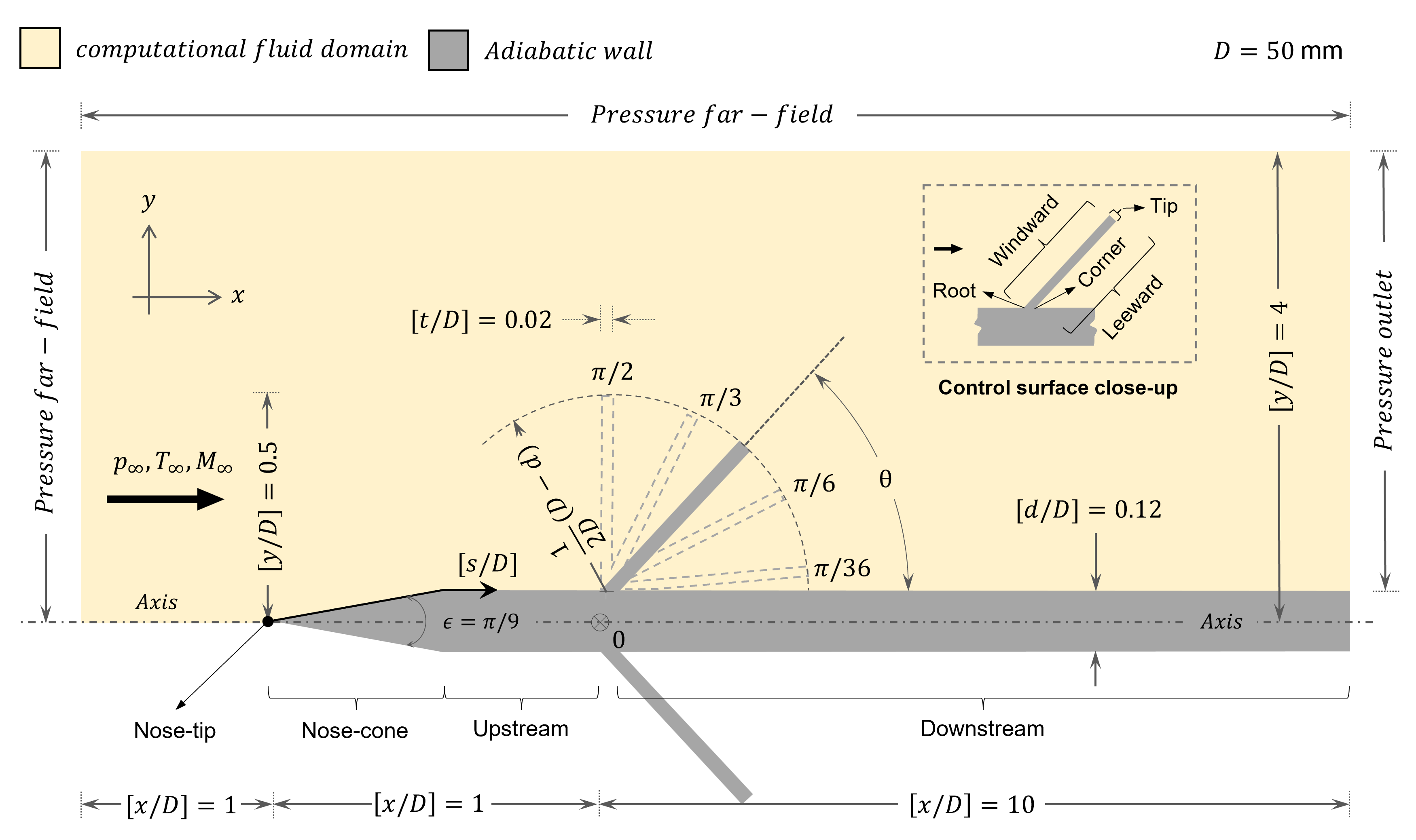}
	\caption{\label{fig:schematic} A typical sectional schematic showing the geometrical features of the axisymmetric cylindrical body with a sharp-tip cone housing the control surface at different orientations ($\theta=\pi/2,\pi/3,\pi/6,\pi/36$, rad), computational fluid domain limits, respective coordinate system (origin is marked as $\otimes$), and the imposed boundary conditions. Flow is from left to right. The schematic is not drawn to the scale.}
\end{figure*}

\section{Problem Statement} \label{sec:prob_stat}
A simple axisymmetric model with a deployed control surface is simulated to study the unsteadiness of the resulting flow. The control surface is simulated all along the azimuth ($\phi$) in an axisymmetric simulation. The resulting configuration is seen in the supersonic spoilers to reduce the vehicle's velocity or in the deployable aerospike to reduce the wave drag. The resulting unsteady flow dynamics from an infinite control surface segment are comparable to the actual case and reduce the computational load. Hence, an axisymmetric simulation is preferred further to explore the problem. An axisymmetric sharp-tip cone having a base body diameter of $[d/D]=0.12$ mm and a cone angle of $\epsilon = [\pi/9]$ is considered, where $D$ is the reference length of 50 mm. A control surface having a thickness of $[t/D]=0.02$ and a radius of $[r/D] = (1/2D)(D-d)$ is placed with its root firmly placed downstream the leading-edge at a length of $[x/D]=1$. The pivot point for the control surface deflection is placed at $[x/D,y/D]=[0,0.06]$, which is just above the origin. The cylindrical base body extends to a length of $[x/D]=10$ downstream of the control surface's root. When the control surface deploys to $\theta = [\pi/2]$, the vertical distance between the sharp nose-cone tip in the leading edge and the blunt control surface tip is set to $[y/D]=0.5$. The rectangular computational domain horizontally extends to a length of $[x/D]=1$ ahead of the cone's sharp tip and vertically spans over a distance of $[y/D]=4$ from the axis. In the present numerical exercise, four control surface deflection angles are considered: $\theta=[\pi/36,\pi/6,\pi/3,\pi/2]$ (rad) to represent a wide range of possible operation. A surface coordinate system ($s$) approach is adopted to monitor the flow variable changes along the wall. If the local wall coordinates are given in $x$ and $y$ for $n$ number of points, then the surface coordinate system about the nose tip is given by Eqns. \ref{eq:surf1} to \ref{eq:surf3}. A typical schematic explaining the problem statement is given in Figure \ref{fig:schematic}.
\begin{align}
\label{eq:surf1}
    \Delta s_i &= s_{i+1} - s_i = \sqrt{(x_{i+1}-x_i)^2 + (y_{i+1}-y_i)^2},\\
    \label{eq:surf2}
    \Delta {s} &= \left[ \Delta s_1, \Delta s_2, \dots, \Delta s_n \right],\\
    \begin{split}
    \label{eq:surf3}
    {s} &= [0,\Delta s_1,\Delta s_1+\Delta s_2, \Delta s_1+\Delta s_2+\Delta s_3,\dotsm,\\
    & \textcolor{white} {\Delta s_1+} \Delta s_1+\Delta s_2+\Delta s_3\dotsm \Delta s_n].
    \end{split}
\end{align}
The flow conditions are carefully considered after reviewing the typical operational altitudes of air-to-air missiles, surface-to-air rockets, and other supersonic payload carriers where a similar model configuration is seen. The computational domain is simulated with a fluid flow having a typical freestream Mach number of $M_\infty=2.01$ and a Reynolds number based on the base body diameter of $Re_D=2.16 \times 10^6$. Air as an ideal gas is used as the working fluid, and the rest of the flow conditions are tabulated in Table \ref{tab:flow_cond}. The freestream conditions are chosen from the experimental work of Sahoo \etal\cite{Sahoo2021}. The decision has arrived in the best interest of studying a practical flow field along with the ease of validating the simulation.

\section{Numerical Methodology} \label{sec:num}

\begin{table}
	\caption{\label{tab:flow_cond} Tabulation of supersonic freestream flow conditions ($\infty$) achieved ahead of the control surface at different deflections ($\theta=[\pi/2,\pi/3,\pi/6,\pi/36$], rad).}
	\begin{ruledtabular}
		\begin{tabular}{lc}
			Parameters\footnote{Flow conditions are taken from the experiments of Sahoo \etal\cite{Sahoo2021} on a similar geometrical configuration} & Values \\ 
			\midrule 
			Reference diameter\footnote{effective diameter seen for the control surface deflecting at $\theta=[\pi/2]$} ($D \times 10^{-3}$, m) & 50\\
			Freestream velocity ($u_\infty$, m/s) & 515.15 \\
			Freestream pressure ($p_\infty\times 10^5$, Pa) & 0.44 \\
			Freestream temperature ($T_\infty$, K) & 163.59 \\
			Freestream density ($\rho_\infty$, kg/m$^3$) & 0.94 \\
			Freestream kinematic viscosity ($\nu_\infty\times 10^{-5}$, m$^2$/s) & 1.12 \\
			Freestream Mach number ($M_\infty$) & 2.01 \\
			Freestream Reynolds number\footnote{defined as $Re_D = [u_\infty D/\nu_\infty]$} ($Re_D \times 10^6$) & 2.16 \\
		\end{tabular}
	\end{ruledtabular}
\end{table}

The basic fluid flow equations, including continuity, momentum, and energy (Navier-Stokes equations), are solved numerically to resolve the flow features around the control surface closely placed near the leading edge. A commercial flow solver like Fluent\textsuperscript{\tiny\textregistered}\cite{Fluent_2013} is used assuming an axisymmetric compressible turbulent fluid flow. The governing conservative form of the equations is averaged using the principles of Favre-Averaging (Favre-averaged Navier-Stokes equations) as shown in Eqs. \ref{eq:sol_1} to \ref{eq:sol_3} in the index notation. The indices increment represents the three flow axes ($x,y,z$). 

\begin{align}
    \frac{\partial \rho}{\partial t} + \frac{\partial}{\partial x_i} (\rho u_i) =& 0,\label{eq:sol_1}\\
    \frac{\partial}{\partial t} (\rho u_i) + \frac{\partial}{\partial x_j} (\rho u_i u_j) =& -\frac{\partial p \delta_{ij}}{\partial x_i} + \frac{\partial \tau^t_{ij}}{\partial x_j},\label{eq:sol_2}\\
    \frac{\partial}{\partial t} (\rho E_0) + \frac{\partial}{\partial x_j}(\rho E_0 u_j) = & -\frac{\partial}{\partial x_j}(pu_j - q^t_j + u_i \tau^t_{ij}),\label{eq:sol_3}
\end{align}

The governing equations contain partial derivatives, and they are discretised by Green-Gauss finite volume schemes using cell-centred formulation in the present solver. During the axisymmetric simulation, the Cartesian coordinates are transformed to the cylindrical coordinates as $[x,y,z]=[r\cos\theta,r\sin\theta,z]$. The term $r$ and $\theta$ represent the radial and angular coordinates, respectively. In an axisymmetric flow, the flow variations in the angular coordinates ($\theta$) and the derivatives ($\partial/\partial \theta$) are considered negligible or zero. Such an assumption in the present simulation is fairly considerable, provided the results are used to extract the first-order flow characteristics and to deduce a contrast between the cases. In reality, the effects of tangential flow also play a vital role, provided the base body rotates about the axis or, in the case of azimuth instability generation.

\begin{align}
    \tau^t_{ij} = & \left[(\mu + \mu_t) \left(\frac{\partial u_i}{\partial x_j}+\frac{\partial u_j}{\partial x_i}-\frac{2}{3}\frac{\partial u_k}{\partial x_k}\delta_{ij} \right)\right] - \frac{2}{3}\rho k\delta_{ij},\label{eq:sol_4}\\
    q^t_{j} = & - \frac{\gamma}{\gamma-1}\frac{\partial}{\partial x_j}\left[\frac{p}{\rho}\right]\left(\frac{\mu}{Pr}+\frac{\mu_t}{Pr_t}\right). \label{eq:sol_5}
\end{align}

Bossinique's hypothesis is used in Eqs. \ref{eq:sol_1} to \ref{eq:sol_3} to resolve the terms like the viscous stress ($\tau^t_{ij}$) and heat flux ($q^t_j$) as shown in Eqs. \ref{eq:sol_4} and \ref{eq:sol_5}, where the superscript $t$ represents the terms arising due to turbulence. The unknown turbulent quantities are resolved using `Spallart-Allmaras' (S-A) closure models\cite{Acquayea,Acquaye2016b,AugustoFontanMoura2015,Sun2013}, where the turbulent kinematic viscosity ($\nu_t$) is modelled as
\begin{equation}
\begin{aligned}
    \frac{\partial}{\partial t}(\rho \nu_t) + \frac{\partial}{\partial x_i}(\rho \nu_t u_i) =& G_\nu + \frac{1}{\sigma_{\nu_t}} \left[\frac{\partial}{\partial x_j}\left\{(\mu+\rho \nu_t)\frac{\partial \nu_t}{\partial x_j}\right\}\right. \\ 
    &+ \left. C_{b2} \rho \left(\frac{\partial \nu_t}{\partial x_j}\right)^2\right]-Y_\nu + S_{\nu_t}, \label{eq:sol_6}
\end{aligned}
\end{equation}
In Eq. \ref{eq:sol_6}, the terms $G_\nu$ and $Y_\nu$ indicate the production and destruction of turbulent viscosity, $\sigma_{\nu_t}$ and $C_{b2}$ are constants, $S_{\nu_t}$ is the user-defined source term, and $k$ is the turbulent kinetic energy. Moreover, a detached eddy simulation (DES) strategy is adopted to resolve the turbulent eddies away from the wall. The production and destruction of the turbulent viscosity terms in Eq. \ref{eq:sol_6} contain an eddy length scale term ($d$), and it is modelled as
\begin{equation}
    d=\textrm{min}(d,C_{des}\Delta). \label{eq:sol_7}
\end{equation}
In Eq. \ref{eq:sol_7}, $\Delta$ is the grid spacing and $C_{des}$ is an empirical constant of 0.65. More detailed explanations of the solver schemes and turbulence closure models are available in the Fluent\textsuperscript{\tiny\textregistered} technical guide\cite{Fluent_2013} and are not further discussed in the manuscript for brevity.

\subsection{Grid generation and solver settings}
The fluid flow domain is meshed using the commercial grid generator like Ansys-ICEM\textsuperscript{\tiny\textregistered}\cite{Icem_2013}. A structured quadrilateral mesh is generated on the fluid domain shown in Figure \ref{fig:schematic}. A very densely packed grid is adopted close to the walls, and it is gradually coarsened towards the fluid domain boundary using a successive cell progression value of 1.2. The generated mesh contains 95\% of cells with an equisize skewness parameter of 0.2, indicating a good quality grid. The near-wall cells are in the size of $\sim 1$ $\mu$m with a turbulence wall parameter of $y^+$ being less than one. The final grid size and density are obtained after an elaborate grid independence study as discussed in Sec. \ref{ssec:grid_indep}. The edges defining the model with the control surface are kept at an adiabatic state, and a no-slip wall boundary condition is enforced. The upstream vertical and the top horizontal fluid domain edges are set with a pressure far-field boundary condition. The vertical fluid domain edge downstream is enforced with a pressure outlet condition.

Air, as an ideal gas, is used as the working fluid with a specific heat ratio of $\gamma=1.4$. Viscosity is modelled using Sutherland's three-coefficient method. A pressure-based coupled solver with compressibility correction in Fluent\textsuperscript{\tiny\textregistered} is used to resolve the flow features. The near-wall turbulence is modelled using unsteady Favre-averaged Navier-Stokes equations with a vorticity-strain-based S-A turbulence model. The turbulent structures away from the wall are modelled using the DES approach with the default constants available in the solver. The gradients are resolved using a Green-Gauss node-based spatial discretisation scheme. Pressure terms are discretised using a second-order scheme. Density, modified viscosity, and energy terms are discretised using a second-order upwind scheme. Momentum terms are discretised using a bounded central difference scheme. Temporal terms are discretised using a second-order implicit scheme. Under relaxation factors (explicit), and Courant numbers are set to the existing default values. Locally scaled residuals from each iteration are monitored while solving the governing equations. A residual of $10^{-5}$ is achieved for all the cases under consideration.

\begin{figure}
	\includegraphics[width=\columnwidth]{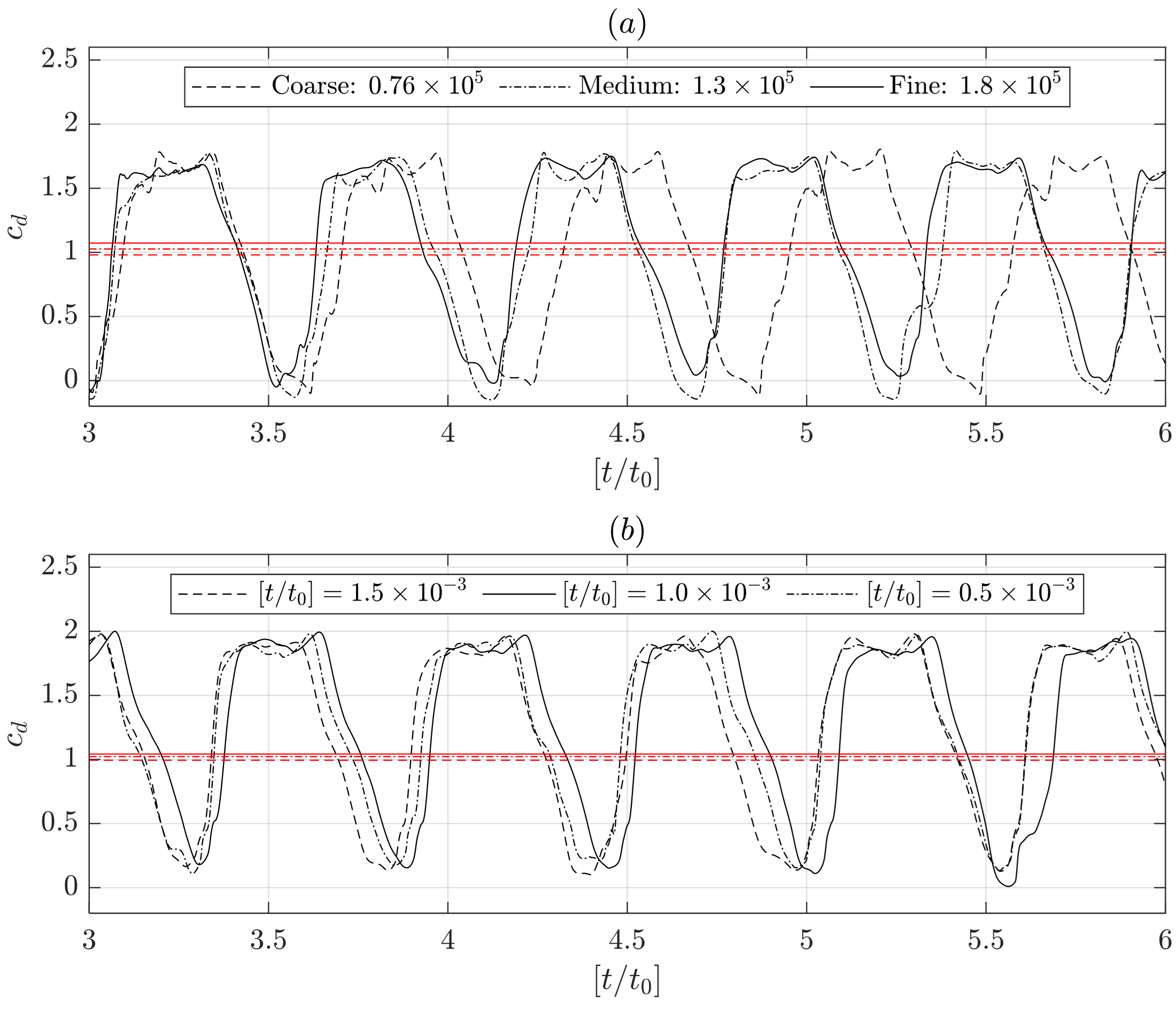}
	\caption{\label{fig:mesh_indep_signal}  Plot showing the variation of $c_d$ for the $\theta=[\pi/2]$ case from the present simulation over a definitive duration of the self-sustained oscillation ($t/t_0$, where $t_0=1$ ms). (a) Plot showing the effect of different mesh cells at a time-step of $[t/t_0]=1.0 \times 10^{-3}$: $0.76 \times 10^5$ (coarse), $1.3 \times 10^5$ (coarse), and $1.8 \times 10^5$ (coarse). The horizontal line in red-color indicates the time-averaged drag coefficient ($\overline{c}_d$) which varies across the meshes as $\overline{c}_d=1.07$ (coarse), $\overline{c}_d=1.02$ (medium), and $\overline{c}_d=0.97$ (fine); (b) Plot showing the effect of time-step changes in a fine mesh environment of $1.8 \times 10^5$ cells: $[t/t_0]=1.5 \times 10^{-3}$ (coarse), $[t/t_0]=1.0 \times 10^{-3}$ (medium), and $[t/t_0]=0.5 \times 10^{-3}$ (fine). The horizontal line in red-color indicates $\overline{c}_d$ varying across the time-steps as $\overline{c}_d=0.99$ (coarse), $\overline{c}_d=1.02$ (medium), and $\overline{c}_d=1.04$ (fine).}
\end{figure}

\subsection{Grid and time independence} \label{ssec:grid_indep}
The drag coefficient in the configuration of $\theta = [\pi/2]$ is monitored for the grid independence study. The selected case produces a strong unsteadiness due to `shock pulsation' and is a robust case for grid and time independence studies. The drag coefficient is given by
\begin{equation}
    c_d = \frac{D_F}{2\rho_\infty u^2_\infty S}, \label{eq:cd}
\end{equation}
where $D_F$ is the total drag force (N), and $S$ is the reference surface area (m$^2$). In the present case, $S=[\pi/4]\times D^2$, where $D=50$ mm. The DES outcomes are extracted after initialising the solution across the domain in a hybrid manner. Later, a self-sustained oscillation zone is identified to calculate the mean drag coefficient ($\overline{c}_d$). In general, a non-dimensional time-stamp of $[t/t_0] \geq 2$ (where $t_0$ is a reference time of 1 ms) stands as a self-sustained oscillation zone once the solution stabilizes after initialization. Three grid sizes are considered: $0.76 \times 10^5$ (coarse), $1.3 \times 10^5$ (medium), and $1.8 \times 10^5$ (fine). The current exercise's grid sizes are changed by modifying the grid density closer to the walls. The variation of $c_d$ across the simulation time with a time-step of $[t/t_0]=1.0 \times 10^{-3}$ is noticeable for the coarse grid in comparison with the fine grid in Figure \ref{fig:mesh_indep_signal}a. However, between medium and fine grids, the variations are almost negligible. The time-average drag coefficient obtained during the self-sustained oscillation also shows almost a constant value as the grid is refined with $\overline{c}_d=0.97$. Given the need to capture the compressible flow field events, including shocks and expansion zones, the fine grid is selected as a suitable choice. Similar to grid independence studies, a series of time-step independence studies is done by varying the non-dimensionalized time-steps as $[t/t_0]=1.5 \times 10^{-3}$, $[t/t_0]=1.0 \times 10^{-3}$, and $[t/t_0]=0.5 \times 10^{-3}$ in a fine mesh environment. Typical plot of $c_d$ variations over the solution duration is shown in Figure \ref{fig:mesh_indep_signal}b. As in the previous exercise, the profile variations between the medium and fine time steps are minimal with $\overline{c}_d \sim 1$. The computational time to resolve the flow field with a fine time step is comparatively larger than the medium case. Hence, a medium-time step is preferred to resolve the unsteady flow field faithfully and to keep the computational cost minimal.

\begin{figure}
	\includegraphics[width=\columnwidth]{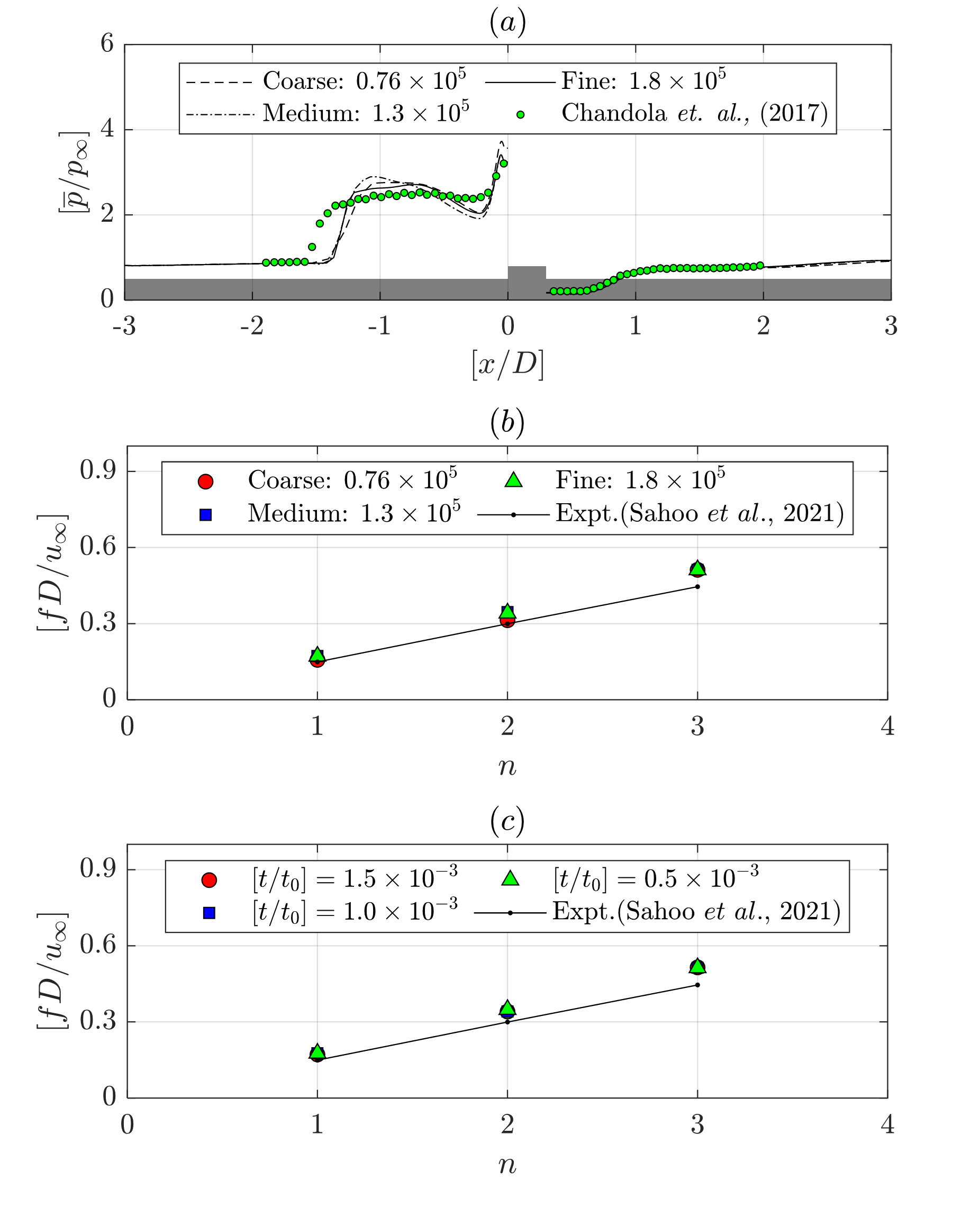}
	\caption{\label{fig:mesh_indep_spectra} (a) Validation plot showing the close match between the experiments\cite{Chandola2017} and the present computations while comparing the mean wall-static pressure ($\overline{p}/p_\infty$) across a square protrusion on an axisymmetric cylinder having a base body diameter of $D=75$ mm. Plots showing the non-dimensionalised frequency values in the first ($n=1$), second ($n=2$) and third ($n=3$) harmonics while taking the pressure fluctuation spectra at a half-distance location about the axis on the control surface or forebody. The plot includes the results from the present computations (shown using solid markers) as well as from the experiments of Sahoo \etal\cite{Sahoo2021} (shown as a solid line) while using different (b) mesh cells and (c) time-steps.}
\end{figure}

\subsection{Solver validation in spatiotemporal events}

The present flow solver is known to represent the compressible flow field close to the experimental conditions in many test cases as provided in the open literature \cite{Sekar2020,Karthick2021}. The findings from the experimental study of a classical axisymmetric step-flow\cite{Chandola2017} and a typical flat-face spiked body in a supersonic flow \cite{Sahoo2021} are considered for validating the present solver. The experimental cases closely resemble the numerical case with the control surface deployed at $\theta = [\pi/3]$ and $[\pi/2]$, at least in terms of the flow features encountered ahead and behind the control surface. Time-averaged quantities like the wall-static pressure ($\overline{p}/p_\infty$) variation around the axisymmetric step-flow is firstly compared between the experiments and a FANS-based computation for different grid sizes as shown in Figure \ref{fig:mesh_indep_spectra}a. The computed pressure jump near the separation bubble ahead of the step shows about 6\% variation from the experiments. The point of separation ahead of the step shows deviance of about 10\% between the experiments. Downstream [$\overline{p}/p_\infty$] variation shows even a better match with the experiments within 0.5\%. Secondly, time-dependent quantities like the mean drag coefficient ($\overline{c}_d$) and the dominant shock oscillating frequency ($fD/u_\infty$) from the spiked-body experiments are compared between the experiments and computational case of $\theta=[\pi/2]$ as shown in Figure \ref{fig:mesh_indep_spectra} b-c. In the experiments, the spike length and base body diameter ratio are $[L/D]=1$. In the computations of $\theta = [\pi/2]$ case, the ratio of the leading-edge distance ($[x/D]=1$) and tip-to-tip control surface diameter ($[2y/D]=1$) is 1. Hence, the expected flow features ahead are similar. In the experiments, the mean forebody drag coefficient is $\overline{c}_{d,e}=1.06$, whereas in the computations (Figure \ref{fig:mesh_indep_signal}b), it is $\overline{c}_{d,c}=1.02$ (fine mesh, medium time-step) which is about 4\% deviation from the experiments. The pressure fluctuations in the forebody are later considered for matching the spectral contents between the experiments and computations. The time-varying pressure in the middle of the control surface above the axis is used to calculate the spectra. The dominant peak is plotted between the experiment and computation along with its harmonics ($n=1$ being the fundamental and other integers being its harmonics). For the fine mesh and medium time-step, the fundamental ($fD/u_\infty \sim$ 0.18) is slightly higher than the experiment ($fD/u_\infty \sim$ 0.13). On the other hand, as shown in Eqn. \ref{eq:kenworthy}, the non-dimensionalised fundamental frequency from the empirical relation of Kenworthy \cite{Kenworthy1978}  ($fD/u_\infty \sim$ 0.18) matches fairly well with the numerical case ($fD/u_\infty \sim$ 0.18). 
\begin{equation}
\label{eq:kenworthy}
    \left[\frac{fD}{u_\infty}\right] = 0.25 - \left(\frac{L}{D}\right)0.067.
\end{equation}

\section{Results and Discussions} \label{sec:res_disc}

This section analyses results from the DES-based simulations and compares the cases under consideration to draw out the driving flow physics. Initially, gross flow feature differences between the cases are identified using the contour plots of numerical schlieren, Mach number, and streamwise velocity component along with the time-varying drag coefficients and their spectra. Moreover, the downstream recirculation bubble characteristics are investigated using the time-averaged streamwise velocity contours. Later, only the contours of static pressure variation are considered for the rest of the analysis. Mean pressure loading, fluctuation intensity, and overall sound pressure level variation on the model surface are computed to highlight the influence of control surface-induced wake at different deflections. The relationship of upstream and downstream pressure fluctuations along the wall and the presence of prominent spatiotemporal events are identified using the cross-correlation estimation and through the construction of $x-t$ and $x-f$ plots. Finally, proper orthogonal and dynamic mode decomposition (POD and DMD) are done on the pressure field to reveal the dominant underlying spatiotemporal modes.

\subsection{Instantaneous and time-averaged flow field} \label{ssec:inst_avg}

\begin{figure*}
	\includegraphics[width=0.7\textwidth]{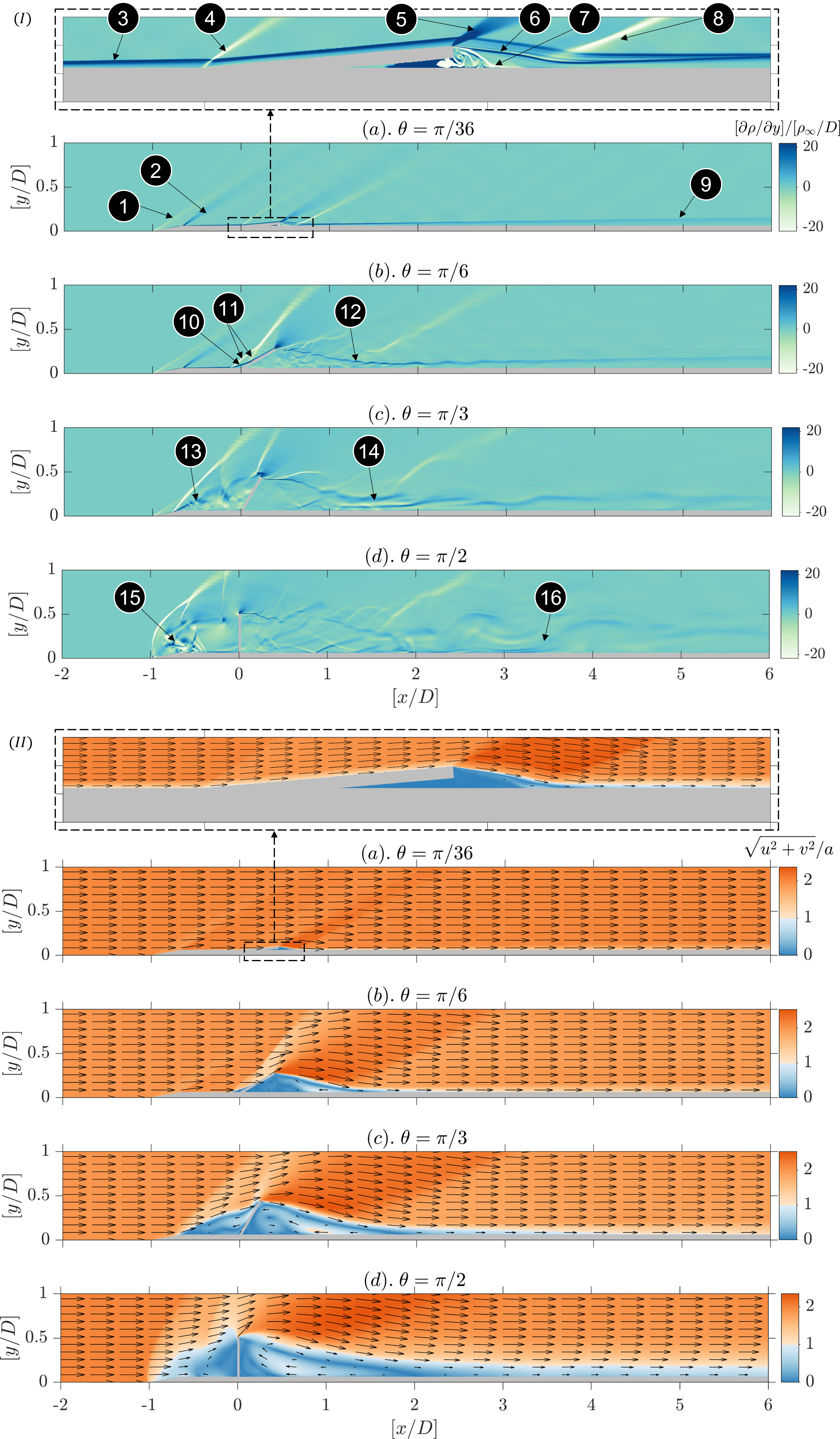}
	\caption{\label{fig:sch_M_contours} \href{https://youtu.be/RSdGcW1ODpA}{(Multimedia View)} I. Contour plots showing the instantaneous non-dimensionalized numerical schlieren about the transverse direction for different control surface deflection angles ($\theta$, a-d). Key flow features: 1. leading-edge shock, 2. expansion waves at the corner, 3. incoming boundary layer, 4. compression shock, 5. expansion fan emanating from the control surface tip, 6. reattaching laminar shear layer, 7. recirculation region in the control surface's wake, 8. recompression shock, 9. developing laminar boundary layer, 10. laminar separation bubble, 11. separation and reattachment shock, 12. reattaching transient shear layer, 13. separated transient shear layer, 14. reattaching turbulent shear layer, 15. unsteady leading pulsating shock systems, 16. destabilised wake region. II. Contour plots showing the time-averaged Mach number with time-averaged velocity vectors for different control surface deflection angles ($\theta$, a-d).}
\end{figure*}

Deflection ($\theta$) of the control surface produces a wide range of unsteady flow fields. Notably, the separation happens ahead, and the recirculation region that forms behind the control surface is unique as $\theta$ varies between $0$ and [$\pi/2$]. In Figure \ref{fig:sch_M_contours} \href{https://youtu.be/RSdGcW1ODpA}{(Multimedia View)}, instantaneous contour plots of non-dimensionalised numerical schlieren (about the transverse direction) and time-averaged Mach number plots with velocity vectors are shown for different $\theta$ to appreciate the apparent flow field differences. In general, across all the configurations, some features are seen in common except with varying intensities. 

At $\theta=[\pi/36]$ (Figure \ref{fig:sch_M_contours} I-a), flow moves past the model with leading-edge shock and expansion wave at corner. The control surface deflected the flow, leading to the formation of compression shock. The presence of an incoming boundary layer is seen with a thickness comparable to that of the control surface. At the control surface's tip, the flow expands due to the formation of an expansion fan. Due to the protrusion height originating from the deflecting control surface, a recirculation region is formed behind it. The flow then deflects across the recompression shock to run in parallel with the freestream. Further downstream, a developing laminar boundary layer is seen as the flow propagates.

At $\theta = [\pi/6]$ (Figure \ref{fig:sch_M_contours} I-b), ahead of the control surface, a laminar separation bubble is formed which induce a separation and a reattachment shock. As the obstacle height due to the control surface deflection is significantly high, as in the previous case, the reattaching shear layer travels a longer distance, comparatively. The separated shear layer is perturbed and undergoes transition due to travel length. Moreover, the developing boundary layer behind the control surface is relaminarised after a certain distance.

At $\theta= [\pi/3]$ (Figure \ref{fig:sch_M_contours} I-c), the separation ahead of the control surface is prominent and extends up to the expansion corner. The separated shear layer is initially laminar and undergoes a transition and later to a turbulent state. The separated shear layer leaves the control surface tip tangentially, leaving a large stationary unsteady separation bubble. The reattaching shear layer carries significant fluctuations from the separated shear layer and breaks down to turbulence before the recompression shock. The developing boundary layer behind the control surface thus is not completely relaminarised as before and carries transient convecting structures. 

At $\theta = [\pi/2]$ (Figure \ref{fig:sch_M_contours} I-d), the leading-edge separation is highly unstable to an extent, the stationary shock system as seen in the previous case starts to oscillate violently. Such an oscillation is called `pulsation' as seen in the spiked-body flows at supersonic and hypersonic speeds \citep{Sahoo2021,Karthick2022}. A typical pulsation cycle contains three phases of shock motion: inflate, with-hold, and collapse. As the separation bubble constantly collapses during pulsation, there is a periodic forcing on the wake flow rendering the reattaching shear layer unstable. The resulting flow further downstream is turbulent with a thickened boundary layer.

In Figure \ref{fig:sch_M_contours} II, the flow path is further visualised using the Mach contours and the overlaid velocity vectors. The colour map for the contour plot is selected so that the orange shade represents the supersonic flow field, whereas the blue shade identifies the subsonic part. In between, the region of whites indicates the sonic line. Let us look into the separation bubble in each case. At $\theta=[\pi/36]$, there is no separation bubble. At $\theta = [\pi/6]$, the separation bubble is confined to the control surface's root. Moreover, the separation bubble is almost stationary and comprises a slow recirculating fluid at subsonic speed. At $\theta = [\pi/3]$, the impinging fluid near the control surface tip bifurcates. A part of the fluid is ejected into the wake region, and the rest is detoured into the separation bubble closer to sonic velocities. The separated shear layer forming several small vortical zones further entrains part of this diverted fluid flow. At $\theta = [\pi/2]$, the separation bubble itself collapses frequently (see the \href{https://youtu.be/RSdGcW1ODpA}{Multimedia View} in Figure \ref{fig:sch_M_contours}). It results in the temporary formation of near sonic jet velocities inside the separation bubble, which is seen as a white patch in the time-averaged image.

Similarly, using Figure \ref{fig:sch_M_contours} II, the recirculation region in the wake is also studied to identify the distinguished flow patterns. At $\theta = [\pi/36]$, the white patch indicating sonic velocity is weakly present. Whereas at $\theta = [\pi/6]$, the extent is comparatively high. Moreover, at $\theta = [\pi/3]$, a supersonic reversed velocity region can be spotted as orange pits amidst the whitish patch in the wake-recirculation zone. At $\theta = [\pi/2]$, a larger portion of the recirculation region is laden with reversed sonic wall-jet. The produced wall jet is further entrained into the reattaching shear layer, causing a larger production of upward velocities compared to the rest of the cases. The severe perturbation in the reattaching shear layer and the resulting transient boundary layer that forms after the recompression shock is attributed to the aforementioned strong entrainment behaviour.

\subsection{Recirculation bubble length and scaling analysis} \label{ssec:bubble_scale}

\begin{figure*}
	\includegraphics[width=0.6\textwidth]{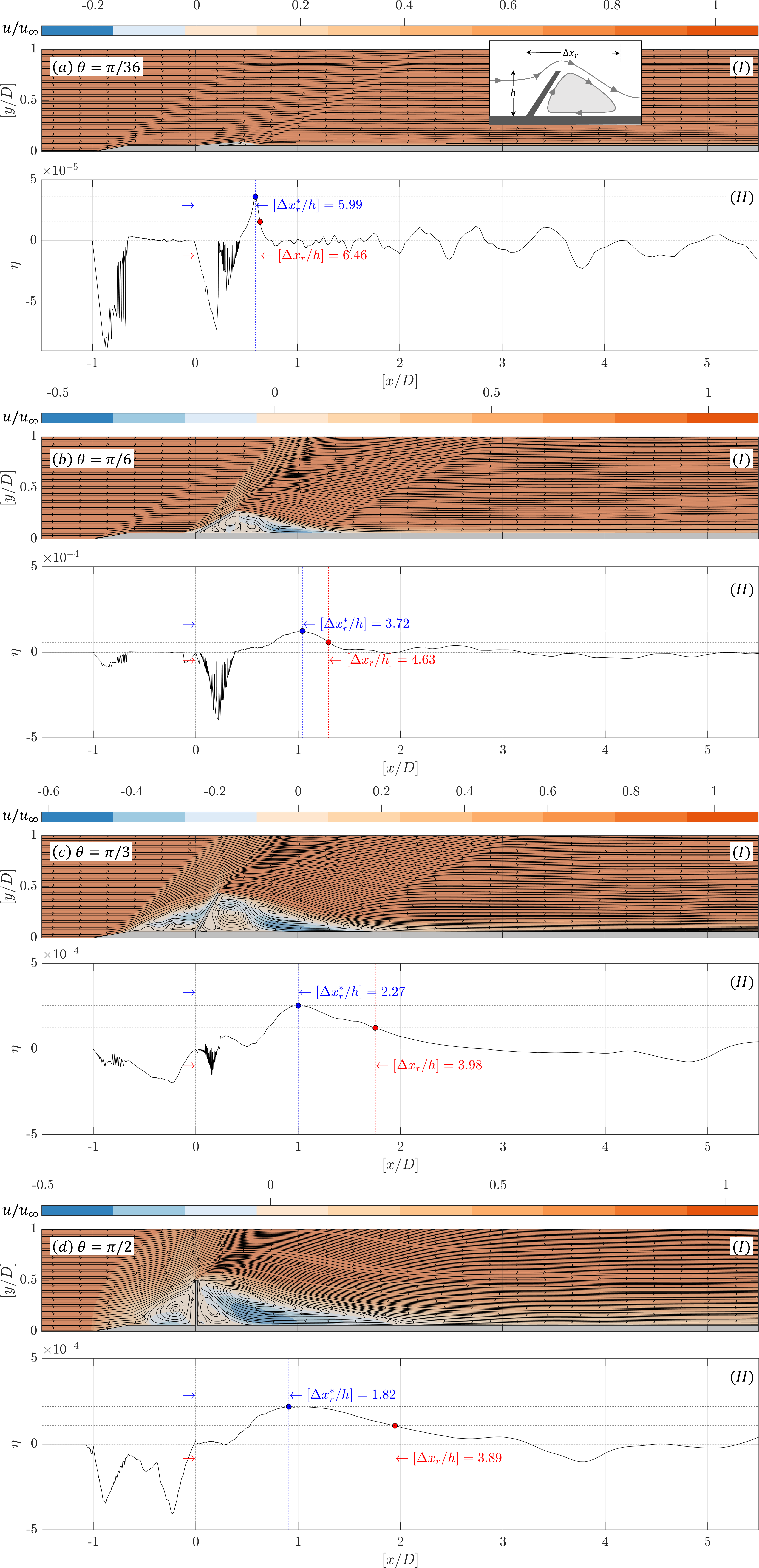}
	\caption{\label{fig:recirc_contour} (I) Non-dimensionalized streamwise velocity contour plots with streamlines for the different control surface deflection angles (a-d); (II) Variation of the recirculation region identification parameter ($\eta$) along the streamwise direction for the different control surface deflection angles (a-d). Non-dimensionalized apparent ($\Delta x_r^*/h$) and actual ($\Delta x_r/h$) recirculation length about the origin are marked in blue and red color, respectively. Values of [$\Delta x_r/h$] are calculated by estimating the downstream half-maximum of $\eta$ between $0 \leq \eta \leq 4$.}
\end{figure*}

\begin{figure}
	\includegraphics[width=0.85\columnwidth]{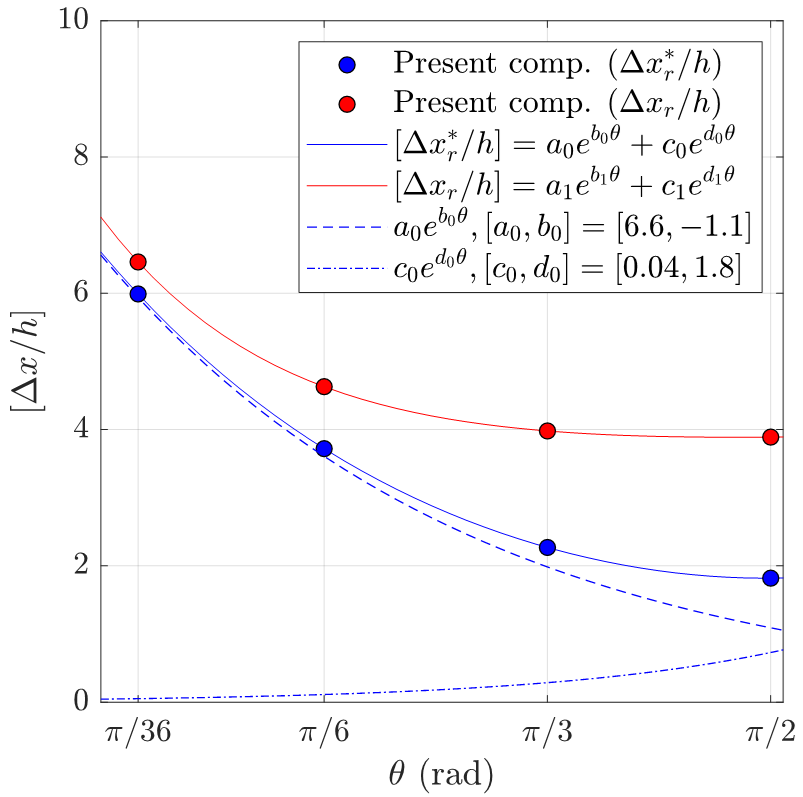}
	\caption{\label{fig:recirc_2d} Plot showing the variation of two types of recirculation length scale obtained from the recirculation region identification parameter ($\eta$) along the streamwise direction for different control surface deflections ($\theta$): apparent ($\Delta x_r^*/h$), blue colored circular markers) and actual ($\Delta x_r/h$), red coloured circular markers) recirculation length. The exponential line fits with two terms corresponding to each length scale are represented as solid lines. The blue-coloured dash and dot-dash lines represent the variation of each exponential term used to fit the corresponding computed apparent recirculation lengths ($\Delta x_r^*/h$) for different $\theta$.}
\end{figure}

Separation bubble forms ahead of the control surface for $\theta \geq [\pi/6]$. It is stationary only for $\theta = [\pi/6]$ and $[\pi/3]$ with separation length varies as $[\Delta x_s/D] \sim \mathcal{O}(d/D)$ and $[\Delta x_s/D] \sim 1$, respectively. However, the separation bubble severely unsteady for $\theta =[\pi/2]$ with $[\Delta x_s/D] \sim 1$. No definitive scaling exists for the separation bubble as the control surface is closer to the leading edge. Moreover, the range of leading-edge induced separation types varies with $\theta$. Hence, a dedicated investigation of the separation bubble variations in the present manuscript is not done.

On the contrary, a definite recirculation region exists in the wake of the control surface for various deflections. The streamwise length of the recirculation region from the origin is defined as $\Delta x_r$. Likewise, the vertical distance between the control surface tip and the origin is computed as $h$. The non-dimensionalized recirculation region's length $(\Delta x_r/h)$ is computed using the streamwise velocity contour plots. A parameter called $\eta$ is defined along every streamwise location. Firstly, the average streamwise velocity across the transverse direction at each streamwise location is computed. As mentioned in Eqn. \ref{eq:eta}, a streamwise gradient of the average sectional velocity is taken later.
\begin{equation}
\label{eq:eta}
    \eta = \left[\frac{1}{u_\infty}\right]\frac{d}{d x}\left[\int_0^y u(y) dy\right].
\end{equation}
The streamwise location between $0 \leq [x/D] \leq 4$ where the maximum value of $\eta$ occurs is defined as $\Delta x_r^*$ which is later non-dimensionalized using the vertical height of the control surface ($h$). The manner in which the streamwise velocity contour plots are used to compute $\eta$ and subsequently $[\Delta x_r^*/h]$ for all the cases of $\theta$ is depicted in Figure \ref{fig:recirc_contour}. The obtained $[\Delta x_r^*/h]$ is an apparent quantity as it is calculated through a definite mathematical routine. Actual recirculation region's length exits at the downstream half-maximum of $\eta$ which is marked as $[\Delta x_r/h]$ in Figure \ref{fig:recirc_contour}-II. From a cursory analysis, an underlying monotonic trend in the $[\Delta x_r/h]$ variations with $\theta$ is seen. Figure \ref{fig:recirc_2d} shows a typical line plot exposing the above variations. The computed variations are indeed in agreement with the exponential fit as described,
\begin{equation}
\label{eq:fit}
\left[\frac{\Delta x_r}{h}\right] = a_1 e^{b_1\theta} +c_1 e^{d_1\theta}.
\end{equation}
In Eqn. \ref{eq:fit}, $\theta$ varies as $[\pi/36]\leq \theta \leq [\pi/2]$ and the constants are computed to be $[a_1,b_1,c_1,d_1]=[3.8,-2.3,3.4,0.08]$ with a goodness of fit close to $R^2 \sim 0.99$. From Figure \ref{fig:recirc_contour} and \ref{fig:recirc_2d}, the values of $[\Delta x_r/h]$ can be considered as a appropriate scale and also as a mere indicator of the recirculation zone's aspect ratio. Qualitatively, for a shallow deflection, $[\Delta x_r/h]$ is substantial. On the other hand, as the control surface is deployed, $[\Delta x_r/h]$ grows smaller. While comparing the cases between $\theta=[\pi/36]$ and $[\pi/2]$, $[\Delta x_r/h]$ decreases by almost 40\%. 

\subsection{Drag coefficient variation and spectra} \label{ssec:drag_var}

\begin{figure*}
	\includegraphics[width=0.9\textwidth]{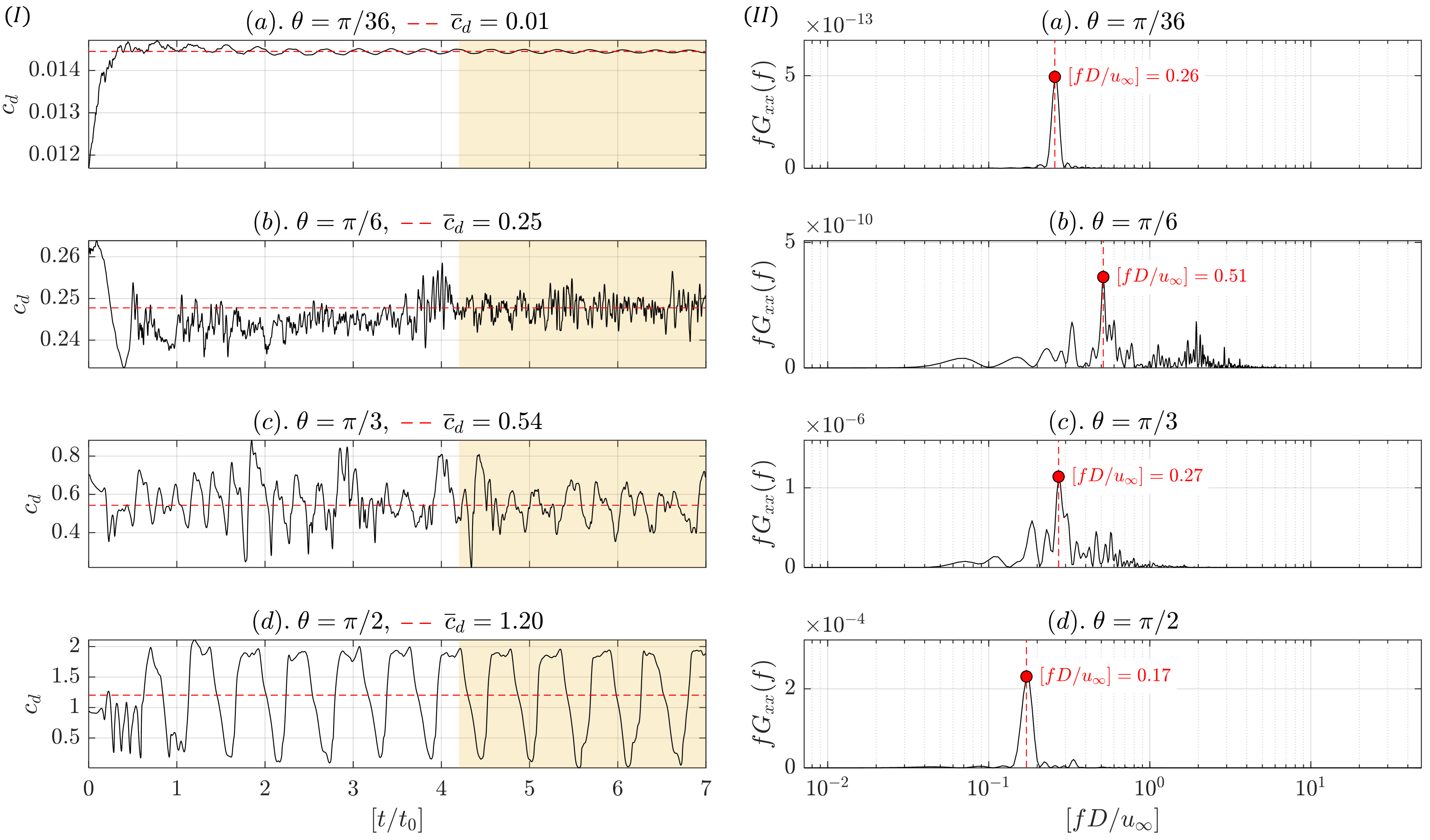}
	\caption{\label{fig:drag_signal}(I) Typical drag coefficient ($c_d$) evolution during the numerical simulation from the initial stages ($t/t_0$) to some part of the self-sustained oscillations (the shaded yellow patch region) for a wide range of control surface deflection angles-$\theta$ (a-d). The horizontal red-coloured dash line represents the mean drag coefficient ($\overline{c}_d$) over the portion of the considered self-sustained oscillation region. (II) Typical plots of power spectra obtained from the self-sustained oscillating signal of $c_d$ showing the presence of dominant non-dimensionalised frequency ($fD/u_\infty$) for a wide range of control surface deflection angles-$\theta$ (a-d). The vertical red-coloured dash line and the red-colour circular marker represent the peak [$fD/u_\infty$] and power.}
\end{figure*}

\begin{figure}
	\includegraphics[width=0.85\columnwidth]{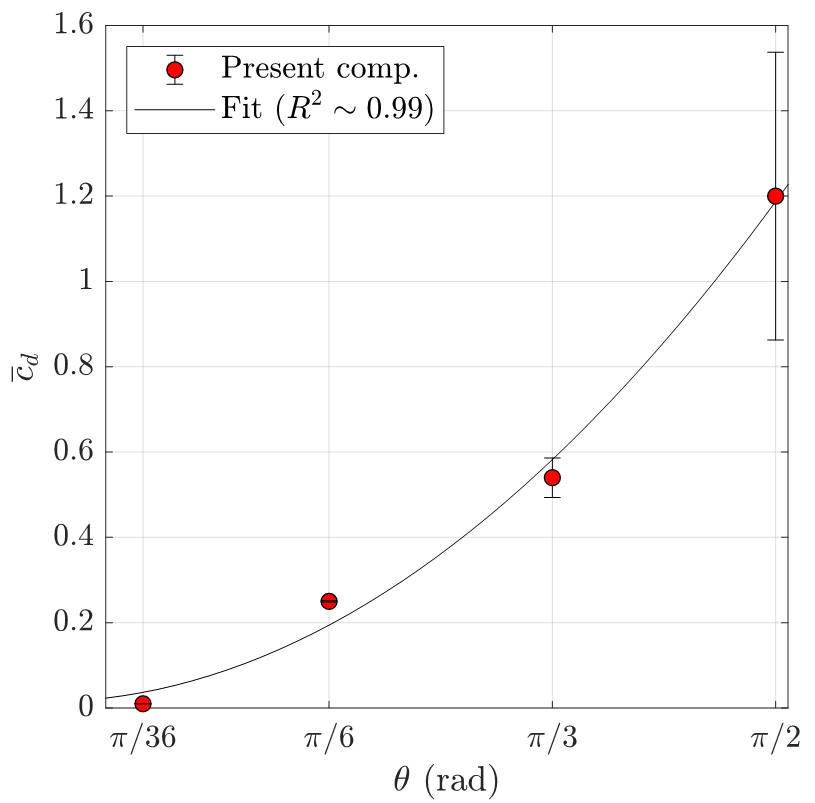}
	\caption{\label{fig:drag_var} Plot showing the variation of the mean drag coefficient ($\overline{c}_d$) for different control surface deflection angles ($\theta$) from the present numerical simulation as red-colored circular markers. The length of the error bar represents the standard deviation from the fluctuating $c_d$ signal in the self-sustained oscillation period, as shown in Figure \ref{fig:drag_signal}. The solid black line represents the adopted rational fit ($R^2\sim 0.99$) with the numerator and the denominator carrying a second and first-order polynomial, respectively.}
\end{figure}

The formation of the recirculation region during the deployment of a control surface is purely unsteady, as we have seen from the initial analysis of the instantaneous and time-averaged contour plots as shown in Figures \ref{fig:sch_M_contours} and \ref{fig:recirc_contour}. An unsteady flow field introduces variation in the total drag acting on the body. The coefficient of drag term ($c_d$) is used to understand the intensity of the unsteadiness forming ahead and behind the control surface as $c_d$ is directly affected by those adverse effects. Typical plots showing the variation $c_d$ for different $\theta$ from the beginning of the numerical solution are shown in Figure \ref{fig:drag_signal}-I. However, for collecting the required statistics, only the self-sustained oscillation zone is considered (marked as the shaded yellow patch). 

From a preliminary analysis, the variation in $c_d$ for different $\theta$ is qualitatively derived. At $\theta = [\pi/36]$, a sinuous wave with low amplitude and frequency is seen. At $\theta = [\pi/6]$, high frequency undulations are seen in $c_d$ variations. At $\theta=[\pi/3]$, the same high-frequency undulations are observed to be mildly riding on top of a low-frequency wave. At $[\pi/2]$, a distinct low-frequency wave with a considerably higher amplitude is seen in $c_d$ variation. In Figure \ref{fig:drag_signal}-II, using Fast Fourier Transform (FFT), the power spectra of those time-varying $c_d$ signals are computed for each cases of $\theta$. As observed in the qualitative analysis, a discrete low frequency tone is present at $[fD/u_\infty] = 0.26$ for $\theta = [\pi/36]$ (Figure \ref{fig:drag_signal} II-a). The production of small vertical velocity as the flow crosses the control surface deflecting at $\theta = [\pi/36]$ and immersed in a thick boundary layer ($\delta/D \sim t/D$) is suspected of producing the low-frequency component with least power. At $\theta = [\pi/6]$ (Figure \ref{fig:drag_signal} II-b), high frequency contents are observed with significant amplitude spread across a wide range of frequency bands. The dominant tone is seen at $[fD/u_\infty] = 0.51$. The reattaching shear layer experiencing a shorter path and a smaller recirculation zone volume produces shear layer instabilities with shorter wavelengths, thereby, higher frequency with significant power. 

At $\theta = [\pi/3]$ (Figure \ref{fig:drag_signal} II-c), the frequency bandwidth decreases and the dominant frequency shifts towards the lower end ($fD/u_\infty =0.27$). The separation and recirculation region volume increases considerably from that of the previous case. The separated shear layer impinges on the control surface tip, and a part of the fluid is expelled into the reattaching shear layer. The expulsion results in the convection of periodically excited shear layer structures further downstream, resulting in the observation of low-frequency contents in the $c_d$ spectra with a comparatively higher power. At $\theta=[\pi/2]$ (Figure \ref{fig:drag_signal} II-d), a discrete low frequency tone is observed at $[fD/u_\infty]=0.17$ with the highest power among all the other cases of $\theta$ in the $c_d$ variations. The unsteady shock oscillation, as observed in the spiked body flows called `pulsation'\cite{Feszty2004a} primarily results in the high-amplitude $c_d$ variation. 

\begin{figure*}
	\includegraphics[width=0.85\textwidth]{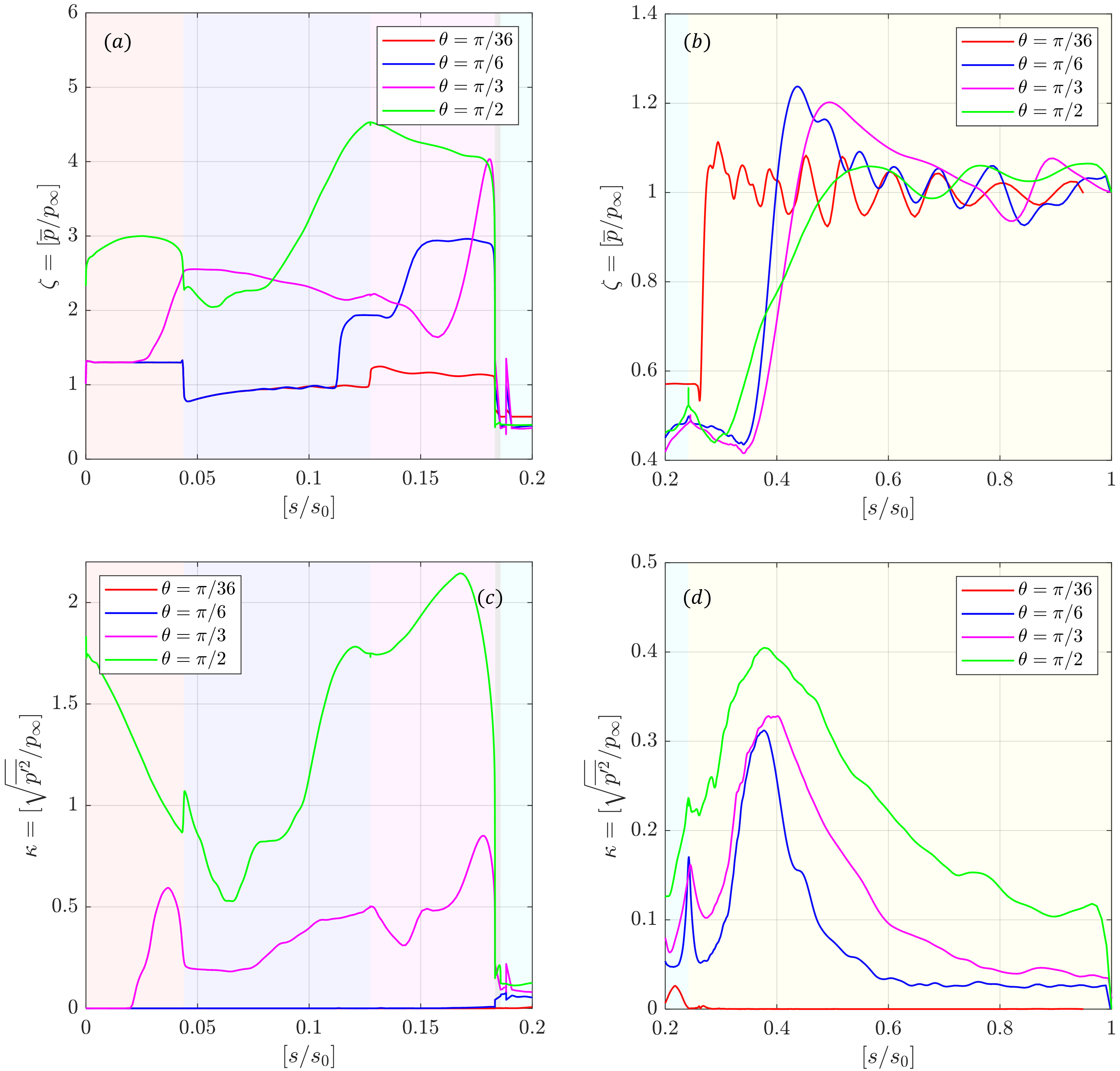}
	\caption{\label{fig:pmean_pstd} Typical plots showing the variation of the mean pressure or pressure loading ($\zeta$, a-b) and the RMS (root-mean-square) of the pressure fluctuations or fluctuation intensity ($\kappa$, c-d) for a wide range of control surface deflections ($\theta$). The variations are plotted in the surface coordinate system ($s$), and the $x$-axis is normalised by a reference surface length which is the maximum surface length of $[s_0/D]=7.89$ m for $\theta=\pi/2$. The shaded pale color zones mark the respective zones of the tested body specifically at $\theta=\pi/2$: $0 \leq [s/s_0] \leq 0.0438$ (red, nose tip region), $0.0438 \leq [s/s_0] \leq 0.1275$ (blue, upstream to control surface), $0.1275 \leq [s/s_0] \leq 0.1833$ (magenta, control surface - windward), $0.1833 \leq [s/s_0] \leq 0.1858$ (black, control surface tip), $0.1858 \leq [s/s_0] \leq 0.2416$ (cyan, control surface - leeward), $0.2416 \leq [s/s_0] \leq 1$ (yellow, downstream to control surface). Plots are segregated across $[s/s_0]$ into two to appreciate the local variations in $\zeta$ and $\kappa$.}
\end{figure*}

The variations of the mean drag coefficient ($\overline{c}_d$) with different control surface deflections ($\theta$) are plotted in Figure \ref{fig:drag_var}. A monotonic trend in $\overline{c}_d$ is seen. A rational type curve fitting as shown in Eqn. \ref{eq:cd_fit} is done to estimate the underlying trend with $R^2 \sim 0.99$. The numerator and the denominator part of the fitting equation carrying a second and a first-order polynomial as,
\begin{equation}
\label{eq:cd_fit}
    \overline{c}_d = \left[\frac{a_2\theta^2 + b_2\theta + c2}{\theta + d_2}\right].
\end{equation}
In Eqn. \ref{eq:cd_fit}, the constants are computed as $[a_2,b_2,c_2,d_2]=[1205,363.4,70.71,3043]$. The error bar's length in Figure \ref{fig:drag_var} represents the standard deviation of the time-varying $c_d$ signals. At $\theta = [\pi/36]$, owing to the low blockage, $\overline{c}_d$ is minimal with negligible deviation. On the other hand, $\theta = [\pi/2]$, $\overline{c}_d$ is computed to be almost 120 times higher than that of the previous case and the standard deviation being four orders larger than $\theta = [\pi/36]$. The reason is the higher surface area exposure followed by the control surface deployment and the resulting pulsation-type shock-oscillation phenomena.

\begin{figure*}
	\includegraphics[width=0.6\textwidth]{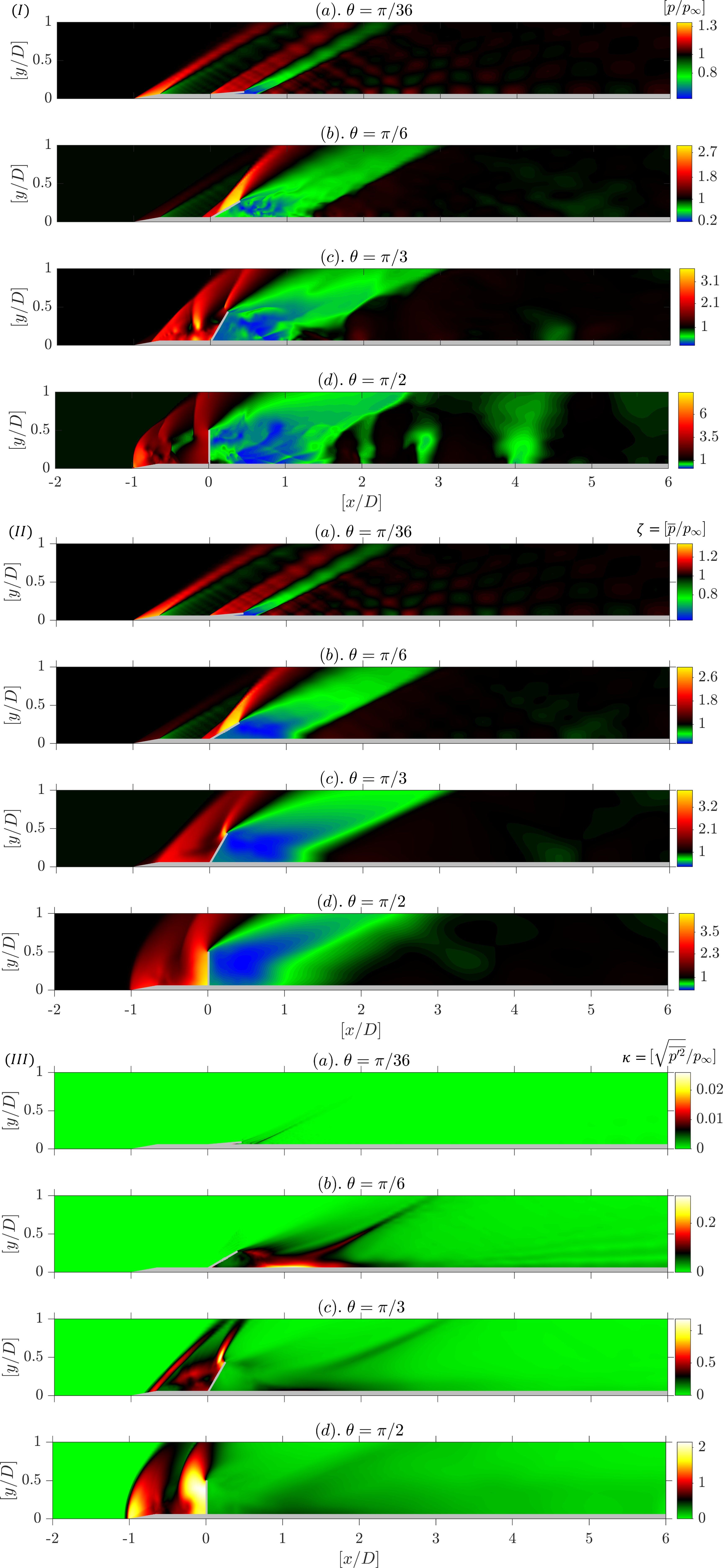}
	\caption{\label{fig:pressure_contours} \href{https://youtu.be/_WhFuGqlhfc}{(Multimedia View)} Typical contour plots showing the (I) instantaneous pressure at an arbitrary time, (II) time-averaged pressure, and (III) root-mean-square of the pressure fluctuations for a wide range of control surface deflection angles ($\theta$,a-d).}
\end{figure*}

\subsection{Mean pressure loading and fluctuation intensity} \label{ssec:press_load}

Unsteady loads experienced by the displacement of the control surface vary as $\theta$ changes. Critically, the control surface placed closer to the leading edge introduces a plethora of unsteady fluid motion as described before using the contour plots of Sec. \ref{ssec:inst_avg}. Unsteady loads acting on the flying body or the wall is accessed using the pressure loading and fluctuating intensity variations as,
\begin{align}
    \zeta =& \frac{\overline{p}}{p_\infty}, \label{eq:zeta}\\
    \kappa =& \frac{\sqrt{\overline{p'^2}}}{p_\infty}.
    \label{eq:kappa}
\end{align}
Calculation of $\zeta$ and $\kappa$ from Eqns. \ref{eq:zeta} and \ref{eq:kappa} merely indicate the average pressure and the root-mean-square (RMS) of the pressure fluctuations on the wall, respectively. In Figure \ref{fig:pmean_pstd}, values of $\zeta$ and $\kappa$ are plotted in two segments along the length. In Figure \ref{fig:pmean_pstd} a and c, only the leading-edge separation zone is highlighted ($0 \leq [s/s_0] \leq 0.2$), whereas in Figure \ref{fig:pmean_pstd} b and d, recirculation region and flow further downstream are emphasized ($0.2 \leq [s/s_0] \leq 1$). The plots background is shaded in different pale-colour patches representing each part of the tested model at $\theta=[\pi/2]$: red-nose tip region, blue-upstream to control surface, magenta-windward control surface, black-control surface tip, cyan-leeward control surface, yellow-downstream to control surface. Figure \ref{fig:pmean_pstd} is better understood with Figure \ref{fig:pressure_contours} which contains the field data of instantaneous, averaged, and RMS pressure for all the $\theta$. In Figure \ref{fig:pressure_contours} I-II, the colour map is selected so that the zones $p_\infty$ are always black, compression zones are red-yellow, and expansion zones are green-blue.

In the upcoming discussions, the readers are directed to refer to Figure \ref{fig:pmean_pstd} while referring to the quantified values of $\zeta$ and $\kappa$, whereas they are to refer Figure \ref{fig:pressure_contours} for locating the discussed flow features. At $\theta=[\pi/36]$, a weak shock generates at the nose-tip based on the nose cone angle ($\epsilon=[\pi/9]$) which rises $\zeta$ to 1.3 on the nose cone. Later, at the shoulder of the nose cone, the expansion fan generates and drops the pressure back to $\zeta=0.045$ and gradually recovers back to the freestream static pressure ($p_\infty$). Just at the beginning of the control surface deflection, a compression shock forms, which rises $\zeta=1.24$ and remains closer to that value until the control surface tip. Across the tip, the flow expands rapidly, dropping $\zeta$ to 0.22. The downstream conical zone where the recirculation bubble exists in the shadow of the deflecting control surface is at a low pressure dictated by the $\zeta$ mentioned above. Downstream the recirculation bubble, the expanding flow impinges on the model surface and turns back to the freestream, forming a recompression shock. The event results in a massive jump in $\zeta$ and recovers the local $\overline{p}$ to $p_\infty$. Further downstream, $\zeta$ oscillates about $p_\infty$ and gradually decays downstream. However, analysing the instantaneous and averaged pressure contours in Figure \ref{fig:pressure_contours} along with the \href{https://youtu.be/_WhFuGqlhfc}{Multimedia View}, standing compression and rarefaction waves are seen. As they remain stationary and inconsistent with the known physics, they are suspected to be an artefact of the higher-order numerical schemes used in the present simulation. While comparing the $\kappa$, values are mostly close to zero except near the shear layer reattachment zone where $\kappa \sim 0.025$.

At $\theta = [\pi/6]$, the upstream $\zeta$ follows as that of the $\theta = [\pi/36]$ case, except near the control surface. A small separation bubble is found ahead of the control surface deflection. It merely resembles a separation bubble ahead of a compression corner which is indicated by a jump in $\zeta \sim 1.9$ locating the separation point at $[s/s_0] \sim 0.11$. The plateau of $\zeta$ indicates the length of the separation bubble ($L_s$) up to $[s/s_0] \sim 0.14$. The second peak in $\zeta$ indicates the end of the separation bubble and the formation of a second compression shock with $\zeta \sim 2.9$. In the literature, the correlation for $L_s$ is given by,
\begin{equation}
\label{eq:sep_length}
    \frac{L_s}{\delta^*} \approx \frac{\theta^2 \sqrt{Re_s/C_w}}{M_s},  
\end{equation}
after approximating the pressure ratio across the bubble using the tangent wedge formula as given in the work of Katzer\cite{Katzer1989}, Mallinson\cite{Mallinson1996} and Anderson\cite{Anderson2019}. In Eqn. \ref{eq:sep_length}, $\delta^*$ is the displacement thickness measured ahead of the separation, $Re_s$ is the freestream Reynolds number based on the length up to the separation point ($Re_s \sim Re_D$), $M_s$ is the freestream Mach number ahead of the separation ($M_s \sim M_\infty$), and $\theta$ (rad) is the control surface deflection angle. The term $C_w$ is given by $(\mu_w/\mu_\infty)(T_\infty/T_w)$, where $\mu$ and $T$ represent dynamic viscosity and temperature. The subscripts $w$ and $\infty$ indicate the wall and freestream condition ahead of the separation, respectively. From the local jump in $\zeta$ (Figure \ref{fig:pmean_pstd}) and the averaged streamwise velocity distribution in Figure \ref{fig:recirc_contour} for $\theta=[\pi/6]$, the separation length and the displacement thickness is calculated as $[L_s/D]=0.2$ and $[\delta^*/D]=0.001$, respectively. Between the computed and estimated $[L_s/\delta^*]$ from the present simulation and Eqn. \ref{eq:sep_length}, a deviation of $\sim 5\%$ is seen which indicates a close prediction of expected flow field in $\theta=[\pi/6]$ case.

While monitoring $\zeta$ variation downstream the control surface, the flow expands more than the case of $\theta=[\pi/36]$ leading to the formulation of a low-pressure recirculation bubble nearly 1.3 times than the previous case. Closer to the reattachment point ($s/s_0 \sim 0.44$), $\zeta$ peaks at 1.23, which is the highest among all the cases indicating the presence of higher dynamic pressure ahead of the stagnation at the reattachment. The flow gradually recovers to $p_\infty$ at $[s/s_0] \sim 0.64$ and oscillates about it as the convecting structures create local compression and rarefaction pockets. While monitoring $\kappa$ variations, the separation bubble zone is laden with significant values of $\kappa$ of about 0.06. Around the conical shadow zone in the downstream wake, $\kappa$ peaks at about 0.24. It is almost the same for the $\theta=[\pi/3]$ case. The peak indicates the fluctuations arising due to the corner or Moffat vortices. Further downstream, the $\kappa$ value increases to about 0.31 in the reattachment zone, which is the least among the rest of the $\theta$ cases, indicating the small-scale fluctuations arising from the convecting structures in the reattaching shear layer. As a result of the convecting structures downstream, after $[s/s_0] \sim 0.64$, the values of $\kappa$ settle down to almost $0.025$.

At $\theta=[\pi/3]$, $\zeta$ values remain the same as in previous cases, only up to the nose-tip region. However, the separation bubble dominates the nose tip and control surface region. At the separation point, $\zeta$ rises to about 2.5 and gradually decreases to about 2.2, close to the control surface. At the reattachment point on the control surface, which is significantly higher (closer to its tip) than the $\theta=[\pi/6]$ case, $\zeta$ rises to about 4. Part of the fluid impinges and re-enters into the separation bubble, whereas the rest escapes through the tip. The resulting ejection of unsteady fluid mass excites the reattaching shear layer forming comparatively large-scale structures than the $\theta=[\pi/6]$ case. The base pressure behind the control surface remains the same as that of the previous case indicating the exact extent of fluid expansion at the control surface tip. The peak value of $\zeta$ at the reattachment point is seen at $[s/s_0] \sim 0.49$ to about 1.2. It is about 2.5\% lesser than the previous case indicating the fluid momentum was lost due to the diversion into the separation bubble. Moreover, $\zeta$ approaches one only far later at $[s/s_0] \sim 0.77$ compared with the previous as the large-scale structures take some considerable distance to equilibrium with the freestream.

Near the separation point, $\kappa$ values peak at about 0.6. It is an indication that there is a severe oscillation of the separation shock about the separation point. Values of $\kappa$ progressively increase while approaching the control surface, signifying the severe interaction of pressure waves inside the separation bubble. Closer to the control surface tip, the flow reattaches and $\kappa$ peaks again to about 0.85, implying the impingement of small and large-scale structures from the separated shear layer. At the downstream corner of the control surface, Moffat vortices introduce another peak in $\kappa$ as seen in the previous case. At the reattachment point, $\kappa$ is significantly more extensive and shifted slightly downstream than the $\theta=[\pi/6]$ case signalling the subsequent interaction and stagnation of the comparatively large-scale structures ejected along the reattaching shear layer. Saturation in downstream $\kappa$ value of about 0.04 is observed at $[s/s_0] \sim 0.8$. The behaviour is due to the bifurcation of fluid mass further into the recirculation zone and eventual expulsion of the residuals to the freestream. As the recirculation bubble length is comparatively longer than in the previous case, the saturation distance is also pushed further downstream.

\begin{figure*}
	\includegraphics[width=0.8\textwidth]{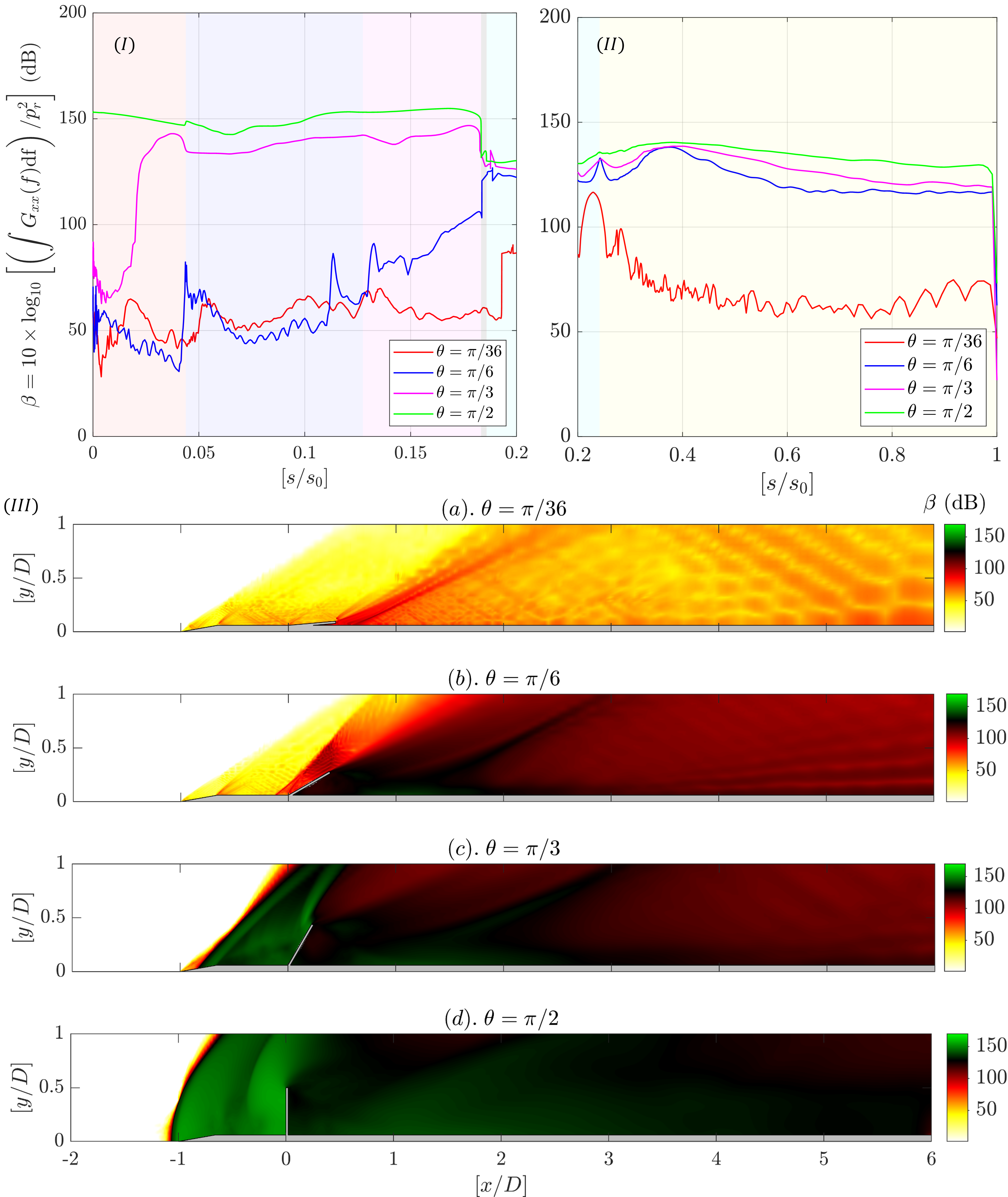}
	\caption{\label{fig:acoustics} (I-II) Plots showing the acoustic loading on the model wall. See caption of Figure \ref{fig:pmean_pstd} to interpret the background color patches. (III) Contour plots of acoustic loading around the model having a wide range of control surface deflections ($\theta$, a-d)}
\end{figure*}

At $\theta=[\pi/2]$, the shock systems ahead are not oscillating about a point but across the control surface and the nose-tip. The pulsation produces a rapid rise in pressure from the nose-tip, leading to a higher $\zeta$ of about 2.7. Eventually, a rise to 4.5 at the bottom of the control surface is also observed. The resulting recirculating fluid due to the ramification of shocks on the control surface is expelled during the `collapse' cycle of the shock pulsation. A resulting expulsion from the control surface tip rises $\zeta$ to almost four as that of the $\theta=[\pi/3]$ case, except the expulsion in the present case, contains a discrete frequency matching the shock pulsation cycle. Between the shock pulsation cycle, the recirculation bubble experiences a wide range of base pressures, as seen in the case of $\theta=[\pi/36]$ and $\theta=[\pi/3]$. Due to the maximum blockage offered by the control surface deflection to $\theta=[\pi/2]$, the recirculation bubble length is the longest. Hence, peak $\zeta$ shifts to $[s/s_0] \sim 0.56$ and carries a meagre value of 1.06 at the reattachment point, indicating a significant total pressure loss in the upstream fluid flow to mostly the shock systems encountered in a typical pulsation cycle. Owing to the less momentum carried by the reattaching fluid flow, pressure equilibrium ($\zeta$ sim 1) of expelled structure is quickly achieved at $[s/s_0] \sim 0.66$.

Due to the pulsating shock oscillation, $\kappa$ is pushed to the highest value among all the other cases, especially towards the control surface tip. At that location, $\kappa$ is about 2.1, almost 11 times higher than the $\theta=[\pi/3]$ case. Downstream the control surface, the corner vortices are stronger near the corner due to the periodic ejection of large fluid mass among all the cases along the reattaching shear layer. Hence, $\kappa$ is pushed to a maximum value of 0.23 around the corner. It is almost 43\% higher than the last two cases. The value of $\kappa$ is observed to be 0.4. It is 25\% larger than the previous case alone, owing to the same reason at the reattachment point. The large-scale structures exhibit integrity and do not deform or decay as much as in the previous cases. It leads to the non-observation of a plateau region in $\kappa$ along $[s/s_0]$, despite a decay in $\kappa$. 

\subsection{Acoustic loading on the model and the freestream} \label{ssec:acoustics}
Some of the key physics associated with separation flow ahead and the recirculating flow behind the deflected control surface are discussed so far with key flow variables. All four types of unsteadiness between $[\pi/36] \leq \theta \leq [\pi/2]$ are categorized as less, moderate, significant, and severe based on the separation bubble length scaling, $c_d$ trend, and $\zeta$ and $\kappa$ variations. One practical requirement from an unsteady flow field is to facilitate an outcome through which the resulting fluid-structure interaction can be measured. Hence, quantifying the pressure-induced fluid-structure interaction on the model and around the unsteady flow field is one way to achieve the objective. It helps develop a suitable protective/redundant system and a stable flying vehicle. Variations in the pressure fluctuation spectra are used to calculate a non-dimensional parameter called acoustic loading ($\beta$) as, 
\begin{equation}
\label{eq:beta}
    \beta = 10\times\log_{10}\left[ \frac{ \left( \int G_{xx}(f) \rm{df} \right)}{p_r^2} \right].
\end{equation}
In Eqn. \ref{eq:beta}, $G_{xx}(f)$ represents the power spectral density in Pa$^2$/Hz, and $p_r$ is the reference pressure of 20 $\mu$Pa. Despite the term $\beta$ being a non-dimensional parameter, it is conventional in the field of applied mechanics to use the term $\beta$ as `Overall Sound Pressure Level (OSPL)' and refer to it with a unit called decibel (dB). Figure \ref{fig:acoustics} presents the variation of calculated $\beta$ along the model wall Figure \ref{fig:acoustics} I-II and also as contour plots  Figure \ref{fig:acoustics}-III.

At $\theta = [\pi/36]$, $\beta$ is at the minimum everywhere along the model wall among all the other cases. On average, $\beta$ is around 50 dB ahead of the control surface. In the recirculation zone where the unsteadiness creep-in, $\beta$ peeks to 115 dB and gradually plateaus to 50 dB. In the contour plot, the spatial field is also laden with a similar noise signature ahead and behind the control surface. A small separation bubble is ahead of the control surface at $\theta = [\pi/6]$, which rises $\beta$ to about 100 dB. At the reattachment point, $\beta$ peaks at about 138 dB and gradually approaches 116 dB. In the contour plots, across the separation-induced shock, expansion fan, and recirculation bubble, $\beta$ increases. At $\theta = [\pi/3]$, $\beta$ is about 135-145 dB between the separation to the reattachment point. The plateaued values of $\beta$ between those two points indicate the presence of strong acoustic wave interactions inside the separation bubble. Behind the control surface, the peak value at the reattachment point remains the same. However, $\beta$ saturates at 120 dB, slightly higher than the previous case. The contour plot of $\beta$ shows that the separation bubble is severe laden interaction than that of the recirculation bubble. On average, the separation bubble field is laden with $\beta=$160 dB, whereas the separation bubble contains mostly $\beta=130$ dB. The drop is attributed to most of the flow energy loss to the upstream interactions.

At $\theta=[\pi/2]$, $\beta$ is about 150 dB everywhere on the wall from the nose-tip to the reattachment point or the control surface tip. Periodic forcing by the pulsating shock is attributed to the observation of almost a constant value of $\beta$ upstream of the control surface. The downstream variation in $\beta$ follows the trend of the previous case, including a similar peak at the reattachment point. However, the saturation of $\beta$ further downstream is observed slightly at a higher value of about 130 dB at a comparatively larger $[s/s_0]$. Investigating the respective contour plot, the pulsating shock zone ahead of the control surface has $\beta$ at about 170 dB. Behind the control surface, the intensity only drops to 160 dB, and the spatial extent is felt at least up to $[x/D] \sim 4$, which was not the case in the previous one. Once again, the periodic ejection of comparatively large-scale structures from the fully-deployed control surface owing to pulsation is the prime reason for a slight reduction in $\beta$ around the recirculation region. 

\begin{figure*}
	\includegraphics[width=0.9\textwidth]{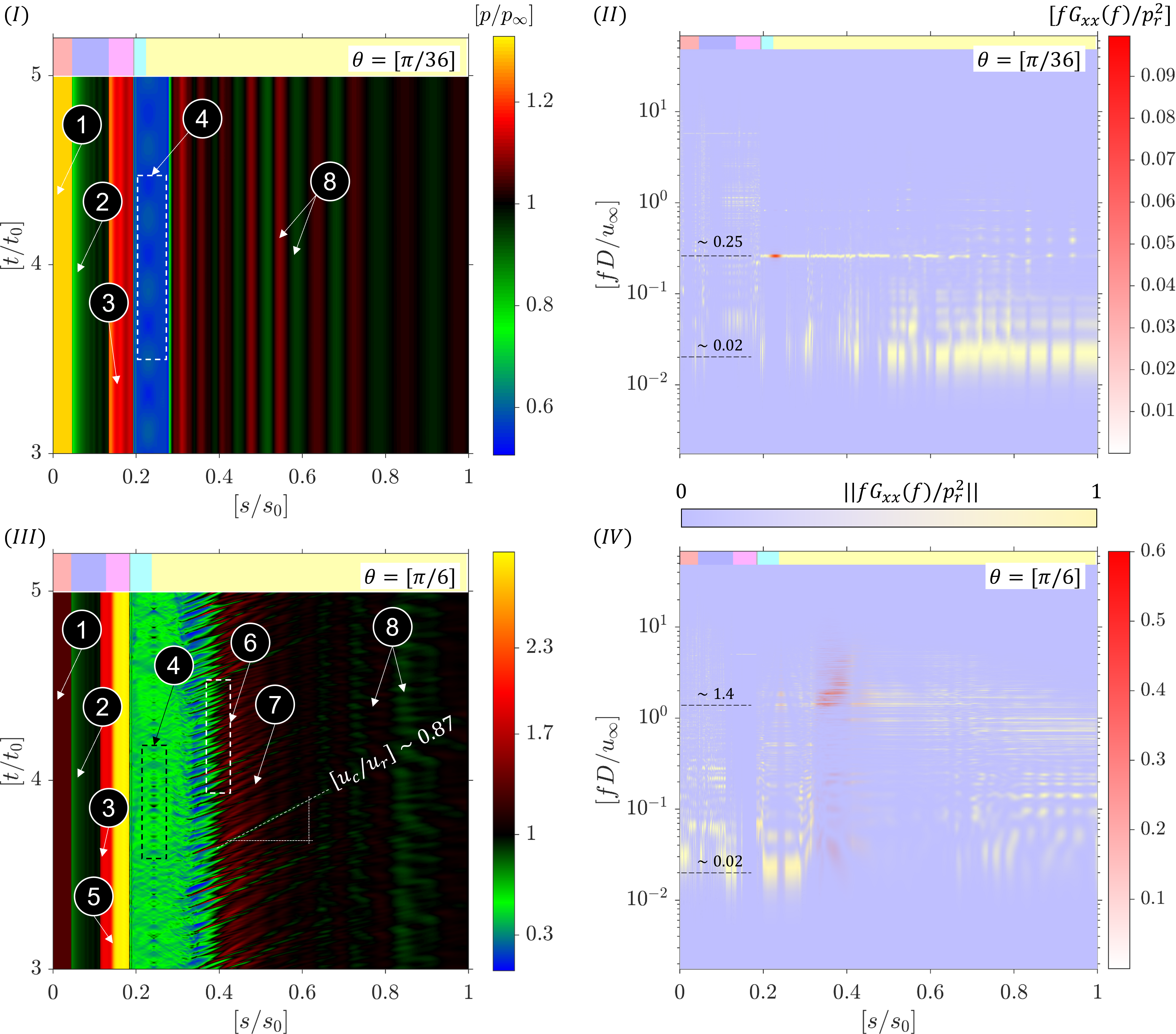}
	\caption{\label{fig:xt_part1} \href{https://youtu.be/y2RTmyok-Yc}{(Multimedia View)} (I, III) Construction of $x-t$ plots by stacking the instantaneous pressure ($p/p_\infty$) observed on the model wall in surface coordinate system ($s/s_0$) over the observed period of time ($t/t_0$) for $\theta=[\pi/36]$ and $[\pi/6]$. The colour map is selected in a manner to reflect freestream (black), compression (red-yellow), and expansion (green-blue). See the caption of Figure \ref{fig:pmean_pstd} to interpret the background colour patches given at the contour plot's top. Key flow features: 1. leading-edge shock, 2. expansion waves at the corner, 3. compression shock or shock ahead of the separation bubble, 4. pressure fluctuation in the corner, 5. reattaching shock behind the separation bubble, 6. reattaching shear layer, 7. convecting compressed fluid structures, and 8. local pockets of compression and rarefaction. (II, IV) Corresponding contour plots showing the Power Spectral Density (PSD) by performing a fast Fourier transform (FFT) on the $x-t$ images along $s/s_0$. Two overlays of PSD are given for analysis: 1. normalized PSD ($\lVert fG_{xx}(f)/p_r^2\rVert$) given by the blue-yellow contours, and 2. absolute PSD ($fG_{xx}(f)/p_r^2$) given by transparent-red contours. The reference pressure is $p_r=0.1$ kPa.}
\end{figure*}

\begin{figure*}
	\includegraphics[width=0.85\textwidth]{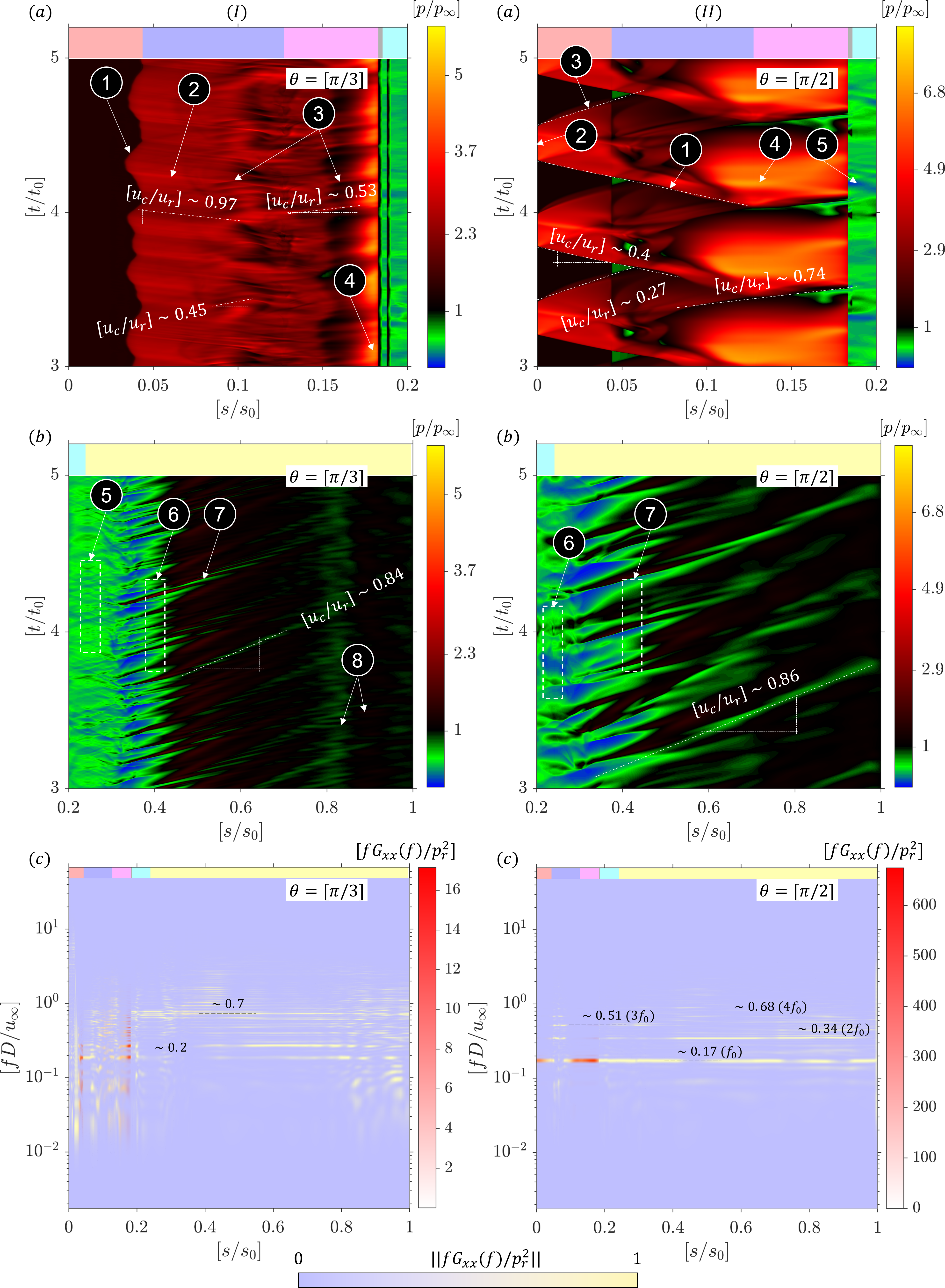}
	\caption{\label{fig:xt_part2} \href{https://youtu.be/y2RTmyok-Yc}{(Multimedia View)} Constructed $x-t$ plots by stacking the instantaneous pressure ($p/p_\infty$) observed on the model wall in surface coordinate system ($s/s_0$) over the observed period of time ($t/t_0$) for $\theta=[\pi/3]$ (I) and $[\pi/2]$ (II). Two $x-t$ plots are given to analyse the upstream (a) and downstream (b) structures. The colour map is selected in a manner to reflect freestream (black), compression (red-yellow), and expansion (green-blue). (I a-b) Key flow features: 1. leading-edge separation, 2-3. left and right running waves, 4. separated shear layer's reattachment point, 5. chaos near the corner, 6. downstream reattaching shear layer, 7. convecting structures, and 8. pocket of rarefaction and compression. (II a-b) Key flow features: 1-3. Phases of pulsation: inflate, with-hold, and collapse, 4. high-pressure zone while stagnating near the control surface, 5. rapid expansion near the tip, and 6-7. as mentioned in (I a-b). (I-c, II-c) Corresponding contour plots showing the Power Spectral Density (PSD) by performing a fast Fourier transform (FFT) on the $x-t$ images along $s/s_0$. Two overlays of PSD are given for analysis: 1. normalized PSD ($\lVert fG_{xx}(f)/p_r^2\rVert$) given by the blue-yellow contours, and 2. absolute PSD ($fG_{xx}(f)/p_r^2$) given by transparent-red contours. The reference pressure is $p_r=0.1$ kPa. See the caption of Figure \ref{fig:pmean_pstd} to interpret the colour patches given at the contour plot's top.}
\end{figure*}

\subsection{Analysis of $x-t$ and $x-f$ plots} \label{ssec:xt_xf}

The last section elaborates on the effects of acoustic loading on the model and the surrounding spatial field. The present section explores the dynamic events happening along the model length using the construction of $x-t$ plots. The resulting $x-t$ plots are subjected to a fast Fourier transform (FFT) along the spatial length to compute the power spectral density (PSD). The resulting contour plots are called $x-f$ plots. As in the previous case, in the present analysis, we are also interested in the time-varying wall-static pressure ($p/p_\infty$) as it determines the total power and dominant frequency acting on the model structure. Firstly, instantaneous $[p/p_\infty]$ is collected along the model wall in the surface coordinate system $[s/s_0]$ over a considerable amount of time ($t/t_0$) where the flow is stationary (excluding the transient period required for the flow to develop). Later, the $x-t$ plot is constructed by stacking the instantaneous $[p/p_\infty]$ along $[s/s_0]$ in the $x$-direction over the considered $[t/t_0]$ along the $y$-direction. Along the $[s/s_0]$ path, the line color mark the $[p/p_\infty]$ variation: freestream (black), compression (red-yellow), and expansion (green-blue), thereby the entire contour developed at the end of the $x-t$ plot construction. Typical $x-t$ plots for the case of $\theta=[\pi/36]$ and $[\pi/6]$ are given in Figure \ref{fig:xt_part1} I and III along with the \href{https://youtu.be/y2RTmyok-Yc}{animated video} explaining the construction of $x-t$ plot itself. 

The obtained $x-t$ plots are converted into $x-f$ plots by performing spectral analysis along $[s/s_0]$. The normalized spectra ($fG_{xx}(f)/p_r^2$) helps in marking the influence of the local dominant spectra and its coupling along $[s/s_0]$ using blue-yellow color map (Figure \ref{fig:xt_part1} II and IV). On top of those plots, absolute spectra and their power are overlaid to track the global variation using a transparent-red colour map (Figure \ref{fig:xt_part1} II and IV). Let's consider $\theta=[\pi/36]$ case in Figure \ref{fig:xt_part1} II to understand the events better. Until, $[s/s_0] \sim 0.19$, there are no dominant local spectra that is continuing from the nose-tip. However, after $[s/s_0] \sim 0.19$, a dominant local spectra at $[fD/u_\infty] \sim 0.25$ is seen up to $[s/s_0] \sim 0.5$. Although the local spectra is identified in the manner mentioned above, the global spectra is only concentrated on the red-spotted region which is bounded between $0.21 \leq [s/s_0] \leq 0.24$, where the flow recirculates in the corner (Figure \ref{fig:xt_part1} I). Thus, in summary, one can say that the corner zones experience the most fluctuating pressure with a dominant spectrum of $[fD/u_\infty] \sim 0.25$. However, the forcing is felt until $[s/s_0] \leq 0.24$ at a low intensity.

The case of $\theta=[\pi/6]$ is analysed using the earlier tools and interpretation methods. Some more interpretations can also be obtained from them. For example, in the $x-t$ plots, if there are no visible streaks persisting along the temporal direction, then the event zone is declared as steady, just as in the case of the separation bubble for $\theta=[\pi/6]$. On the other hand, if the $x-t$ plots contain streaks running left or right, they indicate the structures that are convecting from left to right or vice versa. The inverse slope of the streak from a $x-t$ plot directly gives the speed of the convecting structures. The frequencies in these zones are extracted from the respective $x-f$ plot. From the $x-t$ plot in Figure \ref{fig:xt_part1} III, the unsteady interaction region is isolated only to the recirculation bubble and the separation bubble is found to be merely steady. After the reattachment point, the structures from the reattaching shear layer convects at a speed of $[u_c/u_r] sim 0.87$, where $u_r=[s_0/t_0]$ with $s_0=394$ m/s, and $t_0=1$ ms. From the $x-f$ plot in Figure \ref{fig:xt_part1} IV, the upstream (of the control surface) is laden with a low power spectrum at $[fD/u_\infty] \sim 0.02$. On the other hand, the downstream contains the highest power at $[fD/u_\infty] \sim 1.4$ near the reattachment point.

The $x-t$ and $x-f$ plots for the cases of $\theta=[\pi/3]$ and $[\pi/2]$ are provided separately in Figure \ref{fig:xt_part2} I-II. In particular, the $x-t$ plots are bifurcated into two parts to appreciate the upstream and downstream dynamics as shown in Figure \ref{fig:xt_part2} I a-b and II a-b. At $\theta=[\pi/3]$, leading-edge separation occurs, which is marked by the sudden pressure rise in the $x-t$ plot. Inside the separation region, there are many left and right streaks, indicating the dominance of sound waves-based interaction. The curling-separated shear layer and the subsequent impingement on the control surface are attributed to the production of these sound waves. The bifurcation of left and right running waves originate almost close to the beginning of the upstream control surface. When the reattaching shock is displaced, the right running waves carry the entrapped fluid mass and eject across the reattaching point of the separated shear layer. A typical velocity of these right running waves is roughly about $[u_c/u_r] \sim 0.53$. Many left-running waves originate from the beginning of the control surface or closer to the reattaching separated shear layer. These waves interact with a part of the right-running waves suddenly appearing in the middle of the separation bubble. 

The sudden appearance is attributed to the convection and the growth of the shear layer structures arising from Kelvin-Helmholtz (KH) instabilities. These structures curl, and they produce a sound pulse. A part of the pulse is convected at supersonic speed on the top of the shear layer. At the same time, the rest is radiated into the separation bubble. Due to the speed of these convecting structures, the radiated pulses only reach the base of the separation bubble after a certain length. Hence, a sudden appearance of right-running waves halfway through the separation bubble region. Only a few left-running waves survive the interaction and convect upstream at a velocity of about $[u_c/u_r] \sim 0.97$. The waves interact with the separation point and further perturb the separated shear layer. The perturbed separated shear layer enters the unstable state and curls to form KH structures. These structures produce sound pulses which propagate inside the separation bubble and eventually form a closed feedback loop and a self-sustained oscillation. In the $x-f$ plot of Figure \ref{fig:xt_part2} I-c, a strong frequency at $[fD/u_\infty] \sim 0.2$ is seen. At the separation point ($s/s_0 \sim 0.04$), only a dominant peak is seen, whereas at the reattachment point ($s/s_0 \sim 0.17$), multitude of peaks are seen with $[fD/u_\infty] \sim 0.2$ being the dominant one. The coupling of frequencies from the periodically impinging shear layer structures, locally oscillating reattachment shock and the mass ejection from the separation bubble is attributed to the multitude of peaks occurring in the $x-f$ plot.

\begin{figure*}
	\includegraphics[width=0.9\textwidth]{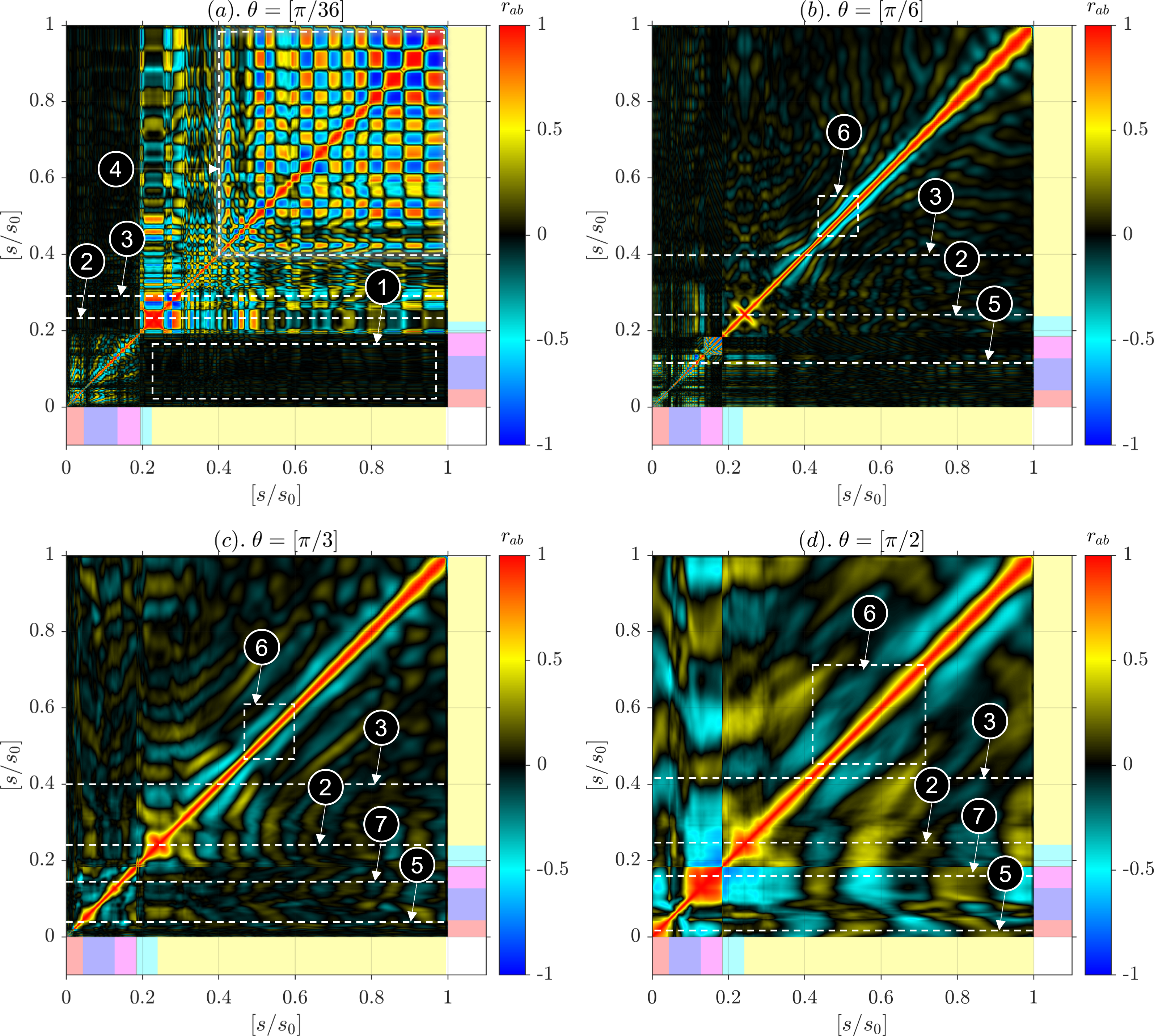}
	\caption{\label{fig:correlation_spectra} Contour plots showing the Pearson's correlation coefficient map obtained by linearly changing the reference point to be correlated with the rest of the time-varying wall-static pressure ($p/p_\infty$) in the surface coordinate system ($s/s_0$). The colour map is selected in a manner that reflects no correlation (black), positive correlation (yellow-red), and negative correlation (cyan-blue). See the caption of Figure \ref{fig:pmean_pstd} to interpret the background colour patches given at the contour plot's bottom and right. Features: 1. no downstream correlation, the dash lines represent the correlation about the following entities: 2. corner, 3. downstream reattaching shear layer, 5. upstream separation bubble, and 7. upstream reattaching shear layer; 4. downstream pockets of compression and rarefaction, and 6. downstream coherent shedding structure size variation.}
\end{figure*}

Downstream the control surface for $\theta = [\pi/3]$ case in Figure \ref{fig:xt_part2} I-b, the $x-t$ plot reveal the presence of chaotic streaks closer to the corner. Further downstream, a series of definite streaks are running on the right side. The origin of the streaks marks the location of the reattaching shear layer on the model wall. A part of the accelerating fluid mass turns into a recirculation bubble. The rest is ejected downstream, resulting in the observation of a strong low-pressure zone (identified as blue colour). However, due to the recompression shock downstream  (which arises as the supersonic flow recovers back to the freestream direction), a sharp streak is also seen closer to the reattachment point. The structures crossing the recompression shock convect at a velocity of about $[u_c/u_r]\sim 0.84$, which is slightly less than in the case of $\theta=[\pi/6]$. By analysing the respective $x-f$ plot in Figure \ref{fig:xt_part2} I-c, a dominant frequency is seen at $[fD/u_\infty] \sim 0.7$ along with its harmonics and some more peaks of their combination.

At $\theta=[\pi/2]$, the upstream is dominated by shock pulsation. In the $x-t$ diagram shown in Figure \ref{fig:xt_part2} II-a, three distinct phases of the pulsation cycle are marked: 1. inflate, 2. with-hold, and 3. collapse. The convective shock velocity during the inflate and collapse cycle is identified from the $x-t$ plot as right and left running waves. Their velocities are about $[u_c/u_r]\sim 0.4$ and $[u_c/u_r]\sim 0.27$, respectively. The collapsing shock interacts with the inflating shock, thereby undergoing refraction and gaining an even higher velocity of about $[u_c/u_r]\sim 0.74$. The refraction of shock and further fluid compression against the control surface raise the static pressure and hence, the highest pressure on the control surface. A similar chaotic pattern is seen close to the downstream control surface's corner (Figure \ref{fig:xt_part2} II-b) as seen in the $\theta=[\pi/3]$ case. The streak patterns close to the reattachment point are coarsely packed due to the low-frequency pulsation, unlike the previous case. The ejected fluid structures are convecting in a similar flow velocity as the previous cases at about $[u_c/u_r] \sim 0.86$. At the same time, analysing the respective $x-f$ plot in Figure \ref{fig:xt_part2} II-c, the pulsating shock in the upstream shows a hot spot only in the leading edge and close to the control surface. The absolute values are indeed an order higher than that of the previous case. Moreover, the distinct pulsation cycle results in the observation of a fundamental ($f_0$) and its harmonics ($2f_0, 3f_0,$ and $4f_0$) on the $x-f$ plot. The harmonics are pronounced stronger in the middle of the upstream zone where the collapsing and inflating shock interacts. In general, the fundamental is observed at $[fD/u_\infty]\sim 0.17$, and it is seen to dominate everywhere, however, with a lower intensity downstream the control surface. 

\subsection{Correlation coefficient of pressure fluctuations} \label{ssec:cross_corr}

The present section discusses how the influence of unsteady events at one location on the model surface is felt everywhere else. Instantaneous wall-static pressure ($p_w=p/p_\infty$) on the model wall in the surface coordinate system ($s/s_0$) is considered for this investigation as shown in Eqn. \ref{eq:pw}. In order to come up with a measure to quantify the influence using $p_w$, Pearson's correlation coefficient ($r_{ab}$) as shown in Eqn. \ref{eq:pearson} is sought. In Eqn. \ref{eq:pearson}, $a$ and $b$ indicate the time-varying pressure fluctuation data set at two locations, and $n$ represents the total number of sample points in each set. Values of $r_{ab}$ are obtained by taking one point as a reference and correlating it with the rest. Later, a contour plot of $r_{ab}$ is constructed by moving the reference point linearly from the nose-tip to the end of the model. Typical plots of $r_{ab}$ are shown in Figure \ref{fig:correlation_spectra} for a wide range of $\theta$. As the reference point for $r_{ab}$ calculation is moved linearly, there is a diagonal line with $r_{ab}=1$ running from the bottom to the top rightwards. Hence, the $r_{ab}$ contour plot is symmetric about the diagonal. However, the upper and lower triangular portions of the $r_ab$ square matrix are retained to avoid complexity. Similarly, as the correlation is happening between the points lying on the same model, $x$ and $y$ axis contain the same $[s/s_0]$. Hence, colour patches are given at the bottom and right-hand sides to indicate the model surface zone. The readers are directed to the caption of Figure \ref{fig:pmean_pstd}, where they are described first, to associate the colour patches with the respective zone on the model surface. 
\begin{equation}
    \label{eq:pw}
    p/p_\infty = f(s/s_0,t/t_0).
\end{equation}

The physics from these plots can be understood in the following manner. Draw a horizontal line corresponding to the $[s/s_0]$ along which the correlation is sought. The equation of the horizontal line describes the reference value against which the rest of the values are correlated. In Figure \ref{fig:correlation_spectra}, the color map indicates the extent of correlation. Red-blue colours represent the extrema (+1 and -1) and hence, a strong correlation, whereas yellow-cyan colours indicate a moderate correlation (+0.5 and -0.5). Black colour points to the zone where there is no correlation at all. Let's consider the case of $\theta=[\pi/36]$ (Figure \ref{fig:correlation_spectra}a). If a horizontal line is drawn at $[s/s_0]=0.1$ (between the nose-cone and windward control surface), then the point $[s/s_0]=0.1$ becomes a reference against which the rest of the values are correlated. Hence, value of $r_{ab}$ becomes 1 along the correlation line at $[s/s_0]=0.1$, indicating auto-correlation. Ahead of $[s/s_0]=0.1$, very minute islands of correlation contours (yellow and cyan colour) sit. However, no contours of correlation with significant magnitude (only black colour) are found downstream. Thus, the resulting variation of $r_{ab}$ along that specific line is interpreted so there is no noticeable correlation ahead and behind the selected point. 
\begin{equation}
    \label{eq:pearson}
    r_{ab} = \displaystyle \frac{ \displaystyle n\left(\sum a b\right) - \left(\sum a\right)\left(\sum b\right)}{ \displaystyle \sqrt{\left[n\sum a^2 - \left(\sum a\right)^2\right]\left[n\sum b^2 - \left(\sum b\right)^2\right]}}.
\end{equation}

With the aforementioned route map, rest of the $r_{ab}$ contours are read in Figure \ref{fig:correlation_spectra}a. Two lines are drawn: one at the downstream control surface's corner ($s/s_0 \sim 0.22$), and the next one close to the downstream reattaching shear layer ($s/s_0 \sim 0.3$). The events happening in the downstream corner are not correlated anywhere upstream. On the other hand, two large-scale structures of size $[\Delta s/s_0] \sim 0.024$ are immediately seen downstream in blue and red colours, indicating the structures in the recirculation zone. Structures further downstream indicate simply the pockets of compression and rarefaction. Analysing the line passing through the reattachment point, one can see that the leeward side of the control surface is strongly correlated with the reattaching point. Similarly, a series of coherent structures with opposite magnitudes is seen after $[s/s_0] \sim 0.4$, indicating the influence of fluctuations in the reattaching point to the local pressure pockets. Checkered flag-like structures are seen at the top right, pointing to the correlation among the local pressure pockets downstream.

At $\theta=[\pi/6]$ (Figure \ref{fig:correlation_spectra}b), a small separation bubble is seen at the beginning of the control surface's windward side. A correlation line is drawn along the head side of the separation bubble at $[s/s_0] \sim 0.11$. Between the separation bubble length, head and tail shocks are present. Hence, the correlation value peaks with extrema between the bubble. However, the trailing shock is moderately correlated with the heading shock. Positive coherent forcing is felt everywhere downstream until the reattachment point of the reattaching shear layer. A strong correlation is seen near the corner, where the red colour peak correlation is spread around it. The upstream separation bubble is moderately correlated with the corner and almost weakly correlated with the rest of the downstream structures. For the correlation line drawn about the downstream reattachment point ($s/s_0 \sim 0.4$), no upstream coherence is seen. However, it produces an immediate negative correlation across the reattaching point. The thickness of the negative correlation strands around the reference indicates the length scale of the shedding structures in the reattaching shear layer. In the present case, the thickness of the strand is measured as $[\Delta s/s_0] \sim 0.02$, which is comparatively smaller than the rest of the three cases.

At $\theta = [\pi/3]$ (Figure \ref{fig:correlation_spectra}c), correlation lines are drawn at four places: leading-edge separation point ($s/s_0 \sim 0.04$), reattachment point of the separated shear layer ($s/s_0 \sim 0.155$), downstream corner ($s/s_0 \sim 0.24$), and downstream reattachment point ($s/s_0 \sim 0.4$). Analysing the first two lines, one can see that the upstream separation and reattachment point are negatively correlated. When one is negative, then the other is positive. The separation and reattachment shock hangs about the point and oscillates, indicating that there must be an out-of-phase shock motion as described in the work of \cite{Karthick2021,Karthick2022}. Downstream the control surface, only weak correlations are seen. Upstream has a mild correlation about the corner, whereas downstream is strongly correlated up until the downstream reattaching point. About the reattachment point, no upstream correlation is seen like before. The immediate negative correlation indicating the length scale of the structures in the reattaching shear layer is seen across the reattachment point. It is now comparatively larger than the previous case with $[\Delta s/s_0] \sim 0.04$, which is two times bigger. These downstream correlations persist with a similar pattern indicating the convection of these structures along the model surface until its end.

\begin{figure*}
	\includegraphics[width=0.9\textwidth]{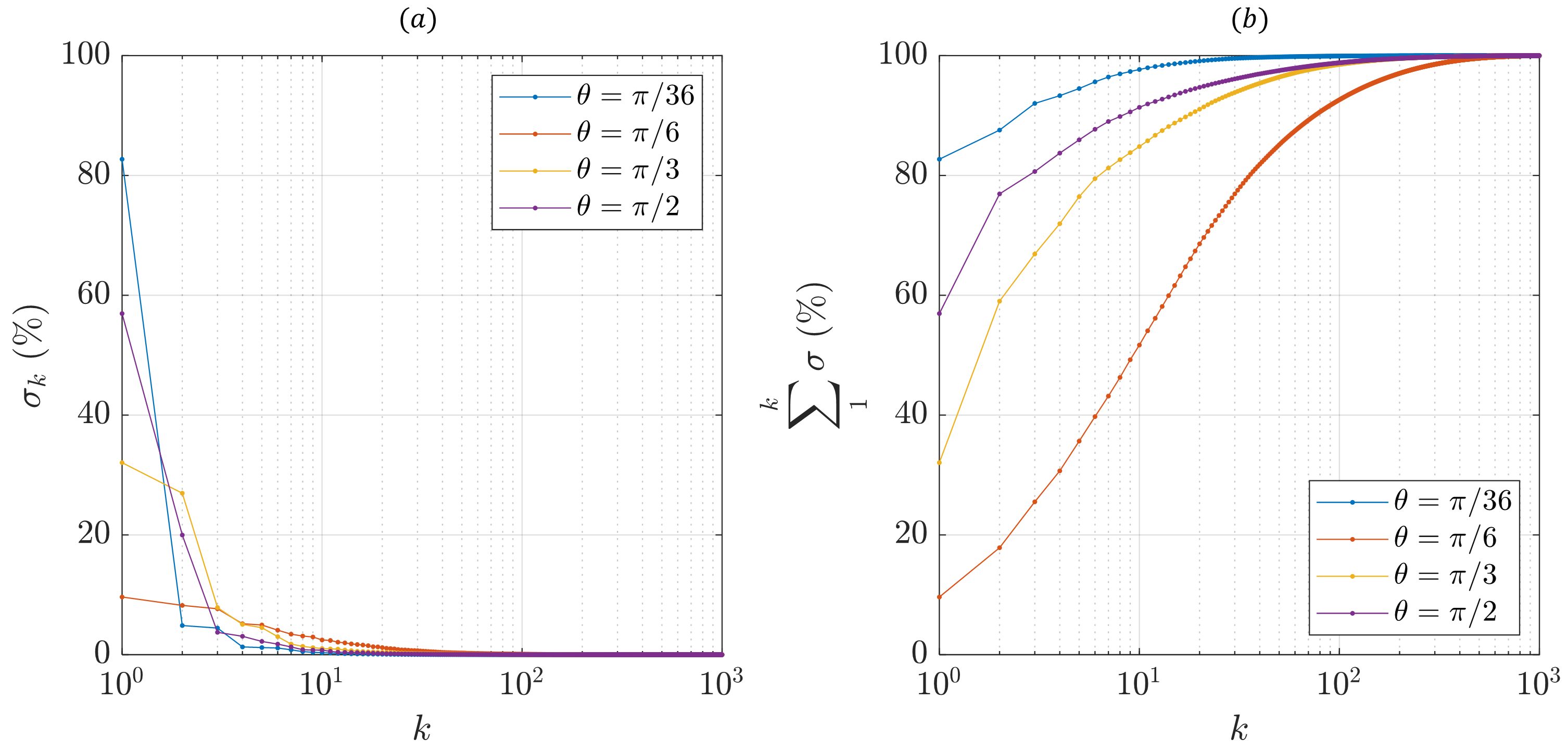}
	\caption{\label{fig:pod_energy} Distribution of (a) diminishing and (b) cumulative energy (in \%) across the ranked-modes ($k$) from the POD based modal decomposition of time-varying static pressure ($p/p_\infty$) variable for a wide range of control surface deflections ($\theta$).}
\end{figure*}

At $\theta = [\pi/2]$ (Figure \ref{fig:correlation_spectra}d), due to low-frequency pulsation, the resulting correlation pattern everywhere looks comparatively bigger than all the other cases. Once again, correlation lines are drawn at four locations: nose tip ($s/s_0 \sim 0$), bottom of control surface's windward side ($s/s_0 \sim 0.19$), downstream corner ($s/s_0 \sim 0.24$), and downstream reattachment point ($s/s_0 \sim 0.4$). While watching the pulsation cycle, one can see that the shock goes to and fro along the upstream control surface. These shock systems also exist in out-of-phase relations. It is again evident by seeing the red and blue colour patch (inverse correlation) at the nose-tip and control surface. The pattern looks the opposite while taking the reference at the control surface's windward side. Physically it is understood like the following: when the shock stays at the nose-tip during the `with-held' phase, the control surface experiences minimum pressure. Likewise, when the shock hangs on the control surface during the `collapse' phase, the control surface experiences a higher pressure than the nose tip. These events are captured in opposing colours on the $r_{ab}$ contour plot. Correlation about the corner is spread around due to strong activity. Correlation about the downstream reattachment point reveals that the immediate negative correlation strand is the thickest among all the other cases with $[\Delta s/s_0] \sim 0.1$. These large-scale structures are attributed to the ejected fluid mass between the pulsation cycle, which convect along the model surface.

\subsection{Modal analysis of pressure field} \label{ssec:modal_analysis}

In the last section, we see the correlating spatial field arising from the static pressure fluctuations only on the model wall. Performing a similar correlation field analysis for the entire spatial field will be computationally heavy and challenging to visualize. Powerful data analytic tools to study the spatiotemporally evolving coherent structures are available, like the Modal Decomposition \cite{Taira2017}. Proper Orthogonal Decomposition (POD) and Dynamic Mode Decomposition (DMD) are popular among the modal analysis tools to study the spatiotemporal modes by ranking the energy and dynamic contents in the developed flow field, respectively. These two techniques are well discussed in the open literature and applied across various flow field investigations. Hence, further treatment of the methodologies in this manuscript is avoided. The readers are referred to the respective papers available in the open literature. 

\begin{figure*}
	\includegraphics[width=0.8\textwidth]{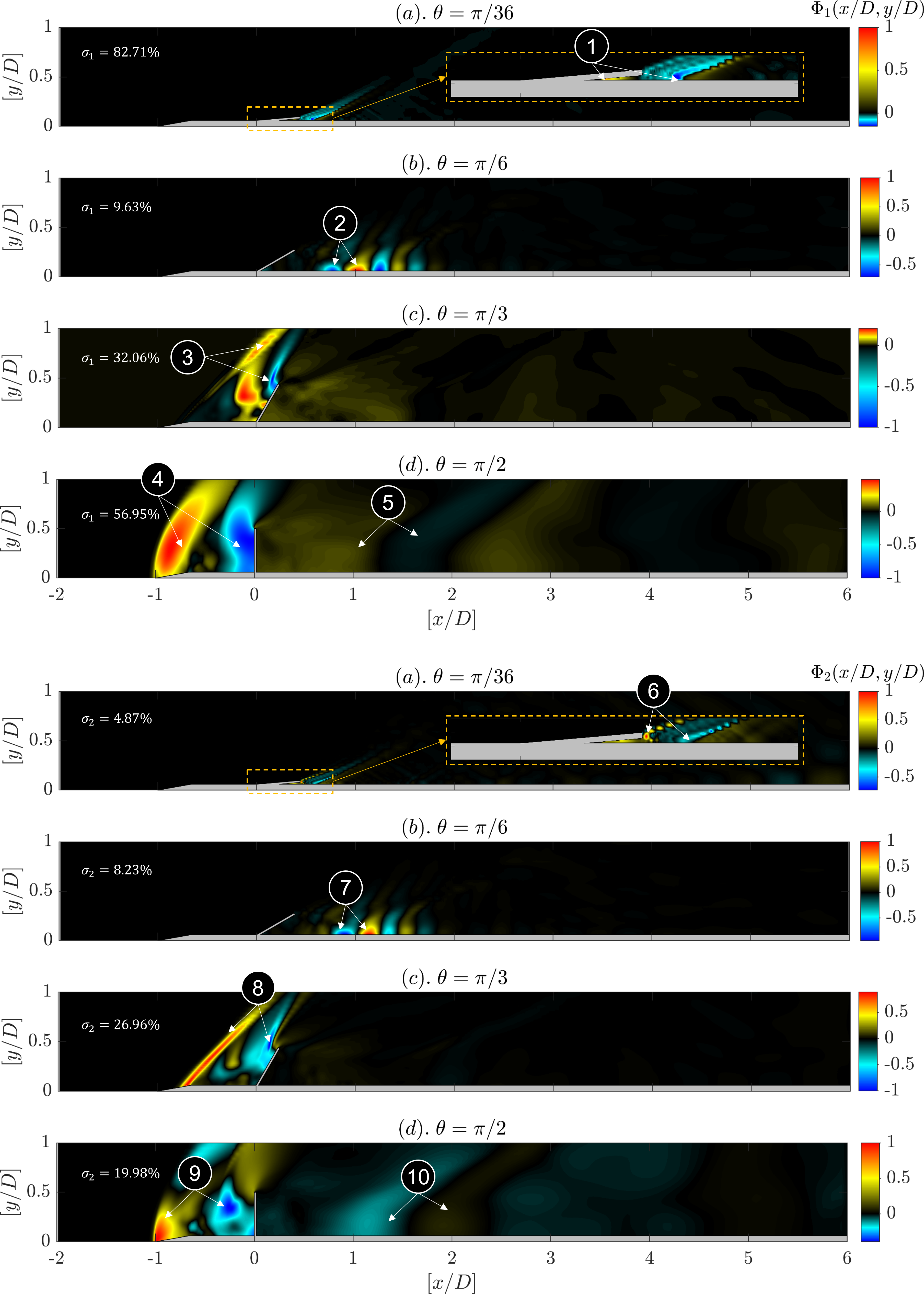}
	\caption{\label{fig:pod_spatial} Contour plots showing the first ($\Phi_1$) and second ($\Phi_2$) dominant energetic spatial modes from POD type modal analysis of the time-varying pressure fluctuation field-$p'(x/D,y/D,t/t_0)$ for a wide range of control surface deflections ($\theta$, a-d). Key features: 1. out-of-phase coherence between the corner vortices and downstream reattachment point, 2. coherent convecting structures creating local pockets of compression and rarefaction, 3. weak out-of-phase motion of the upstream separation and reattachment shock, 4. pulsating shock moving in out-of-phase between nose-tip and control surface, 5. large-scale oscillation zones, 6. inverse correlation between the expansion corner and reattaching shock, 7. convecting coherent structures - mode pair of $\Phi_1$, 8. strong out-of-phase motion of the upstream separation and reattachment shock, 9. inverse correlation between nose-tip vortex and $\lambda$ shock vortex, 10. comparatively small-scale oscillation zones than in $\Phi_1$.}
\end{figure*}

The POD-based modal analysis ranks the spatiotemporal modes based on the energy contents arising from pressure fluctuations or $p'^2$. From the simulation, instantaneous pressure fluctuations ($p'$) observed everywhere on the fluid domain are converted into a column vector. The temporal spacing for the selection of $p'(x/D,y/D)$ is taken as $[\Delta t/t_0] \sim 5$ and nearly 1000 subsequent time steps are considered for this analysis. Later, the sampled column vectors are stacked side-by-side to construct a column matrix. Hence, this method is also called a `snapshot' POD. The prepared column matrix is then reduced into a square or covariance matrix by multiplying it with its transpose matrix. The energy contents ($\sigma$) are further calculated by subjecting the covariance matrix through an eigenvalue decomposition. As mentioned earlier, the resulting ranked eigenvalues from the decomposition are plotted in Figure \ref{fig:pod_energy} for a wide range of control surface deflections ($\theta$). Diminishing (Figure \ref{fig:pod_energy}a) and cumulative (Figure \ref{fig:pod_energy}b) energy contents across the observed mode numbers ($k$) are plotted to better understand the physics. 

From Figure \ref{fig:pod_energy}a, the highest energy for a given $\theta$ is contained in the first mode and those values exhibit non-monotonic variations between $[\pi/36] \leq \theta \leq [\pi/2]$. The energy content in the first mode ($k=1$) is the largest for the $\theta=[\pi/36]$ case with $\sigma_1=82.7\%$ which is almost 0.9 times higher than that of the $\theta=[\pi/6]$ case. Such a high energy level indicates an underlying strong driving mode representing the whole flow field around the model. It is like that for the $\theta=[\pi/2]$ case as it contains $\sigma_1 \sim 57\%$. Cumulative energy plot (Figure \ref{fig:pod_energy}b) provides better insights into identifying the total number of energetic modes representing at least 50\% of the flow energy. As can be seen in the previous discussions, $\theta=[\pi/36]$ and $[\pi/2]$ already have more than 50\% of energy in the first mode itself. The situation is different for $\theta=[\pi/3]$ and $[\pi/6]$ as they need at least $k=3$ modes and $k=10$ modes to represent 50\% of the total flow energy. From the current situation, it can be said that there are multitudes of events (spatiotemporal scales or even fluid turbulence) driving the flow for $\theta=[\pi/3]$ and $[\pi/6]$ cases, whereas, for the cases of $\theta=[\pi/36]$ and $[\pi/2]$, it is by a dominant underlying event.

In Figure \ref{fig:pod_spatial}, the first and second dominant energetic spatial modes are shown from the POD analysis. The first and second mode for $\theta=[\pi/36]$ contain spatial structures located in the corner, expansion-tip, and downstream reattachment point (Figure \ref{fig:pod_spatial}a I-II). In the first mode, a correlation exists between the corner and reattachment point with $\sigma_1=82.7\%$. The left running structures from the reattachment point and the reflected right running structures from the corner is represented in the first mode. The multimedia view of the instantaneous numerical schlieren and Mach number contour in Figure \ref{fig:sch_M_contours} shows that the locations, as mentioned earlier, are active zones. In the second mode, weak coherence is seen between the reattachment point and the expansion corner on the control surface tip with $\sigma_1=4.87\%$. In these zones, the minimum activity observed in the multimedia view of Figure \ref{fig:sch_M_contours} is thus justified from the weak second mode. No dynamic events are seen anywhere else in $\theta=[\pi/36]$; thus, the cumulative energy approaches 90\% of the total flow energy just in $k=3$ modes.

In Figure \ref{fig:pod_spatial}b I-II, dominant spatial structures for $\theta=[\pi/6]$ are given. In both modes, coherent structures are seen close tot he reattachment point ($s/s_0 \sim 0.4$). The shedding structures convecting along the reattaching shear layer hit the model and created a reattachment point. Part of the structures turn into the recirculation, and the rest convects downstream, creating pockets of compression and rarefaction. These structures are observed in alternative colours with inverse correlation. Later, they diminish once they achieve equilibrium with the freestream pressure. As the structures are convective, modes one and two are spatially displaced contours. The respective mode's energy content remains almost the same ($\sigma_1 \sim \sigma_2 \sim 9\%$) owing to these behaviours. Hence, they are called a `mode pair'. A similar event is expected in the flapping jet flow or a vortex shedding wake flow \cite{Rao_2020,George2022}. As the dominant part of the flow involves shedding structures from the shear layer, which are comparatively small and decaying, $k=18$ modes are needed to represent 90\% of the total flow energy.

In Figure \ref{fig:pod_spatial}c I-II, dominant spatial structures for $\theta=[\pi/3]$ are given. In the first mode, inverse coherence is seen between the KH structure (halfway between the upstream separation and reattachment point) and the upstream reattachment shock. The KH structure is also coupled with the movements detected in the farther portion of the separation shock. However, in the second mode, the clear inverse coherence between the separation and reattachment shock is clear. Similar energy contents between the modes once again ensure the coupled relation ($\sigma_1 \sim \sigma_2 \sim 28\%$). It indicates the out-of-phase shock motion about the separation and reattachment point. Such a motion of the shocks displace the entrapped fluid mass in the separation bubble causing it to shrink or expand. For example, the downstream motion of the separation shock ejects the fluid mass by pushing the reattachment shock upstream and eventually shrinks the separation bubble volume and vice versa. A similar event is seen in the leading-edge separation of the spiked body flows in supersonic\cite{Sahoo2020} and hypersonic flow field\cite{Karthick2022}. It is also seen in the classic forward-facing step in a supersonic flow \cite{Chandola2017,EstruchSamper2018}.

In Figure \ref{fig:pod_spatial}d I-II, dominant spatial structures for $\theta=[\pi/2]$ are given. The flow field is dominated by the well-known `pulsation' type of shock oscillation. The driving frequency is in the low to mid-frequency ($fD/u_\infty \sim 0.17$), involving the to-and-fro motion of the curved shock. The first mode shows the extreme position of the shocks. The inverse correlation explains the switching pressure field: when the shock increases the compression near the model surface, the flow recovers to the freestream conditions near the nose-tip. Such an event is seen during the `collapse' phase of the pulsation cycle. The second mode shows the generation of structures like the leading-edge vortex and the vortex from the triple point (behind $\lambda$ shock) during the `inflate' phase of the pulsation cycle\cite{Sahoo2021}. In both modes, large-scale convecting structures corresponding to the pulsation cycle frequency are seen downstream. However, the first mode creates noticeable pressure fluctuations; hence, it contains $\sigma_1 \sim 57\%$, whereas the next mode contains only $\sigma_1 \sim 20\%$.
 
\begin{figure}
	\includegraphics[width=0.9\columnwidth]{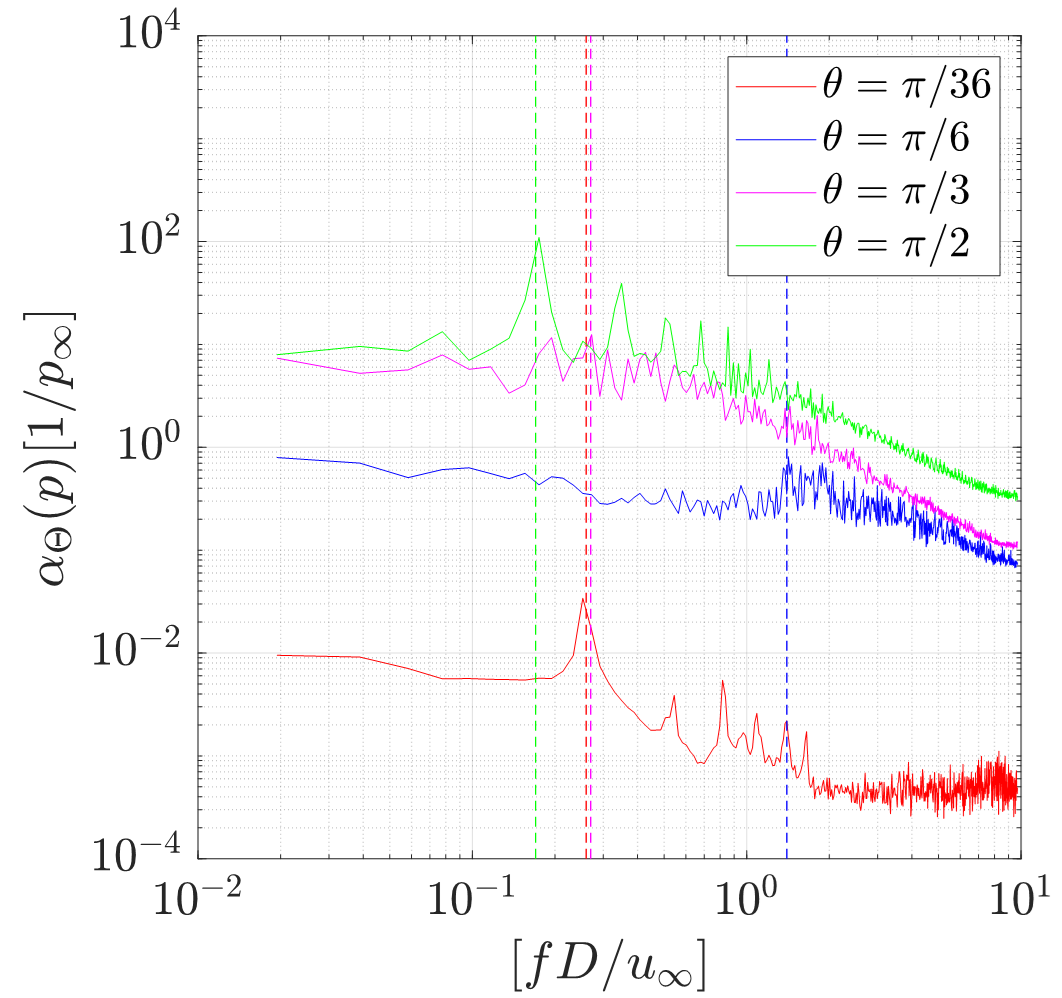}
	\caption{\label{fig:dmd_spectra} Plot showing the non-dimensionalized amplitude and frequency variation from the DMD based modal decomposition of time-varying static pressure ($p/p_\infty$) variable for a wide range of control surface deflections ($\theta$).}
\end{figure}

The spectral contents from the dynamic events are extracted using DMD, which involves preparing a column matrix. Later, it is subjected to a singular value decomposition and some more algebraic operations as mentioned in the book of Kutz\cite{Kutz2016}. Figure \ref{fig:dmd_spectra} presents a typical spectral plot from the DMD analysis. The highest amplitude is seen for the $\theta=[\pi/2]$ case with the non-dimensionalized amplitude approaching 110 with $[fD/u_\infty] \sim 0.17$ indicating the pulsation cycle's frequency. The next dynamically significant event is observed for the $\theta=[\pi/3]$ case with $[fD/u_\infty] \sim 0.27$ and amplitude about 12.5. The presence of leading-edge separation and a large recirculation bubble downstream are attributed to those observations. For the  $\theta=[\pi/6]$ case, $[fD/u_\infty] \sim 1.4$ and the amplitude is about 0.8, which corresponds to the presence of small-scale shedding structures near the downstream recirculation bubble. The case of $\theta=[\pi/36]$ contains the least dynamics with its dominant spectral peak amplitude dunking to 0.034 at $[fD/u_\infty] \sim 0.26$. The reason is attributed to very few active zones behind the control surface. The observed frequency is indeed in correspondence with the spectral plots generated from the drag coefficient (Figure \ref{fig:drag_signal}) and wall-static pressure variations (Figure \ref{fig:xt_part1} and \ref{fig:xt_part2}).

\section{Conclusions}\label{sec:conclusions}

The deflection of the control surface placed close to the leading-edge results in a plethora of flow dynamics investigated in the present manuscript using numerical simulation. The problem is simplified through the axisymmetric assumption and further reduced by modelling the separated turbulent structures using detached eddy simulation. Four control surface deflection angles ($\theta$, rad) are considered which are found to represent the different forms of flow unsteadiness: 1. $[\pi/36]$, 2. $[\pi/6]$, 3. $[\pi/3]$, and 4. $[\pi/2]$. The model is immersed in a computational fluid domain with a freestream supersonic Mach number of $M_\infty=2.0$ with a freestream Reynolds number of $Re_D=2\times10^6$ ($D=50$ mm, base body diameter). The resulting dynamic flow field is understood using different analysis techniques to quantify the different forms of unsteadiness. Some of the major conclusions from those analyses are summarized below:

\begin{itemize}
    \item \textbf{Gross flow field analysis}: Instantaneous and time-averaged snapshots of the flow reveal the following: 1. there is no noticeable unsteadiness seen in $\theta=[\pi/36]$ except around the recirculation region sitting in the vicinity of control surface tip, 2. almost steady separation bubble and an unsteady recirculation region are seen ahead and behind the control surface for $\theta=[\pi/6]$, 3. strong unsteady stationary separation from the leading-edge and unsteady recirculation region are observed upstream and downstream the control surface for $\theta=[\pi/3]$, and 4. the entire flow field around the control surface (from nose-tip till the downstream base) is seen chaotic with shocks moving to-and-from in the upstream for $\theta=[\pi/2]$.
    
    \item \textbf{Recirculation bubble scaling}: The length of the recirculation bubble indicates the drag force acting on a model. The transverse integral of the streamwise velocity along the $x$-direction is calculated first. A gradient is taken later to establish a parameter called $\eta$ to identify the downstream reattachment point, which is later used to scale the recirculation bubble's aspect ratio ($\Delta x_r/h$). The scaling indicates that $\Delta x_r/h$ exponentially decays as $\theta$ increases between $[\pi/36] \theta \leq [\pi/2]$.
    
    \item \textbf{Drag coefficient studies}: Larger values of $\theta$ introduce unsteadiness and cause variations in the resulting forces acting on the model. The spectral analysis of the drag coefficient ($c_d$) variation indicates a `bucket' trend while plotting $\theta$ and the frequency carrying the highest amplitude. The dominant frequency is discrete for $\theta = [\pi/36,\pi/2]$ and appears on the lower end of the spectra. On the other hand, $\theta = [\pi/6,\pi/3]$ show broadened spectra at a higher octave in a regressive manner. A scaling exists between the average drag coefficient ($\overline{c}_d$) and $\theta$ with exponential increment in the fluctuation intensity of $c_d$.
    
    \item \textbf{Wall-static pressure variations}: Statistics from the wall-static pressure fluctuations over the model surface offer insights into the shock-related unsteadiness. Peak pressures are seen at the point of separation and reattachment. The highest mean pressure is seen at the control surface's upstream corner for $\theta=[\pi/2]$ due to shock pulsation, which is 4.5 times higher than the freestream pressure. Similarly, the pressure fluctuation intensity is also seen to be the highest for the same case, which peaks at two times that of the freestream static pressure near the control surface tip. Fluctuation intensity is 0.4 times higher than the freestream near the downstream reattachment point due to large-scale structures impinging on the model wall and the rapid motion of the recompression shock.
    
    \item \textbf{Acoustic loading}: The pressure fluctuations on the model surface cause periodic forcing resulting in fluid-structure interactions. Identifying acoustic loading through the definition of an overall sound pressure level called $\beta$ helps monitor the noise. Smaller deflections ($\theta=[\pi/36]$) lead to the observation of $\beta \sim 50$ dB, whereas larger deflections ($\theta=[\pi/2]$) produce $\beta \sim 150$ dB (almost twice). Monitoring the variation of $\beta$ reveals the concentration of acoustic loading closer to shedding structures at near sonic speed around the separating and reattaching shear layer.
    
    \item \textbf{Outcomes of $x-t$ and $x-f$ plots}: The trail of the unsteady structures ahead and behind the control surface is studied using the $x-t$ and $x-f$ plots. The former helps identify the speed of the convecting events, and the latter reveals the underlying frequency. Discrete fluctuations at $[fD/u_\infty] \sim 0.25$ are seen closer to the reattaching shear layer for $\theta=[\pi/36]$. No noticeable convecting structures are seen. For the rest of the cases, the reattaching shear layer is observed to contain a convection velocity of about $[u_c/u_r] \sim 0.86$ as the shedding structures mingle with the freestream. However, qualitatively the length-scales are the smallest for $\theta=[\pi/6]$ and the largest for $\theta=[\pi/2]$ with underlying driving frequencies identified as $[fD/u_\infty] \sim 1.4$ and $[fD/u_\infty] \sim 0.17$, respectively.
    
    \item \textbf{Correlation coefficient}: Regions of correlating field from the wall-static pressure measurement indicate the influencing events across the flow field along with the coherent structure's length. The separation and reattachment shock ahead of the control surface in $\theta=[\pi/6]$ indicates an out-of-phase motion about the separation and reattachment point, which is identified as the driving source of the separation bubble unsteadiness. An inverse correlation between the nose-tip and the control surface's windward side indicates the pulsating type of shock oscillation. At the reattachment side, the smallest and the largest length-scales for $\theta=[\pi/6]$ and $[\pi/2]$ are calculated to be $[\Delta s/s_0] \sim 0.02$ and $0.1$ (almost 20\% larger), respectively.
    
    \item \textbf{Modal analysis}: POD and DMD modal analysis are done on the pressure fluctuating flow field snapshots to extract the dominant spatiotemporal modes. The driving dominant energetic modes for each of the cases are: $\theta = [\pi/36]$ - corner vortex and reattaching shear layer, $\theta = [\pi/6]$ - convecting coherent structures from the reattaching shear layer, $\theta = [\pi/3]$ - out-of-phase motion of the separation and reattachment shock about the respective separation and reattachment points, and $\theta = [\pi/2]$ - to-and-fro oscillation of the curved shock representing the well-known `pulsation' type of shock oscillation. The spectra of the dominant temporal modes reveal the underlying frequency. They are similar to the spectra from the $x-f$ analysis. 
\end{itemize}

\section*{Supplementary material}
See supplementary material for a video showing the time-varying spatial contours of $[u/a]$ at different $\theta$. The respective file is available under the name \href{https://youtu.be/UVvCu-u-I1U}{`video.mp4'}. Moreover, video showing the instantaneous $x$ and $y$ velocity component variations as contour plots are available at \href{https://youtu.be/f3jiLU9QffE}{Multimedia View: $x$-velocity} and
\href{https://youtu.be/s6MDfFvDATI}{Multimedia View: $y$-velocity}, respectively.

\section*{Acknowledgment}
 The authors acknowledge the support of the Computer Center at the Indian Institute of Technology Kanpur (IITK) for providing the necessary resources to carry out the simulation studies and data analysis. The first author thanks his post-doctoral supervisor Prof. Jacob Cohen for his support, encouragement, and inspiration. The first author also would like to thank his wife, Dr E. Hemaprabha and his colleagues, Dr N. Purushothaman and Dr Soumya R. Nanda, for their insights in creating intuitive plots and constructive reviews of the manuscript. The second author thanks the SURGE-2021 fellowship for providing an opportunity to do a research project at IIT Kanpur. 

\section*{Authors' Declaration}
 All authors have contributed equally to this work and report no conflicts.

\section*{Data availability statement}
The data supporting this study's findings are available from the corresponding author upon reasonable request.

\section*{References}
\bibliography{References}

\begin{thebibliography}{62}%
\makeatletter
\providecommand \@ifxundefined [1]{%
 \@ifx{#1\undefined}
}%
\providecommand \@ifnum [1]{%
 \ifnum #1\expandafter \@firstoftwo
 \else \expandafter \@secondoftwo
 \fi
}%
\providecommand \@ifx [1]{%
 \ifx #1\expandafter \@firstoftwo
 \else \expandafter \@secondoftwo
 \fi
}%
\providecommand \natexlab [1]{#1}%
\providecommand \enquote  [1]{``#1''}%
\providecommand \bibnamefont  [1]{#1}%
\providecommand \bibfnamefont [1]{#1}%
\providecommand \citenamefont [1]{#1}%
\providecommand \href@noop [0]{\@secondoftwo}%
\providecommand \href [0]{\begingroup \@sanitize@url \@href}%
\providecommand \@href[1]{\@@startlink{#1}\@@href}%
\providecommand \@@href[1]{\endgroup#1\@@endlink}%
\providecommand \@sanitize@url [0]{\catcode `\\12\catcode `\$12\catcode
  `\&12\catcode `\#12\catcode `\^12\catcode `\_12\catcode `\%12\relax}%
\providecommand \@@startlink[1]{}%
\providecommand \@@endlink[0]{}%
\providecommand \url  [0]{\begingroup\@sanitize@url \@url }%
\providecommand \@url [1]{\endgroup\@href {#1}{\urlprefix }}%
\providecommand \urlprefix  [0]{URL }%
\providecommand \Eprint [0]{\href }%
\providecommand \doibase [0]{https://doi.org/}%
\providecommand \selectlanguage [0]{\@gobble}%
\providecommand \bibinfo  [0]{\@secondoftwo}%
\providecommand \bibfield  [0]{\@secondoftwo}%
\providecommand \translation [1]{[#1]}%
\providecommand \BibitemOpen [0]{}%
\providecommand \bibitemStop [0]{}%
\providecommand \bibitemNoStop [0]{.\EOS\space}%
\providecommand \EOS [0]{\spacefactor3000\relax}%
\providecommand \BibitemShut  [1]{\csname bibitem#1\endcsname}%
\let\auto@bib@innerbib\@empty
\bibitem [{\citenamefont {Deshpande}\ and\ \citenamefont
  {Poggie}(2020)}]{Deshpande2020}%
  \BibitemOpen
  \bibfield  {author} {\bibinfo {author} {\bibfnamefont {A.~S.}\ \bibnamefont
  {Deshpande}}\ and\ \bibinfo {author} {\bibfnamefont {J.}~\bibnamefont
  {Poggie}},\ }\bibfield  {title} {\enquote {\bibinfo {title} {Unsteady
  characteristics of compressible reattaching shear layers},}\ }\href
  {https://doi.org/10.1063/5.0008752} {\bibfield  {journal} {\bibinfo
  {journal} {Physics of Fluids}\ }\textbf {\bibinfo {volume} {32}},\ \bibinfo
  {pages} {066103} (\bibinfo {year} {2020})}\BibitemShut {NoStop}%
\bibitem [{\citenamefont {Kumar}\ and\ \citenamefont {De}(2021)}]{Gaurav2021}%
  \BibitemOpen
  \bibfield  {author} {\bibinfo {author} {\bibfnamefont {G.}~\bibnamefont
  {Kumar}}\ and\ \bibinfo {author} {\bibfnamefont {A.}~\bibnamefont {De}},\
  }\bibfield  {title} {\enquote {\bibinfo {title} {Modes of unsteadiness in
  shock wave and separation region interaction in hypersonic flow over a double
  wedge geometry},}\ }\href {https://doi.org/10.1063/5.0053949} {\bibfield
  {journal} {\bibinfo  {journal} {Physics of Fluids}\ }\textbf {\bibinfo
  {volume} {33}},\ \bibinfo {pages} {076107} (\bibinfo {year}
  {2021})}\BibitemShut {NoStop}%
\bibitem [{\citenamefont {Vatansever}\ and\ \citenamefont
  {Celik}(2021)}]{Vatansever2021}%
  \BibitemOpen
  \bibfield  {author} {\bibinfo {author} {\bibfnamefont {D.}~\bibnamefont
  {Vatansever}}\ and\ \bibinfo {author} {\bibfnamefont {B.}~\bibnamefont
  {Celik}},\ }\bibfield  {title} {\enquote {\bibinfo {title} {Unsteady shock
  interaction mechanisms of high enthalpy reacting flows over double wedges at
  mach 7},}\ }\href {https://doi.org/10.1063/5.0050202} {\bibfield  {journal}
  {\bibinfo  {journal} {Physics of Fluids}\ }\textbf {\bibinfo {volume} {33}},\
  \bibinfo {pages} {056110} (\bibinfo {year} {2021})}\BibitemShut {NoStop}%
\bibitem [{\citenamefont {Tumuklu}, \citenamefont {Levin},\ and\ \citenamefont
  {Theofilis}(2019)}]{Tumuklu2019}%
  \BibitemOpen
  \bibfield  {author} {\bibinfo {author} {\bibfnamefont {O.}~\bibnamefont
  {Tumuklu}}, \bibinfo {author} {\bibfnamefont {D.~A.}\ \bibnamefont {Levin}},\
  and\ \bibinfo {author} {\bibfnamefont {V.}~\bibnamefont {Theofilis}},\
  }\bibfield  {title} {\enquote {\bibinfo {title} {Kinetic modeling of unsteady
  hypersonic flows over a tick geometry},}\ }\href
  {https://doi.org/10.1063/1.5090341} {\bibfield  {journal} {\bibinfo
  {journal} {Physics of Fluids}\ }\textbf {\bibinfo {volume} {31}},\ \bibinfo
  {pages} {056108} (\bibinfo {year} {2019})}\BibitemShut {NoStop}%
\bibitem [{\citenamefont {Hickel}\ and\ \citenamefont {van
  Oudheusden}(2020)}]{Hickel2020}%
  \BibitemOpen
  \bibfield  {author} {\bibinfo {author} {\bibfnamefont {S.}~\bibnamefont
  {Hickel}}\ and\ \bibinfo {author} {\bibfnamefont {B.}~\bibnamefont {van
  Oudheusden}},\ }\bibfield  {title} {\enquote {\bibinfo {title} {Influence of
  upstream disturbances on the primary and secondary instabilities in a
  supersonic separated flow over a backward-facing step},}\ }\href
  {https://doi.org/10.1063/5.0005431} {\bibfield  {journal} {\bibinfo
  {journal} {Physics of Fluids}\ }\textbf {\bibinfo {volume} {32}},\ \bibinfo
  {pages} {056102} (\bibinfo {year} {2020})}\BibitemShut {NoStop}%
\bibitem [{\citenamefont {Tumuklu}, \citenamefont {Levin},\ and\ \citenamefont
  {Theofilis}(2018)}]{Tumuklu2018}%
  \BibitemOpen
  \bibfield  {author} {\bibinfo {author} {\bibfnamefont {O.}~\bibnamefont
  {Tumuklu}}, \bibinfo {author} {\bibfnamefont {D.~A.}\ \bibnamefont {Levin}},\
  and\ \bibinfo {author} {\bibfnamefont {V.}~\bibnamefont {Theofilis}},\
  }\bibfield  {title} {\enquote {\bibinfo {title} {Investigation of unsteady,
  hypersonic, laminar separated flows over a double cone geometry using a
  kinetic approach},}\ }\href {https://doi.org/10.1063/1.5022598} {\bibfield
  {journal} {\bibinfo  {journal} {Physics of Fluids}\ }\textbf {\bibinfo
  {volume} {30}},\ \bibinfo {pages} {046103} (\bibinfo {year}
  {2018})}\BibitemShut {NoStop}%
\bibitem [{\citenamefont {Duck}, \citenamefont {Lasseigne},\ and\ \citenamefont
  {Hussaini}(1997)}]{Duck1997}%
  \BibitemOpen
  \bibfield  {author} {\bibinfo {author} {\bibfnamefont {P.~W.}\ \bibnamefont
  {Duck}}, \bibinfo {author} {\bibfnamefont {D.~G.}\ \bibnamefont
  {Lasseigne}},\ and\ \bibinfo {author} {\bibfnamefont {M.~Y.}\ \bibnamefont
  {Hussaini}},\ }\bibfield  {title} {\enquote {\bibinfo {title} {The effect of
  three-dimensional freestream disturbances on the supersonic flow past a
  wedge},}\ }\href {https://doi.org/10.1063/1.869140} {\bibfield  {journal}
  {\bibinfo  {journal} {Physics of Fluids}\ }\textbf {\bibinfo {volume} {9}},\
  \bibinfo {pages} {456--467} (\bibinfo {year} {1997})}\BibitemShut {NoStop}%
\bibitem [{\citenamefont {Porter}\ and\ \citenamefont
  {Poggie}(2019)}]{Porter2019}%
  \BibitemOpen
  \bibfield  {author} {\bibinfo {author} {\bibfnamefont {K.~M.}\ \bibnamefont
  {Porter}}\ and\ \bibinfo {author} {\bibfnamefont {J.}~\bibnamefont
  {Poggie}},\ }\bibfield  {title} {\enquote {\bibinfo {title} {Selective
  upstream influence on the unsteadiness of a separated turbulent compression
  ramp flow},}\ }\href {https://doi.org/10.1063/1.5078938} {\bibfield
  {journal} {\bibinfo  {journal} {Physics of Fluids}\ }\textbf {\bibinfo
  {volume} {31}},\ \bibinfo {pages} {016104} (\bibinfo {year}
  {2019})}\BibitemShut {NoStop}%
\bibitem [{\citenamefont {Bourdon}\ and\ \citenamefont
  {Dutton}(1999)}]{Bourdon1999}%
  \BibitemOpen
  \bibfield  {author} {\bibinfo {author} {\bibfnamefont {C.~J.}\ \bibnamefont
  {Bourdon}}\ and\ \bibinfo {author} {\bibfnamefont {J.~C.}\ \bibnamefont
  {Dutton}},\ }\bibfield  {title} {\enquote {\bibinfo {title} {Planar
  visualizations of large-scale turbulent structures in axisymmetric supersonic
  separated flows},}\ }\href {https://doi.org/10.1063/1.869913} {\bibfield
  {journal} {\bibinfo  {journal} {Physics of Fluids}\ }\textbf {\bibinfo
  {volume} {11}},\ \bibinfo {pages} {201--213} (\bibinfo {year}
  {1999})}\BibitemShut {NoStop}%
\bibitem [{\citenamefont {Dolling}(2001)}]{Dolling2001}%
  \BibitemOpen
  \bibfield  {author} {\bibinfo {author} {\bibfnamefont {D.~S.}\ \bibnamefont
  {Dolling}},\ }\bibfield  {title} {\enquote {\bibinfo {title} {Fifty years of
  shock-wave/boundary-layer interaction research: What next?}}\ }\href
  {https://doi.org/10.2514/2.1476} {\bibfield  {journal} {\bibinfo  {journal}
  {{AIAA} Journal}\ }\textbf {\bibinfo {volume} {39}},\ \bibinfo {pages}
  {1517--1531} (\bibinfo {year} {2001})}\BibitemShut {NoStop}%
\bibitem [{\citenamefont {Gaitonde}(2015)}]{Gaitonde2015}%
  \BibitemOpen
  \bibfield  {author} {\bibinfo {author} {\bibfnamefont {D.~V.}\ \bibnamefont
  {Gaitonde}},\ }\bibfield  {title} {\enquote {\bibinfo {title} {Progress in
  shock wave/boundary layer interactions},}\ }\href
  {https://doi.org/10.1016/j.paerosci.2014.09.002} {\bibfield  {journal}
  {\bibinfo  {journal} {Progress in Aerospace Sciences}\ }\textbf {\bibinfo
  {volume} {72}},\ \bibinfo {pages} {80--99} (\bibinfo {year}
  {2015})}\BibitemShut {NoStop}%
\bibitem [{\citenamefont {Clemens}\ and\ \citenamefont
  {Narayanaswamy}(2014)}]{Clemens2014}%
  \BibitemOpen
  \bibfield  {author} {\bibinfo {author} {\bibfnamefont {N.~T.}\ \bibnamefont
  {Clemens}}\ and\ \bibinfo {author} {\bibfnamefont {V.}~\bibnamefont
  {Narayanaswamy}},\ }\bibfield  {title} {\enquote {\bibinfo {title}
  {Low-frequency unsteadiness of shock wave/turbulent boundary layer
  interactions},}\ }\href {https://doi.org/10.1146/annurev-fluid-010313-141346}
  {\bibfield  {journal} {\bibinfo  {journal} {Annual Review of Fluid
  Mechanics}\ }\textbf {\bibinfo {volume} {46}},\ \bibinfo {pages} {469--492}
  (\bibinfo {year} {2014})}\BibitemShut {NoStop}%
\bibitem [{\citenamefont {Settles}, \citenamefont {Fitzpatrick},\ and\
  \citenamefont {Bogdonoff}(1979)}]{Settles1979}%
  \BibitemOpen
  \bibfield  {author} {\bibinfo {author} {\bibfnamefont {G.~S.}\ \bibnamefont
  {Settles}}, \bibinfo {author} {\bibfnamefont {T.~J.}\ \bibnamefont
  {Fitzpatrick}},\ and\ \bibinfo {author} {\bibfnamefont {S.~M.}\ \bibnamefont
  {Bogdonoff}},\ }\bibfield  {title} {\enquote {\bibinfo {title} {Detailed
  study of attached and separated compression corner flowfields in high
  reynolds number supersonic flow},}\ }\href {https://doi.org/10.2514/3.61180}
  {\bibfield  {journal} {\bibinfo  {journal} {{AIAA} Journal}\ }\textbf
  {\bibinfo {volume} {17}},\ \bibinfo {pages} {579--585} (\bibinfo {year}
  {1979})}\BibitemShut {NoStop}%
\bibitem [{\citenamefont {Sugarno}\ \emph {et~al.}(2022)\citenamefont
  {Sugarno}, \citenamefont {Sriram}, \citenamefont {Karthick},\ and\
  \citenamefont {Jagadeesh}}]{Sugarno2022}%
  \BibitemOpen
  \bibfield  {author} {\bibinfo {author} {\bibfnamefont {M.~I.}\ \bibnamefont
  {Sugarno}}, \bibinfo {author} {\bibfnamefont {R.}~\bibnamefont {Sriram}},
  \bibinfo {author} {\bibfnamefont {S.~K.}\ \bibnamefont {Karthick}},\ and\
  \bibinfo {author} {\bibfnamefont {G.}~\bibnamefont {Jagadeesh}},\ }\bibfield
  {title} {\enquote {\bibinfo {title} {Unsteady pulsating flowfield over spiked
  axisymmetric forebody at hypersonic flows},}\ }\href
  {https://doi.org/10.1063/5.0075583} {\bibfield  {journal} {\bibinfo
  {journal} {Physics of Fluids}\ }\textbf {\bibinfo {volume} {34}},\ \bibinfo
  {pages} {016104} (\bibinfo {year} {2022})}\BibitemShut {NoStop}%
\bibitem [{\citenamefont {Narayana}\ and\ \citenamefont
  {Selvaraj}(2020)}]{Narayana2020}%
  \BibitemOpen
  \bibfield  {author} {\bibinfo {author} {\bibfnamefont {G.}~\bibnamefont
  {Narayana}}\ and\ \bibinfo {author} {\bibfnamefont {S.}~\bibnamefont
  {Selvaraj}},\ }\bibfield  {title} {\enquote {\bibinfo {title} {Attenuation of
  pulsation and oscillation using a disk at mid-section of spiked blunt
  body},}\ }\href {https://doi.org/10.1063/5.0024649} {\bibfield  {journal}
  {\bibinfo  {journal} {Physics of Fluids}\ }\textbf {\bibinfo {volume} {32}},\
  \bibinfo {pages} {116106} (\bibinfo {year} {2020})}\BibitemShut {NoStop}%
\bibitem [{\citenamefont {Tekure}, \citenamefont {Pophali},\ and\ \citenamefont
  {Venkatasubbaiah}(2021)}]{Tekure2021}%
  \BibitemOpen
  \bibfield  {author} {\bibinfo {author} {\bibfnamefont {V.}~\bibnamefont
  {Tekure}}, \bibinfo {author} {\bibfnamefont {P.~S.}\ \bibnamefont
  {Pophali}},\ and\ \bibinfo {author} {\bibfnamefont {K.}~\bibnamefont
  {Venkatasubbaiah}},\ }\bibfield  {title} {\enquote {\bibinfo {title}
  {Numerical investigation of aerospike semi-cone angle and a small bump on the
  spike stem in reducing the aerodynamic drag and heating of spiked blunt-body:
  New correlations for drag and surface temperature},}\ }\href
  {https://doi.org/10.1063/5.0066028} {\bibfield  {journal} {\bibinfo
  {journal} {Physics of Fluids}\ }\textbf {\bibinfo {volume} {33}},\ \bibinfo
  {pages} {116108} (\bibinfo {year} {2021})}\BibitemShut {NoStop}%
\bibitem [{\citenamefont {Allen}\ and\ \citenamefont
  {Eggers}(1953)}]{allenstudy}%
  \BibitemOpen
  \bibfield  {author} {\bibinfo {author} {\bibfnamefont {H.~J.}\ \bibnamefont
  {Allen}}\ and\ \bibinfo {author} {\bibfnamefont {J.}~\bibnamefont {Eggers},
  \bibfnamefont {A.~J.}},\ }\href@noop {} {\emph {\bibinfo {title} {A study of
  the motion and aerodynamic heating of ballistic missiles entering the earth's
  atmosphere at high supersonic speeds}}}\ (\bibinfo  {publisher} {{UNT}
  {Libraries} {Government} {Documents} {Department}},\ \bibinfo {year}
  {1953})\BibitemShut {NoStop}%
\bibitem [{\citenamefont {Stanbrook}(1957)}]{Stanbrook1957}%
  \BibitemOpen
  \bibfield  {author} {\bibinfo {author} {\bibfnamefont {A.}~\bibnamefont
  {Stanbrook}},\ }\bibfield  {title} {\enquote {\bibinfo {title} {The flow
  upstream of finite span spoilers at supersonic speeds},}\ }\href@noop {}
  {\bibfield  {journal} {\bibinfo  {journal} {ARC Technical Report}\ }\textbf
  {\bibinfo {volume} {2526}},\ \bibinfo {pages} {14} (\bibinfo {year}
  {1957})}\BibitemShut {NoStop}%
\bibitem [{\citenamefont {Kaattari}, \citenamefont {Hill~Jr},\ and\
  \citenamefont {Nielsen}(1955)}]{Kattari1955}%
  \BibitemOpen
  \bibfield  {author} {\bibinfo {author} {\bibfnamefont {G.~E.}\ \bibnamefont
  {Kaattari}}, \bibinfo {author} {\bibfnamefont {W.~A.}\ \bibnamefont
  {Hill~Jr}},\ and\ \bibinfo {author} {\bibfnamefont {J.~N.}\ \bibnamefont
  {Nielsen}},\ }\bibfield  {title} {\enquote {\bibinfo {title} {Controls for
  supersonic missiles},}\ }\href@noop {} {\bibfield  {journal} {\bibinfo
  {journal} {NACA Technical Report}\ }\textbf {\bibinfo {volume}
  {NACA-RM-A55D12}} (\bibinfo {year} {1955})}\BibitemShut {NoStop}%
\bibitem [{\citenamefont {Kuehn}(1959)}]{kuehn1959}%
  \BibitemOpen
  \bibfield  {author} {\bibinfo {author} {\bibfnamefont {D.~M.}\ \bibnamefont
  {Kuehn}},\ }\bibfield  {title} {\enquote {\bibinfo {title} {Experimental
  investigation of the pressure rise required for the incipient separation of
  turbulent boundary layers in two-dimensional supersonic flow},}\ }\href@noop
  {} {\bibfield  {journal} {\bibinfo  {journal} {NASA Memorandum}\ }\textbf
  {\bibinfo {volume} {12159A}},\ \bibinfo {pages} {48} (\bibinfo {year}
  {1959})}\BibitemShut {NoStop}%
\bibitem [{\citenamefont {Wilcox}(1990)}]{Wilcox1990}%
  \BibitemOpen
  \bibfield  {author} {\bibinfo {author} {\bibfnamefont {D.~C.}\ \bibnamefont
  {Wilcox}},\ }\bibfield  {title} {\enquote {\bibinfo {title} {Supersonic
  compression-corner applications of a multiscale model forturbulent flows},}\
  }\href {https://doi.org/10.2514/3.25191} {\bibfield  {journal} {\bibinfo
  {journal} {{AIAA} Journal}\ }\textbf {\bibinfo {volume} {28}},\ \bibinfo
  {pages} {1194--1198} (\bibinfo {year} {1990})}\BibitemShut {NoStop}%
\bibitem [{\citenamefont {Nilavarasan}\ \emph {et~al.}(2022)\citenamefont
  {Nilavarasan}, \citenamefont {Joshi}, \citenamefont {Misra}, \citenamefont
  {Manisankar},\ and\ \citenamefont {Verma}}]{Nilavarasan2022}%
  \BibitemOpen
  \bibfield  {author} {\bibinfo {author} {\bibfnamefont {T.}~\bibnamefont
  {Nilavarasan}}, \bibinfo {author} {\bibfnamefont {G.}~\bibnamefont {Joshi}},
  \bibinfo {author} {\bibfnamefont {A.}~\bibnamefont {Misra}}, \bibinfo
  {author} {\bibfnamefont {C.}~\bibnamefont {Manisankar}},\ and\ \bibinfo
  {author} {\bibfnamefont {S.}~\bibnamefont {Verma}},\ }\bibfield  {title}
  {\enquote {\bibinfo {title} {Control of flow separation over an axisymmetric
  flared body using ramped vanes},}\ }\href
  {https://doi.org/10.1016/j.euromechflu.2022.04.010} {\bibfield  {journal}
  {\bibinfo  {journal} {European Journal of Mechanics - B/Fluids}\ }\textbf
  {\bibinfo {volume} {95}},\ \bibinfo {pages} {160--177} (\bibinfo {year}
  {2022})}\BibitemShut {NoStop}%
\bibitem [{\citenamefont {Amsden}\ and\ \citenamefont
  {Harlow}(1965)}]{Amsden1965}%
  \BibitemOpen
  \bibfield  {author} {\bibinfo {author} {\bibfnamefont {A.~A.}\ \bibnamefont
  {Amsden}}\ and\ \bibinfo {author} {\bibfnamefont {F.~H.}\ \bibnamefont
  {Harlow}},\ }\bibfield  {title} {\enquote {\bibinfo {title} {Numerical
  calculation of supersonic wake flow.}}\ }\href
  {https://doi.org/10.2514/3.3318} {\bibfield  {journal} {\bibinfo  {journal}
  {{AIAA} Journal}\ }\textbf {\bibinfo {volume} {3}},\ \bibinfo {pages}
  {2081--2086} (\bibinfo {year} {1965})}\BibitemShut {NoStop}%
\bibitem [{\citenamefont {Hui}(1972)}]{Hui1972}%
  \BibitemOpen
  \bibfield  {author} {\bibinfo {author} {\bibfnamefont {W.~H.}\ \bibnamefont
  {Hui}},\ }\bibfield  {title} {\enquote {\bibinfo {title} {Effects of upstream
  unsteadiness on hypersonic flow past a wedge},}\ }\href
  {https://doi.org/10.1063/1.1693771} {\bibfield  {journal} {\bibinfo
  {journal} {Physics of Fluids}\ }\textbf {\bibinfo {volume} {15}},\ \bibinfo
  {pages} {1747} (\bibinfo {year} {1972})}\BibitemShut {NoStop}%
\bibitem [{\citenamefont {Kazakov}(1982)}]{Kazakov1982}%
  \BibitemOpen
  \bibfield  {author} {\bibinfo {author} {\bibfnamefont {A.~V.}\ \bibnamefont
  {Kazakov}},\ }\bibfield  {title} {\enquote {\bibinfo {title} {Unsteady
  viscous supersonic flow over a short flap},}\ }\href
  {https://doi.org/10.1007/bf01094603} {\bibfield  {journal} {\bibinfo
  {journal} {Fluid Dynamics}\ }\textbf {\bibinfo {volume} {16}},\ \bibinfo
  {pages} {579--585} (\bibinfo {year} {1982})}\BibitemShut {NoStop}%
\bibitem [{\citenamefont {Rao}(1974)}]{RAO1974}%
  \BibitemOpen
  \bibfield  {author} {\bibinfo {author} {\bibfnamefont {P.~P.}\ \bibnamefont
  {Rao}},\ }\bibfield  {title} {\enquote {\bibinfo {title} {Inviscid supersonic
  far wake flow past pointed bodies},}\ }\href
  {https://doi.org/10.2514/3.49264} {\bibfield  {journal} {\bibinfo  {journal}
  {{AIAA} Journal}\ }\textbf {\bibinfo {volume} {12}},\ \bibinfo {pages}
  {421--422} (\bibinfo {year} {1974})}\BibitemShut {NoStop}%
\bibitem [{\citenamefont {Sebastian}, \citenamefont {Suryan},\ and\
  \citenamefont {Kim}(2016)}]{Sebastian2016}%
  \BibitemOpen
  \bibfield  {author} {\bibinfo {author} {\bibfnamefont {J.~J.}\ \bibnamefont
  {Sebastian}}, \bibinfo {author} {\bibfnamefont {A.}~\bibnamefont {Suryan}},\
  and\ \bibinfo {author} {\bibfnamefont {H.~D.}\ \bibnamefont {Kim}},\
  }\bibfield  {title} {\enquote {\bibinfo {title} {Numerical analysis of
  hypersonic flow past blunt bodies with aerospikes},}\ }\href
  {https://doi.org/10.2514/1.a33414} {\bibfield  {journal} {\bibinfo  {journal}
  {Journal of Spacecraft and Rockets}\ }\textbf {\bibinfo {volume} {53}},\
  \bibinfo {pages} {669--677} (\bibinfo {year} {2016})}\BibitemShut {NoStop}%
\bibitem [{\citenamefont {Sahoo}\ \emph {et~al.}(2020)\citenamefont {Sahoo},
  \citenamefont {Karthick}, \citenamefont {Das},\ and\ \citenamefont
  {Cohen}}]{Sahoo2020}%
  \BibitemOpen
  \bibfield  {author} {\bibinfo {author} {\bibfnamefont {D.}~\bibnamefont
  {Sahoo}}, \bibinfo {author} {\bibfnamefont {S.~K.}\ \bibnamefont {Karthick}},
  \bibinfo {author} {\bibfnamefont {S.}~\bibnamefont {Das}},\ and\ \bibinfo
  {author} {\bibfnamefont {J.}~\bibnamefont {Cohen}},\ }\bibfield  {title}
  {\enquote {\bibinfo {title} {Parametric experimental studies on supersonic
  flow unsteadiness over a hemispherical spiked body},}\ }\href
  {https://doi.org/10.2514/1.j059369} {\bibfield  {journal} {\bibinfo
  {journal} {{AIAA} Journal}\ }\textbf {\bibinfo {volume} {58}},\ \bibinfo
  {pages} {3446--3463} (\bibinfo {year} {2020})}\BibitemShut {NoStop}%
\bibitem [{\citenamefont {Sahoo}\ \emph {et~al.}(2021)\citenamefont {Sahoo},
  \citenamefont {Karthick}, \citenamefont {Das},\ and\ \citenamefont
  {Cohen}}]{Sahoo2021}%
  \BibitemOpen
  \bibfield  {author} {\bibinfo {author} {\bibfnamefont {D.}~\bibnamefont
  {Sahoo}}, \bibinfo {author} {\bibfnamefont {S.~K.}\ \bibnamefont {Karthick}},
  \bibinfo {author} {\bibfnamefont {S.}~\bibnamefont {Das}},\ and\ \bibinfo
  {author} {\bibfnamefont {J.}~\bibnamefont {Cohen}},\ }\bibfield  {title}
  {\enquote {\bibinfo {title} {Shock-related unsteadiness of axisymmetric
  spiked bodies in supersonic flow},}\ }\href
  {https://doi.org/10.1007/s00348-020-03130-2} {\bibfield  {journal} {\bibinfo
  {journal} {Experiments in Fluids}\ }\textbf {\bibinfo {volume} {62}}
  (\bibinfo {year} {2021})}\BibitemShut {NoStop}%
\bibitem [{\citenamefont {Degani}(1992)}]{Degani1992}%
  \BibitemOpen
  \bibfield  {author} {\bibinfo {author} {\bibfnamefont {D.}~\bibnamefont
  {Degani}},\ }\bibfield  {title} {\enquote {\bibinfo {title} {Numerical
  simulation of vortex unsteadiness on a slender body at high incidence},}\
  }\href {https://doi.org/10.2514/3.10995} {\bibfield  {journal} {\bibinfo
  {journal} {{AIAA} Journal}\ }\textbf {\bibinfo {volume} {30}},\ \bibinfo
  {pages} {841--843} (\bibinfo {year} {1992})}\BibitemShut {NoStop}%
\bibitem [{\citenamefont {Dolling}\ and\ \citenamefont
  {Murphy}(1983)}]{Dolling1983}%
  \BibitemOpen
  \bibfield  {author} {\bibinfo {author} {\bibfnamefont {D.~S.}\ \bibnamefont
  {Dolling}}\ and\ \bibinfo {author} {\bibfnamefont {M.~T.}\ \bibnamefont
  {Murphy}},\ }\bibfield  {title} {\enquote {\bibinfo {title} {Unsteadiness of
  the separation shock wave structure in a supersonic compression ramp
  flowfield},}\ }\href {https://doi.org/10.2514/3.60163} {\bibfield  {journal}
  {\bibinfo  {journal} {{AIAA} Journal}\ }\textbf {\bibinfo {volume} {21}},\
  \bibinfo {pages} {1628--1634} (\bibinfo {year} {1983})}\BibitemShut {NoStop}%
\bibitem [{\citenamefont {Park}, \citenamefont {Chung},\ and\ \citenamefont
  {Sung}(1994)}]{Park1994}%
  \BibitemOpen
  \bibfield  {author} {\bibinfo {author} {\bibfnamefont {S.~O.}\ \bibnamefont
  {Park}}, \bibinfo {author} {\bibfnamefont {Y.~M.}\ \bibnamefont {Chung}},\
  and\ \bibinfo {author} {\bibfnamefont {H.~J.}\ \bibnamefont {Sung}},\
  }\bibfield  {title} {\enquote {\bibinfo {title} {Numerical study of unsteady
  supersonic compression ramp flows},}\ }\href
  {https://doi.org/10.2514/3.11973} {\bibfield  {journal} {\bibinfo  {journal}
  {{AIAA} Journal}\ }\textbf {\bibinfo {volume} {32}},\ \bibinfo {pages}
  {216--218} (\bibinfo {year} {1994})}\BibitemShut {NoStop}%
\bibitem [{\citenamefont {Erengil}\ and\ \citenamefont
  {Dolling}(1991)}]{Erengil1991}%
  \BibitemOpen
  \bibfield  {author} {\bibinfo {author} {\bibfnamefont {M.~E.}\ \bibnamefont
  {Erengil}}\ and\ \bibinfo {author} {\bibfnamefont {D.~S.}\ \bibnamefont
  {Dolling}},\ }\bibfield  {title} {\enquote {\bibinfo {title} {Unsteady wave
  structure near separation in a mach 5 compression rampinteraction},}\ }\href
  {https://doi.org/10.2514/3.10647} {\bibfield  {journal} {\bibinfo  {journal}
  {{AIAA} Journal}\ }\textbf {\bibinfo {volume} {29}},\ \bibinfo {pages}
  {728--735} (\bibinfo {year} {1991})}\BibitemShut {NoStop}%
\bibitem [{\citenamefont {Erengil}\ and\ \citenamefont
  {Dolling}(1993)}]{Erengil1993}%
  \BibitemOpen
  \bibfield  {author} {\bibinfo {author} {\bibfnamefont {M.~E.}\ \bibnamefont
  {Erengil}}\ and\ \bibinfo {author} {\bibfnamefont {D.~S.}\ \bibnamefont
  {Dolling}},\ }\bibfield  {title} {\enquote {\bibinfo {title} {Effects of
  sweepback on unsteady separation in mach 5 compression ramp interactions},}\
  }\href {https://doi.org/10.2514/3.60176} {\bibfield  {journal} {\bibinfo
  {journal} {{AIAA} Journal}\ }\textbf {\bibinfo {volume} {31}},\ \bibinfo
  {pages} {302--311} (\bibinfo {year} {1993})}\BibitemShut {NoStop}%
\bibitem [{\citenamefont {Gramann}\ and\ \citenamefont
  {Dolling}(1992)}]{GRAMANN1992}%
  \BibitemOpen
  \bibfield  {author} {\bibinfo {author} {\bibfnamefont {R.}~\bibnamefont
  {Gramann}}\ and\ \bibinfo {author} {\bibfnamefont {D.}~\bibnamefont
  {Dolling}},\ }\bibfield  {title} {\enquote {\bibinfo {title} {A preliminary
  study of the turbulent structures associated with unsteady separation shock
  motion in a mach 5 compression ramp interaction},}\ }in\ \href
  {https://doi.org/10.2514/6.1992-744} {\emph {\bibinfo {booktitle} {30th
  Aerospace Sciences Meeting and Exhibit}}}\ (\bibinfo  {publisher} {American
  Institute of Aeronautics and Astronautics},\ \bibinfo {year}
  {1992})\BibitemShut {NoStop}%
\bibitem [{\citenamefont {Park}, \citenamefont {Lee},\ and\ \citenamefont
  {Kang}(2001)}]{Park2001}%
  \BibitemOpen
  \bibfield  {author} {\bibinfo {author} {\bibfnamefont {S.}~\bibnamefont
  {Park}}, \bibinfo {author} {\bibfnamefont {C.}~\bibnamefont {Lee}},\ and\
  \bibinfo {author} {\bibfnamefont {K.}~\bibnamefont {Kang}},\ }\bibfield
  {title} {\enquote {\bibinfo {title} {Navier-stokes analysis of a supersonic
  flow over moving compression ramp},}\ }in\ \href
  {https://doi.org/10.2514/6.2001-568} {\emph {\bibinfo {booktitle} {39th
  Aerospace Sciences Meeting and Exhibit}}}\ (\bibinfo  {publisher} {American
  Institute of Aeronautics and Astronautics},\ \bibinfo {year}
  {2001})\BibitemShut {NoStop}%
\bibitem [{\citenamefont {Deshpande}, \citenamefont {Eshpuniyani},\ and\
  \citenamefont {Sanghi}(2015)}]{Deshpande2015}%
  \BibitemOpen
  \bibfield  {author} {\bibinfo {author} {\bibfnamefont {V.}~\bibnamefont
  {Deshpande}}, \bibinfo {author} {\bibfnamefont {B.}~\bibnamefont
  {Eshpuniyani}},\ and\ \bibinfo {author} {\bibfnamefont {S.}~\bibnamefont
  {Sanghi}},\ }\bibfield  {title} {\enquote {\bibinfo {title} {Computational
  study of supersonic flow past non-stationary obstructions part-i - moving
  ramp},}\ }\href {https://doi.org/10.1504/pcfd.2015.069580} {\bibfield
  {journal} {\bibinfo  {journal} {Progress in Computational Fluid Dynamics, An
  International Journal}\ }\textbf {\bibinfo {volume} {15}},\ \bibinfo {pages}
  {144} (\bibinfo {year} {2015})}\BibitemShut {NoStop}%
\bibitem [{\citenamefont {Saad}\ \emph {et~al.}(2012)\citenamefont {Saad},
  \citenamefont {Zare-Behtash}, \citenamefont {Che-Idris},\ and\ \citenamefont
  {Kontis}}]{Saad2012}%
  \BibitemOpen
  \bibfield  {author} {\bibinfo {author} {\bibfnamefont {M.~R.}\ \bibnamefont
  {Saad}}, \bibinfo {author} {\bibfnamefont {H.}~\bibnamefont {Zare-Behtash}},
  \bibinfo {author} {\bibfnamefont {A.}~\bibnamefont {Che-Idris}},\ and\
  \bibinfo {author} {\bibfnamefont {K.}~\bibnamefont {Kontis}},\ }\bibfield
  {title} {\enquote {\bibinfo {title} {Micro-ramps for hypersonic flow
  control},}\ }\href {https://doi.org/10.3390/mi3020364} {\bibfield  {journal}
  {\bibinfo  {journal} {Micromachines}\ }\textbf {\bibinfo {volume} {3}},\
  \bibinfo {pages} {364--378} (\bibinfo {year} {2012})}\BibitemShut {NoStop}%
\bibitem [{\citenamefont {Stratford}(1956)}]{Stratford1956}%
  \BibitemOpen
  \bibfield  {author} {\bibinfo {author} {\bibfnamefont {B.~S.}\ \bibnamefont
  {Stratford}},\ }\bibfield  {title} {\enquote {\bibinfo {title} {Mixing and
  the jet flap},}\ }\href {https://doi.org/10.1017/s0001925900010155}
  {\bibfield  {journal} {\bibinfo  {journal} {Aeronautical Quarterly}\ }\textbf
  {\bibinfo {volume} {7}},\ \bibinfo {pages} {85--105} (\bibinfo {year}
  {1956})}\BibitemShut {NoStop}%
\bibitem [{\citenamefont {Raman}\ \emph {et~al.}(2020)\citenamefont {Raman},
  \citenamefont {Kexin}, \citenamefont {Kim}, \citenamefont {Suryan},\ and\
  \citenamefont {Kim}}]{Raman2020}%
  \BibitemOpen
  \bibfield  {author} {\bibinfo {author} {\bibfnamefont {S.~K.}\ \bibnamefont
  {Raman}}, \bibinfo {author} {\bibfnamefont {W.}~\bibnamefont {Kexin}},
  \bibinfo {author} {\bibfnamefont {T.~H.}\ \bibnamefont {Kim}}, \bibinfo
  {author} {\bibfnamefont {A.}~\bibnamefont {Suryan}},\ and\ \bibinfo {author}
  {\bibfnamefont {H.~D.}\ \bibnamefont {Kim}},\ }\bibfield  {title} {\enquote
  {\bibinfo {title} {Effects of flap on the reentry aerodynamics of a blunt
  cone in the supersonic flow},}\ }\href
  {https://doi.org/10.1016/j.ijmecsci.2019.105396} {\bibfield  {journal}
  {\bibinfo  {journal} {International Journal of Mechanical Sciences}\ }\textbf
  {\bibinfo {volume} {176}},\ \bibinfo {pages} {105396} (\bibinfo {year}
  {2020})}\BibitemShut {NoStop}%
\bibitem [{\citenamefont {Kandil}\ and\ \citenamefont
  {Salman}(1991)}]{KANDIL1991}%
  \BibitemOpen
  \bibfield  {author} {\bibinfo {author} {\bibfnamefont {O.}~\bibnamefont
  {Kandil}}\ and\ \bibinfo {author} {\bibfnamefont {A.}~\bibnamefont
  {Salman}},\ }\bibfield  {title} {\enquote {\bibinfo {title} {Unsteady
  supersonic flow around delta wings with symmetric and asymmetric flaps
  oscillation},}\ }in\ \href {https://doi.org/10.2514/6.1991-1105} {\emph
  {\bibinfo {booktitle} {32nd Structures, Structural Dynamics, and Materials
  Conference}}}\ (\bibinfo  {publisher} {American Institute of Aeronautics and
  Astronautics},\ \bibinfo {year} {1991})\BibitemShut {NoStop}%
\bibitem [{\citenamefont {Coon}\ and\ \citenamefont
  {Chapman}(1995)}]{Coon1995}%
  \BibitemOpen
  \bibfield  {author} {\bibinfo {author} {\bibfnamefont {M.~D.}\ \bibnamefont
  {Coon}}\ and\ \bibinfo {author} {\bibfnamefont {G.~T.}\ \bibnamefont
  {Chapman}},\ }\bibfield  {title} {\enquote {\bibinfo {title} {Experimental
  study of flow separation on an oscillating flap at mach 2.4},}\ }\href
  {https://doi.org/10.2514/3.12440} {\bibfield  {journal} {\bibinfo  {journal}
  {{AIAA} Journal}\ }\textbf {\bibinfo {volume} {33}},\ \bibinfo {pages}
  {282--288} (\bibinfo {year} {1995})}\BibitemShut {NoStop}%
\bibitem [{Flu(2013)}]{Fluent_2013}%
  \BibitemOpen
  \href@noop {} {\emph {\bibinfo {title} {ANSYS Fluent Theory Guide}}}\
  (\bibinfo {year} {2013})\BibitemShut {NoStop}%
\bibitem [{\citenamefont {Acquaye}(2016)}]{Acquayea}%
  \BibitemOpen
  \bibfield  {author} {\bibinfo {author} {\bibfnamefont {F.~K.}\ \bibnamefont
  {Acquaye}},\ }\bibfield  {title} {\enquote {\bibinfo {title} {Evaluation of
  various turbulence models for shock-wave boundary layer interaction flows},}\
  }\href {https://doi.org/10.7936/K7RJ4GS6} {\  (\bibinfo {year} {2016}),\
  10.7936/K7RJ4GS6}\BibitemShut {NoStop}%
\bibitem [{\citenamefont {Acquaye}\ \emph {et~al.}(2016)\citenamefont
  {Acquaye}, \citenamefont {Li}, \citenamefont {Wray},\ and\ \citenamefont
  {Agarwal}}]{Acquaye2016b}%
  \BibitemOpen
  \bibfield  {author} {\bibinfo {author} {\bibfnamefont {F.}~\bibnamefont
  {Acquaye}}, \bibinfo {author} {\bibfnamefont {J.}~\bibnamefont {Li}},
  \bibinfo {author} {\bibfnamefont {T.}~\bibnamefont {Wray}},\ and\ \bibinfo
  {author} {\bibfnamefont {R.~K.}\ \bibnamefont {Agarwal}},\ }\bibfield
  {title} {\enquote {\bibinfo {title} {Validation of the wray-agarwal
  turbulence model for shock-wave boundary layer interaction flows},}\ }in\
  \href {https://doi.org/10.2514/6.2016-3477} {\emph {\bibinfo {booktitle}
  {46th {AIAA} Fluid Dynamics Conference}}}\ (\bibinfo  {publisher} {American
  Institute of Aeronautics and Astronautics},\ \bibinfo {year}
  {2016})\BibitemShut {NoStop}%
\bibitem [{\citenamefont {Moura}\ and\ \citenamefont
  {Rosa}(2015)}]{AugustoFontanMoura2015}%
  \BibitemOpen
  \bibfield  {author} {\bibinfo {author} {\bibfnamefont {A.~F.}\ \bibnamefont
  {Moura}}\ and\ \bibinfo {author} {\bibfnamefont {M.~A.~P.}\ \bibnamefont
  {Rosa}},\ }\bibfield  {title} {\enquote {\bibinfo {title} {A numerical
  analysis of boundary layer/shock wave interactions in the compression ramps
  of scramjet intakes},}\ }in\ \href
  {https://doi.org/10.20906/cps/cob-2015-0209} {\emph {\bibinfo {booktitle}
  {23rd {ABCM} International Congress of Mechanical Engineering}}}\ (\bibinfo
  {publisher} {{ABCM} Brazilian Society of Mechanical Sciences and
  Engineering},\ \bibinfo {year} {2015})\BibitemShut {NoStop}%
\bibitem [{\citenamefont {Sun}\ \emph {et~al.}(2013)\citenamefont {Sun},
  \citenamefont {Gao}, \citenamefont {Zhong},\ and\ \citenamefont
  {Yang}}]{Sun2013}%
  \BibitemOpen
  \bibfield  {author} {\bibinfo {author} {\bibfnamefont {P.}~\bibnamefont
  {Sun}}, \bibinfo {author} {\bibfnamefont {H.}~\bibnamefont {Gao}}, \bibinfo
  {author} {\bibfnamefont {J.}~\bibnamefont {Zhong}},\ and\ \bibinfo {author}
  {\bibfnamefont {M.}~\bibnamefont {Yang}},\ }\bibfield  {title} {\enquote
  {\bibinfo {title} {Study on the influence of total pressure distortion on end
  wall flow field in a supersonic compressor},}\ }in\ \href
  {https://doi.org/10.1115/gt2013-94729} {\emph {\bibinfo {booktitle} {Volume
  6C: Turbomachinery}}}\ (\bibinfo  {publisher} {American Society of Mechanical
  Engineers},\ \bibinfo {year} {2013})\BibitemShut {NoStop}%
\bibitem [{Ice(2013)}]{Icem_2013}%
  \BibitemOpen
  \href@noop {} {\emph {\bibinfo {title} {ANSYS Meshing User's Guide}}}\
  (\bibinfo {year} {2013})\BibitemShut {NoStop}%
\bibitem [{\citenamefont {Chandola}, \citenamefont {Huang},\ and\ \citenamefont
  {Estruch-Samper}(2017)}]{Chandola2017}%
  \BibitemOpen
  \bibfield  {author} {\bibinfo {author} {\bibfnamefont {G.}~\bibnamefont
  {Chandola}}, \bibinfo {author} {\bibfnamefont {X.}~\bibnamefont {Huang}},\
  and\ \bibinfo {author} {\bibfnamefont {D.}~\bibnamefont {Estruch-Samper}},\
  }\bibfield  {title} {\enquote {\bibinfo {title} {Highly separated
  axisymmetric step shock-wave/turbulent-boundary-layer interaction},}\ }\href
  {https://doi.org/10.1017/jfm.2017.522} {\bibfield  {journal} {\bibinfo
  {journal} {Journal of Fluid Mechanics}\ }\textbf {\bibinfo {volume} {828}},\
  \bibinfo {pages} {236--270} (\bibinfo {year} {2017})}\BibitemShut {NoStop}%
\bibitem [{\citenamefont {Sekar}\ \emph {et~al.}(2020)\citenamefont {Sekar},
  \citenamefont {Karthick}, \citenamefont {Jegadheeswaran},\ and\ \citenamefont
  {Kannan}}]{Sekar2020}%
  \BibitemOpen
  \bibfield  {author} {\bibinfo {author} {\bibfnamefont {K.~R.}\ \bibnamefont
  {Sekar}}, \bibinfo {author} {\bibfnamefont {S.~K.}\ \bibnamefont {Karthick}},
  \bibinfo {author} {\bibfnamefont {S.}~\bibnamefont {Jegadheeswaran}},\ and\
  \bibinfo {author} {\bibfnamefont {R.}~\bibnamefont {Kannan}},\ }\bibfield
  {title} {\enquote {\bibinfo {title} {On the unsteady throttling dynamics and
  scaling analysis in a typical hypersonic inlet{\textendash}isolator flow},}\
  }\href {https://doi.org/10.1063/5.0032740} {\bibfield  {journal} {\bibinfo
  {journal} {Physics of Fluids}\ }\textbf {\bibinfo {volume} {32}},\ \bibinfo
  {pages} {126104} (\bibinfo {year} {2020})}\BibitemShut {NoStop}%
\bibitem [{\citenamefont {Karthick}(2021)}]{Karthick2021}%
  \BibitemOpen
  \bibfield  {author} {\bibinfo {author} {\bibfnamefont {S.~K.}\ \bibnamefont
  {Karthick}},\ }\bibfield  {title} {\enquote {\bibinfo {title} {Shock and
  shear layer interactions in a confined supersonic cavity flow},}\ }\href
  {https://doi.org/10.1063/5.0050822} {\bibfield  {journal} {\bibinfo
  {journal} {Physics of Fluids}\ }\textbf {\bibinfo {volume} {33}},\ \bibinfo
  {pages} {066102} (\bibinfo {year} {2021})}\BibitemShut {NoStop}%
\bibitem [{\citenamefont {Kenworthy}(1978)}]{Kenworthy1978}%
  \BibitemOpen
  \bibfield  {author} {\bibinfo {author} {\bibfnamefont {M.}~\bibnamefont
  {Kenworthy}},\ }\href@noop {} {\enquote {\bibinfo {title} {A study of
  unstable axisymmetric separation in high speed flows},}\ } (\bibinfo {year}
  {1978}),\ \bibinfo {note} {ph.D. Dissertation, Dept. of Aerospace and Ocean
  Engineering, Virginia Polytechnic Inst. and State Univ}\BibitemShut {NoStop}%
\bibitem [{\citenamefont {Karthick}, \citenamefont {Nanda},\ and\ \citenamefont
  {Cohen}(2022)}]{Karthick2022}%
  \BibitemOpen
  \bibfield  {author} {\bibinfo {author} {\bibfnamefont {S.~K.}\ \bibnamefont
  {Karthick}}, \bibinfo {author} {\bibfnamefont {S.~R.}\ \bibnamefont
  {Nanda}},\ and\ \bibinfo {author} {\bibfnamefont {J.}~\bibnamefont {Cohen}},\
  }\bibfield  {title} {\enquote {\bibinfo {title} {Hypersonic leading-edge
  unsteadiness},}\ }\href {https://arxiv.org/abs/2208.05682} {\bibfield
  {journal} {\bibinfo  {journal} {ArXiv (Fluid Dynamics)}\ } (\bibinfo {year}
  {2022})}\BibitemShut {NoStop}%
\bibitem [{\citenamefont {Feszty}, \citenamefont {Badcock},\ and\ \citenamefont
  {Richards}(2004)}]{Feszty2004a}%
  \BibitemOpen
  \bibfield  {author} {\bibinfo {author} {\bibfnamefont {D.}~\bibnamefont
  {Feszty}}, \bibinfo {author} {\bibfnamefont {K.~J.}\ \bibnamefont
  {Badcock}},\ and\ \bibinfo {author} {\bibfnamefont {B.~E.}\ \bibnamefont
  {Richards}},\ }\bibfield  {title} {\enquote {\bibinfo {title} {Driving
  mechanisms of high-speed unsteady spiked body flows, part i: Pulsation
  mode},}\ }\href {https://doi.org/10.2514/1.9034} {\bibfield  {journal}
  {\bibinfo  {journal} {{AIAA} Journal}\ }\textbf {\bibinfo {volume} {42}},\
  \bibinfo {pages} {95--106} (\bibinfo {year} {2004})}\BibitemShut {NoStop}%
\bibitem [{\citenamefont {Katzer}(1989)}]{Katzer1989}%
  \BibitemOpen
  \bibfield  {author} {\bibinfo {author} {\bibfnamefont {E.}~\bibnamefont
  {Katzer}},\ }\bibfield  {title} {\enquote {\bibinfo {title} {On the
  lengthscales of laminar shock/boundary-layer interaction},}\ }\href
  {https://doi.org/10.1017/s0022112089002375} {\bibfield  {journal} {\bibinfo
  {journal} {Journal of Fluid Mechanics}\ }\textbf {\bibinfo {volume} {206}},\
  \bibinfo {pages} {477--496} (\bibinfo {year} {1989})}\BibitemShut {NoStop}%
\bibitem [{\citenamefont {Mallinson}, \citenamefont {Gai},\ and\ \citenamefont
  {Mudford}(1996)}]{Mallinson1996}%
  \BibitemOpen
  \bibfield  {author} {\bibinfo {author} {\bibfnamefont {S.~G.}\ \bibnamefont
  {Mallinson}}, \bibinfo {author} {\bibfnamefont {S.~L.}\ \bibnamefont {Gai}},\
  and\ \bibinfo {author} {\bibfnamefont {N.~R.}\ \bibnamefont {Mudford}},\
  }\bibfield  {title} {\enquote {\bibinfo {title} {High-enthalpy, hypersonic
  compression corner flow},}\ }\href {https://doi.org/10.2514/3.13203}
  {\bibfield  {journal} {\bibinfo  {journal} {{AIAA} Journal}\ }\textbf
  {\bibinfo {volume} {34}},\ \bibinfo {pages} {1130--1137} (\bibinfo {year}
  {1996})}\BibitemShut {NoStop}%
\bibitem [{\citenamefont {Anderson}(2019)}]{Anderson2019}%
  \BibitemOpen
  \bibfield  {author} {\bibinfo {author} {\bibfnamefont {J.~D.}\ \bibnamefont
  {Anderson}},\ }\href {https://doi.org/10.2514/4.105142} {\emph {\bibinfo
  {title} {Hypersonic and High-Temperature Gas Dynamics, Third Edition}}}\
  (\bibinfo  {publisher} {American Institute of Aeronautics and Astronautics,
  Inc.},\ \bibinfo {year} {2019})\BibitemShut {NoStop}%
\bibitem [{\citenamefont {Taira}\ \emph {et~al.}(2017)\citenamefont {Taira},
  \citenamefont {Brunton}, \citenamefont {Dawson}, \citenamefont {Rowley},
  \citenamefont {Colonius}, \citenamefont {McKeon}, \citenamefont {Schmidt},
  \citenamefont {Gordeyev}, \citenamefont {Theofilis},\ and\ \citenamefont
  {Ukeiley}}]{Taira2017}%
  \BibitemOpen
  \bibfield  {author} {\bibinfo {author} {\bibfnamefont {K.}~\bibnamefont
  {Taira}}, \bibinfo {author} {\bibfnamefont {S.~L.}\ \bibnamefont {Brunton}},
  \bibinfo {author} {\bibfnamefont {S.~T.~M.}\ \bibnamefont {Dawson}}, \bibinfo
  {author} {\bibfnamefont {C.~W.}\ \bibnamefont {Rowley}}, \bibinfo {author}
  {\bibfnamefont {T.}~\bibnamefont {Colonius}}, \bibinfo {author}
  {\bibfnamefont {B.~J.}\ \bibnamefont {McKeon}}, \bibinfo {author}
  {\bibfnamefont {O.~T.}\ \bibnamefont {Schmidt}}, \bibinfo {author}
  {\bibfnamefont {S.}~\bibnamefont {Gordeyev}}, \bibinfo {author}
  {\bibfnamefont {V.}~\bibnamefont {Theofilis}},\ and\ \bibinfo {author}
  {\bibfnamefont {L.~S.}\ \bibnamefont {Ukeiley}},\ }\bibfield  {title}
  {\enquote {\bibinfo {title} {Modal analysis of fluid flows: An overview},}\
  }\href {https://doi.org/10.2514/1.j056060} {\bibfield  {journal} {\bibinfo
  {journal} {{AIAA} Journal}\ }\textbf {\bibinfo {volume} {55}},\ \bibinfo
  {pages} {4013--4041} (\bibinfo {year} {2017})}\BibitemShut {NoStop}%
\bibitem [{\citenamefont {Rao}, \citenamefont {Karthick},\ and\ \citenamefont
  {Anand}(2020)}]{Rao_2020}%
  \BibitemOpen
  \bibfield  {author} {\bibinfo {author} {\bibfnamefont {S.~M.~V.}\
  \bibnamefont {Rao}}, \bibinfo {author} {\bibfnamefont {S.~K.}\ \bibnamefont
  {Karthick}},\ and\ \bibinfo {author} {\bibfnamefont {A.}~\bibnamefont
  {Anand}},\ }\bibfield  {title} {\enquote {\bibinfo {title} {Elliptic
  supersonic jet morphology manipulation using sharp-tipped lobes},}\ }\href
  {https://doi.org/10.1063/5.0015035} {\bibfield  {journal} {\bibinfo
  {journal} {Physics of Fluids}\ }\textbf {\bibinfo {volume} {32}},\ \bibinfo
  {pages} {086107} (\bibinfo {year} {2020})}\BibitemShut {NoStop}%
\bibitem [{\citenamefont {George}\ \emph {et~al.}(2022)\citenamefont {George},
  \citenamefont {Karthick}, \citenamefont {Srikrishnan},\ and\ \citenamefont
  {Kannan}}]{George2022}%
  \BibitemOpen
  \bibfield  {author} {\bibinfo {author} {\bibfnamefont {L.~K.}\ \bibnamefont
  {George}}, \bibinfo {author} {\bibfnamefont {S.~K.}\ \bibnamefont
  {Karthick}}, \bibinfo {author} {\bibfnamefont {A.~R.}\ \bibnamefont
  {Srikrishnan}},\ and\ \bibinfo {author} {\bibfnamefont {R.}~\bibnamefont
  {Kannan}},\ }\bibfield  {title} {\enquote {\bibinfo {title} {Unsteady
  dynamics in a subsonic duct flow with a bluff body},}\ }\href
  {https://doi.org/10.1063/5.0097235} {\bibfield  {journal} {\bibinfo
  {journal} {Physics of Fluids}\ }\textbf {\bibinfo {volume} {34}},\ \bibinfo
  {pages} {067114} (\bibinfo {year} {2022})}\BibitemShut {NoStop}%
\bibitem [{\citenamefont {Estruch-Samper}\ and\ \citenamefont
  {Chandola}(2018)}]{EstruchSamper2018}%
  \BibitemOpen
  \bibfield  {author} {\bibinfo {author} {\bibfnamefont {D.}~\bibnamefont
  {Estruch-Samper}}\ and\ \bibinfo {author} {\bibfnamefont {G.}~\bibnamefont
  {Chandola}},\ }\bibfield  {title} {\enquote {\bibinfo {title} {Separated
  shear layer effect on shock-wave/turbulent-boundary-layer
  interaction~unsteadiness},}\ }\href {https://doi.org/10.1017/jfm.2018.350}
  {\bibfield  {journal} {\bibinfo  {journal} {Journal of Fluid Mechanics}\
  }\textbf {\bibinfo {volume} {848}},\ \bibinfo {pages} {154--192} (\bibinfo
  {year} {2018})}\BibitemShut {NoStop}%
\bibitem [{\citenamefont {Kutz}\ \emph {et~al.}(2016)\citenamefont {Kutz},
  \citenamefont {Brunton}, \citenamefont {Brunton},\ and\ \citenamefont
  {Proctor}}]{Kutz2016}%
  \BibitemOpen
  \bibfield  {author} {\bibinfo {author} {\bibfnamefont {J.~N.}\ \bibnamefont
  {Kutz}}, \bibinfo {author} {\bibfnamefont {S.~L.}\ \bibnamefont {Brunton}},
  \bibinfo {author} {\bibfnamefont {B.~W.}\ \bibnamefont {Brunton}},\ and\
  \bibinfo {author} {\bibfnamefont {J.~L.}\ \bibnamefont {Proctor}},\ }\href
  {https://doi.org/10.1137/1.9781611974508} {\emph {\bibinfo {title} {Dynamic
  Mode Decomposition}}}\ (\bibinfo  {publisher} {Society for Industrial and
  Applied Mathematics},\ \bibinfo {year} {2016})\BibitemShut {NoStop}%
\end{thebibliography}%
\
\onecolumngrid
\PRLsep
\end{document}